\documentclass[preprint,aps,prd,showpacs,nofootinbib,floatfix]{revtex4}
\usepackage{graphicx}
\usepackage{amsmath}
\usepackage{amssymb}
\usepackage{bm}
\usepackage{courier}
\usepackage{verbatim}
\usepackage{ulem}
\usepackage{soul}

\newcommand{\li}{\textrm{Li}_2}

\newcommand{\madgraph}{{\sc MadGraph5~}}
\newcommand{\aloha}{{\sc Aloha~}}

\newcommand{\stkout}[1]{\ifmmode\text{\sout{\ensuremath{#1}}}\else\sout{#1}\fi}

\begin{document}

\title{\mbox{}\\[10pt]
Gluon fragmentation into quarkonium at next-to-leading order\\
using FKS subtraction}

\author{Pierre  Artoisenet}
\email[]{pierreartoisenet@gmail.com}
\affiliation{Centre for Cosmology, Particle Physics and Phenomenology (CP3), Universit\'e catholique de Louvain, Chemin du Cyclotron 2, B-1348 Louvain-la-Neuve, Belgium}
\author{Eric Braaten}
\email[]{braaten@mps.ohio-state.edu}
\affiliation{Department of Physics, The Ohio State University, Columbus,
Ohio 43210, USA}
 \begin{abstract}
\noindent
We present the calculation at next-to-leading  order (NLO) in $\alpha_s$
of the fragmentation function of a gluon into heavy quarkonium 
in the color-octet spin-singlet S-wave channel.
To calculate the real NLO corrections, we adapt
a subtraction scheme introduced by Frixione, Kunszt, and Signer.
Ultraviolet and infrared divergences in the real NLO corrections
are calculated analytically by evaluating the phase-space integrals 
of the subtraction terms using dimensional regularization.
The subtracted phase-space integrals are then evaluated in 4 space-time dimensions. 
The divergences in the virtual NLO corrections are also calculated analytically.
After renormalization, all the divergences cancel. 
The NLO corrections significantly increase the fragmentation probability
for a gluon into the spin-singlet quarkonium states $\eta_c$ and $\eta_b$. \\~

\end{abstract}
\pacs{12.38.Bx,14.40.Pq,13.87.Fh}
\date{\today}
\preprint{CP3-16-45}
\maketitle

\section{Introduction}

The production of a hadron with large transverse momentum $p_T$ can be simplified
by using a factorization theorem for inclusive hadron production
at large $p_T$~\cite{Collins:1981uw}.  It states that 
the leading power in the expansion of the inclusive cross section in powers of $1/p_T$ 
can be expressed as a sum of perturbative QCD (pQCD) cross sections 
for producing a parton convolved with {\it fragmentation functions}:
\begin{equation}
d \sigma[H + X] = \sum_i d\hat\sigma[i+X]\otimes D_{i\to H}(z).
\label{LP-frag}
\end{equation}
The sum extends over the types of partons $i$ (gluons, quarks, and antiquarks). 
The symbol ``$\otimes$'' in Eq.~\eqref{LP-frag} represents a convolution integral over 
the longitudinal momentum fraction $z$ of the hadron $H$ relative to the parton.
The pQCD cross section $d\hat\sigma$  
 for producing the 
parton $i$  can be expanded in powers of $\alpha_s(p_T)$,   
 and it includes convolutions with parton distributions if the colliding particles are hadrons. 
The nonperturbative factor $D_{i\to H}(z)$   is a {\it fragmentation function} that gives
the probability distribution for $z$. 
We refer to Eq.~\eqref{LP-frag} as the  {\it leading-power (LP) factorization formula}.
It was derived by Collins and Soper in 1981
for the case of a light hadron $H$ at a transverse momentum satisfying 
$p_T \gg \Lambda_{\rm QCD}$~\cite{Collins:1981uw}.

The LP factorization formula in Eq.~\eqref{LP-frag} applies equally well to 
heavy quarkonium at a transverse momentum satisfying $p_T \gg m$,
where $m$ is the mass of the heavy quark.
A proof of this  factorization theorem that deals 
with issues specific to heavy quarkonium production
was first sketched by Nayak, Qiu, and Sterman in 2005 \cite{Nayak:2005rt}. 
The LP factorization formula gives the leading power in the expansion in powers of $m/p_T$.
In the case of cross sections summed over quarkonium spins,
the corrections are suppressed by $m^2/p_T^2$. 
The LP factorization formula has limited predictive power,
because the nonperturbative factors 
$D_{i\to H}(z)$ are functions of $z$ 
that must be determined from experiment.

In 1994, Bodwin, Braaten, and Lepage proposed the {\it NRQCD factorization formula}
for cross sections for heavy quarkonium production \cite{Bodwin:1994jh}.
It uses an effective field theory called nonrelativistic QCD
to separate momentum scales of order $m$ and larger 
from momentum scales of order $mv$ and smaller,
where $v$ is the typical relative velocity of the $Q$ or $\bar Q$ in the quarkonium.
The theoretical status of the NRQCD factorization conjecture 
is discussed in Ref.~\cite{Bodwin:2013nua}.
The predictive power of the LP factorization formula for heavy quarkonium in Eq.~\eqref{LP-frag} 
can be increased by applying the 
NRQCD factorization formula to the fragmentation functions,
reducing these nonperturbative functions of $z$ to multiplicative constants.
The fragmentation function for the parton $i$ to produce 
the quarkonium $H$ is expressed as a sum of  pQCD fragmentation functions
multiplied by NRQCD matrix elements:
\begin{equation}
D_{i\to H}(z) = \sum_n \hat D_{i \to Q \bar Q[n]}(z)~
\langle  {\cal O}_n^{H} \rangle.
\label{D-fact}
\end{equation}
The sum extends over the color and angular-momentum 
channels $n$ of a nonrelativistic $Q \bar Q$ pair. 
The pQCD fragmentation  function $\hat D_{i \to Q \bar Q[n]}(z)$
for producing the $Q \bar Q$ pair in the 
channel $n$ can be expanded in powers of $\alpha_s(m)$.
The NRQCD matrix element
$\langle{\cal O}_n^{H} \rangle$ is  proportional to the probability for a
$Q\bar Q$ pair created in the channel $n$ to evolve into a final state 
that includes the quarkonium $H$.
The nonperturbative constants $\langle{\cal O}_n^{H} \rangle$ 
scale as definite powers of  $v$ \cite{Bodwin:1994jh}.

The factorization theorem for inclusive production of heavy quarkonium at large $p_T$
has been extended to the next-to-leading power (NLP) of $m^2/p_T^2$. 
The {\it NLP factorization theorem} was proven diagrammatically by 
Kang, Qiu, and Sterman \cite{Kang:2011mg,Kang:2014tta,Kang:2014pya},
and it was derived using soft collinear effective theory by 
Fleming, Leibovich, Mehen, and Rothstein  \cite{Fleming:2012wy,Fleming:2013qu}.
In addition to  corrections to the terms in the LP factorization formula,
the {\it NLP  factorization formula} has additional terms suppressed by $m^2/p_T^2$  that are expressed as
a sum of pQCD cross sections for producing a pair of collinear partons
convolved with {\it double-parton fragmentation functions}.
In the case of cross sections summed over quarkonium spins,
the corrections to the NLP fragmentation formula are suppressed by a power of $m^4/p_T^4$.
The predictive power of the NLP fragmentation formula 
can be dramatically increased by applying the NRQCD factorization formula 
to the double-parton fragmentation functions,
reducing these nonperturbative functions to multiplicative constants.

The NLP/NRQCD factorization formula 
opens the door to dramatic improvements in the accuracy of 
theoretical predictions for quarkonium production at large $p_T$.  
The factorization formula can be expressed as a triple expansion 
in powers of $\alpha_s$, $v$, and $m/p_T$.
NLP factorization incorporates subleading powers of $m/p_T$.
The NRQCD expansion includes subleading powers of $v$.
Accurate predictions also require calculating all the pQCD factors 
to next-to-leading order (NLO) in $\alpha_s$.
The pQCD factors are the cross sections for producing single partons, 
the  cross sections for producing collinear parton pairs, 
the single-parton fragmentation functions for producing $Q\bar Q$ pairs,
the double-parton fragmentation functions for producing $Q\bar Q$ pairs,
and the evolution kernels for both sets of fragmentation functions.

The first fragmentation function for quarkonium production to be calculated 
to next-to-leading order (NLO) in $\alpha_s$ was for gluon fragmentation into $Q \bar Q$ 
in the color-octet $^3S_1$ channel  \cite{Braaten:2000pc,Ma:2013yla}. 
This calculation is particularly simple, 
because the LO fragmentation function is proportional to $\delta(1-z)$.
The first NLO calculation of a fragmentation function that at LO 
is a nontrivial function of $z$ was that for gluon fragmentation into $Q \bar Q$ 
in the color-singlet $^1S_0$ channel \cite{Artoisenet:2014lpa}.
It would be useful to have all the phenomenologically relevant fragmentation functions 
calculated to NLO in $\alpha_s$.

In NLO QCD calculations, the most challenging step is often the 
calculation of the phase-space integrals from real-gluon emission.
A strategy that is often effective is to design subtractions that cancel 
the infrared divergences in the phase-space integrals,
calculate the integrals of the subtraction terms analytically,
and then calculate the subtracted phase-space integrals numerically.
In the case of fragmentation functions, the phase-space integrals also have ultraviolet divergences.
It is therefore necessary to design subtractions that also cancel these ultraviolet divergences.
The NLO calculation of Ref.~\cite{Artoisenet:2014lpa} was carried out using 
a subtraction procedure for fragmentation functions
that was adapted from the dipole subtraction procedure 
for parton cross sections introduced by Catani and Seymour \cite{Catani:1996vz}.
An alternative subtraction scheme for parton cross sections that has some advantages
was introduced by Frixione, Kunszt, and Signer (FKS)  \cite{Frixione:1995ms}.
The FKS subtraction method has been used in the automation of next-to-leading order 
computations of parton cross sections in QCD \cite{Frederix:2009yq}.

In this paper, we adapt the FKS subtraction method to the NLO calculation 
of fragmentation functions.
We illustrate the method by applying it to gluon fragmentation 
into a $Q \bar Q$ pair in the color-octet $^1S_0$ channel.
This  fragmentation function is of phenomenological importance
for the production of $J^{PC} = 0^{-+}$ quarkonium states, such as the $\eta_b$ or $\eta_c$.
The color-singlet $^1S_0$ channel is leading order in $v$.
The color-octet $^1S_0$ channel is one of three color-octet channels suppressed by only $v^4$.
In the case of production of the $\eta_c$ at large $p_T$ at the Large Hadron Collider, 
the color-octet $^1S_0$ channel is numerically the most important 
of the three \cite{Butenschoen:2014dra,Han:2014jya}.

The outline of our paper is as follows.
In Section~\ref{sec:LO}, we present the LO fragmentation function
for gluon fragmentation into $Q \bar Q$ in the color-octet $^1S_0$ channel
and define some quantities that are useful in the NLO calculation.
In Section~\ref{sec:NLOreal}, we introduce the FKS subtraction terms that cancel all the
ultraviolet and infrared divergences in the real NLO corrections. 
The subtracted phase-space integrals are calculated in 4 dimensions,
and their insensitivity to the cut parameters in the FKS subtractions is verified.
In Section~\ref{sec:NLOvirtual}, we present analytic results for the 
ultraviolet and infrared divergences from loop integrals in the virtual NLO corrections.
In Section~\ref{sec:NLOrenorm}, we verify that all the divergences 
from phase-space integrals and from loop integrals are canceled by renormalization 
of the parameters of QCD and by renormalization of the 
operator whose matrix element defines the fragmentation function.
Some numerical illustrations of our results are
presented in Section~\ref{sec:numres}.
Our results are summarized in Section~\ref{sec:summary}.
In Appendix~\ref{sec:PhaseSpace}, we derive parametrizations of 
massless two-parton phase-space integrals that are used
to integrate the subtractions terms for the real NLO corrections.
In Appendix~\ref{sec:integralpoles},
we calculate the pole terms in the dimensionally regularized phase-space integrals 
of the subtraction terms analytically.
In Appendix~\ref{app:madgraph},
we show how \madgraph \cite{Alwall:2011uj} can be used to generate helicity amplitudes 
for cut diagrams with a heavy-quark pair and two light partons in the final state.


\section{Leading-order fragmentation function}
\label{sec:LO}

In this section, we present the perturbative fragmentation function for 
$g \to Q \bar Q$, with the $Q \bar Q$ pair in a color-octet $^1S_0$ state,
at leading order in $\alpha_s$.
We also introduce some related expressions that are useful in the calculation 
of the real radiative corrections at next-to-leading order in $\alpha_s$.

\subsection{Feynman rules}

\begin{figure}
\center
\includegraphics[scale=0.4]{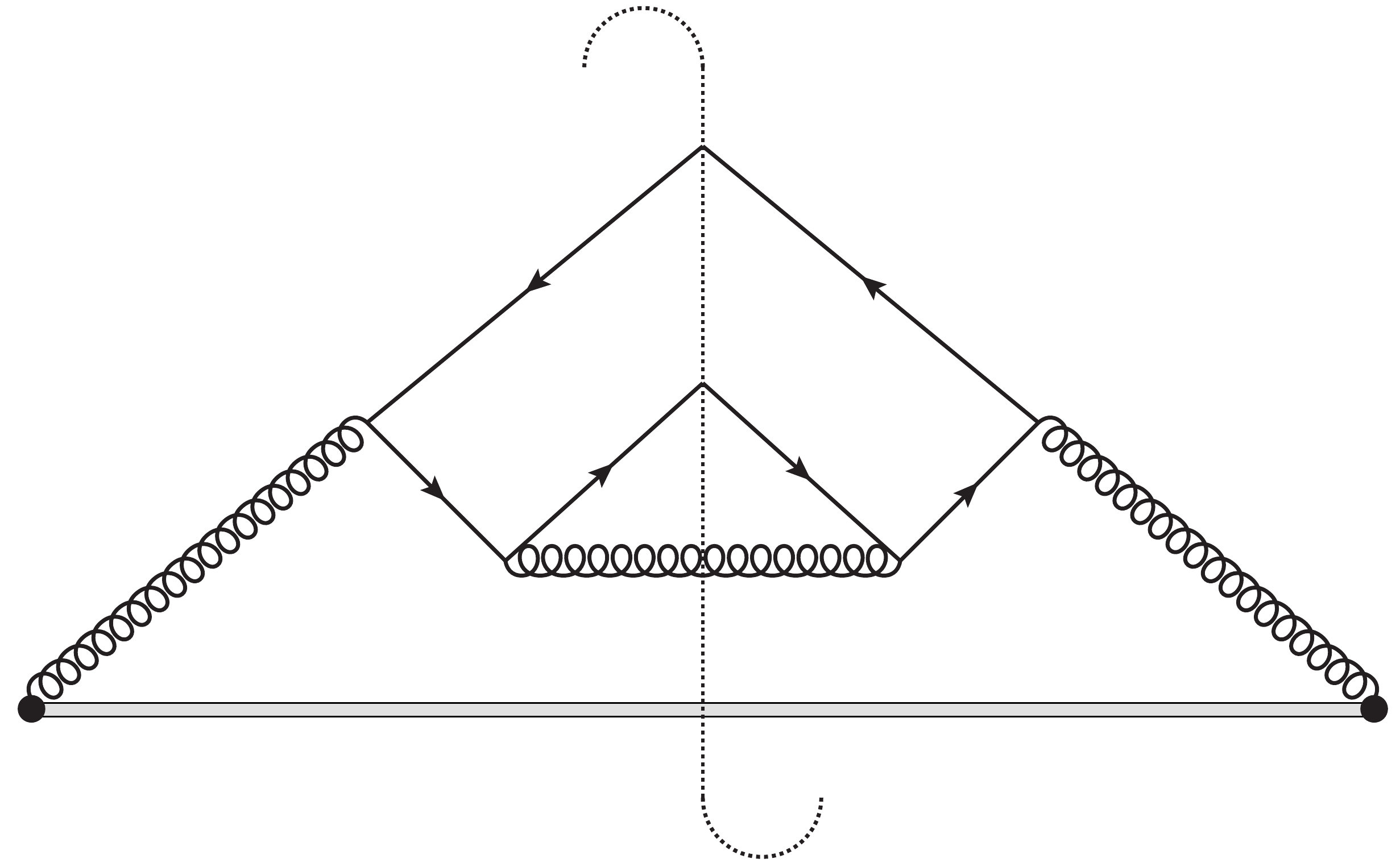}
\caption{
One of the 4 cut diagrams for gluon fragmentation into a 
 color-octet $^1S_0$ $Q \bar Q$ pair
at leading order in $\alpha_s$.
The eikonal line is represented by a double solid line. 
The dotted line is the cut.
The other 3 cut diagrams at leading order 
are obtained by interchanging the two gluon vertices on each side of the cut.}
\label{fig:cutdiagram}
\end{figure}

Gluon fragmentation functions can be calculated using  Feynman rules
derived by Collins and Soper in 1981~\cite{Collins:1981uw}.
The fragmentation function is expressed as the sum of all possible cut diagrams
with an eikonal line that
extends from a gluon-field-strength operator on the left side of the cut
to a gluon-field-strength operator on the right side.
Single virtual gluon lines are attached to the operators on the left side 
and on the right side. The two virtual gluon lines from the operators are 
connected to each other by gluon and quark lines 
produced by QCD interactions, with 
possibly additional gluon lines attached to the eikonal line.
The cut passes through the eikonal line, the line for the particle
into which the gluon is fragmenting, and possibly additional gluon and quark lines.
An example of a cut diagram with the cut passing through the lines of
a heavy quark and antiquark 
and an additional gluon is shown in Figure~\ref{fig:cutdiagram}.

The Feynman rules for the cut diagrams are relatively simple \cite{Collins:1981uw}:
they are summarized in Ref.~\cite{Artoisenet:2014lpa}.
The 4-momentum $K$ of the gluon that is fragmenting
enters the diagram through the operator vertex on the left side of the eikonal line
and it exits through the operator on the right side.
Some of that momentum flows through the single  gluon line attached to the operator
and the remainder flows through the eikonal line.
The particle into which the gluon is fragmenting has a specified 4-momentum.
In the case of fragmentation of a gluon into 
a $Q \bar Q$ pair with zero relative momentum,
it is convenient to express the 4-momentum of the $Q \bar Q$ pair as $2p$.
The longitudinal momentum fraction $z$ of the $Q \bar Q$ pair is: 
\begin{equation}
\label{eq:defz}
 z =\frac{(2p).n}{K.n}.
\end{equation}
The operator at the left end of the eikonal line is labelled by a Lorentz index
$\mu$ and a color index $a$.
The Feynman rule for the operator vertex is 
\begin{equation}
\label{eq:opvertex}
-i \left( K.n g^{\mu \lambda} - q^\mu n^\lambda \right) \delta^{ac},
\end{equation}
where $q$, $\lambda$ and $c$ are the 4-momentum, Lorentz index, 
and color index of the gluon line attached to the vertex.
The operator at the right end of the eikonal line is labelled by a Lorentz index
$\nu$ and a color index $b$.  
The fragmentation function is the sum of all cut diagrams 
contracted with $-g_{\mu\nu}$ and $\delta_{ab}$
and multiplied by the {\it Collins-Soper prefactor}~\cite{Collins:1981uw}:
\begin{equation}
\label{eq:overalfac}
N_{\rm CS} = 
\frac{1}{(N_c^2-1)(2-2\epsilon)} \frac{z^{1-2\epsilon}}{2\pi K.n},
\end{equation} 
where $N_c=3$ is the number of colors of a quark 
and $D=4 - 2 \epsilon$ is the space-time dimension.

\subsection{LO Fragmentation Function}

The NRQCD factorization formula in Eq.~\eqref{D-fact}
for the fragmentation function $D_{i\to H}(z)$ for producing the quarkonium state $H$
expresses it as a sum of pQCD fragmentation functions 
multiplied by NRQCD matrix elements.
The NRQCD matrix elements $\langle  {\cal O}_n^{H} \rangle$ scale as definite powers of the 
relative velocity $v$ of the heavy quark in the quarkonium.
The pQCD fragmentation functions  $\hat D_{i \to Q \bar Q[n]}(z)$
can be calculated as power series in $\alpha_s(m)$, 
where $m$ is the heavy-quark mass.  
The fragmentation function for a gluon into a $0^{-+}$ quarkonium state  $\eta_Q$,
including the color-singlet $^1S_0$ and color-octet $^1S_0$ terms explicitly, has the form
\begin{eqnarray}
\label{eq:DNLOeta}
D_{g \to \eta_Q}(z) &=& 
\langle{\cal O}_1(^1S_0)\rangle^{\eta_Q} 
\left[ \alpha_s^2 \, D_{\rm LO}^{(1)}(z) + \alpha_s^3 \, D_{\rm NLO}^{(1)}(z) + \ldots \right]
\nonumber \\
&&
+\, \langle{\cal O}_8(^1S_0)\rangle^{\eta_Q} 
\left[ \alpha_s^2 \, D_{\rm LO}^{(8)}(z) + \alpha_s^3 \, D_{\rm NLO}^{(8)}(z) + \ldots \right] + \ldots.
\end{eqnarray}
The color-singlet NRQCD matrix element $\langle{\cal O}_1(^1S_0)\rangle^{\eta_Q}$ 
is leading order in $v$, and it can be expressed in terms of the wavefunction at the origin for the $\eta_Q$.
The matrix element $\langle{\cal O}_8(^1S_0)\rangle^{\eta_Q}$ is one of three
color-octet matrix elements that are suppressed by only $v^4$.
It is related by heavy-quark spin symmetry to the NRQCD matrix element
$\langle{\cal O}_8(^3S_1)\rangle^{\psi_Q}$ for a $1^{--}$ quarkonium state  $\psi_Q$,
which can be determined phenomenologically 
by fitting cross sections for production of $\psi_Q$.
In Eq.~\eqref{eq:DNLOeta}, the PQCD fragmentation functions multiplying 
$\langle{\cal O}_1(^1S_0)\rangle^{\eta_Q}$ and $\langle{\cal O}_8(^1S_0)\rangle^{\eta_Q}$
have been expanded to next-to-leading order (NLO) in $\alpha_s$.
The color-singlet $^1S_0$ fragmentation function 
was calculated at leading order (LO) in $\alpha_s$ by Braaten and Yuan in 1993 \cite{Braaten:1993rw}:
\begin{equation}
\label{eq:dLO1-z}
D_{\rm LO}^{(1)}(z) = 
\frac{1}{4 N_c^2 m^3}
\left[2(1-z) \log(1-z) +3z-2z^2 \right].
\end{equation}
The NLO term $D_{\rm NLO}^{(1)}(z)$  in 
this fragmentation function was calculated in Ref.~\cite{Artoisenet:2014lpa}.
Our goal is to calculate the color-octet $^1S_0$ fragmentation function to NLO.

The pQCD fragmentation functions in Eq.~\eqref{eq:DNLOeta} can be determined 
from perturbative QCD calculations of the fragmentation function for producing a $Q \bar Q$ pair.
Fragmentation functions for producing $Q \bar Q$ in a $^1S_0$ state can be determined most easily 
by taking the $Q \bar Q$ pair to be in a spin-singlet state with zero relative momentum.
To determine the color-singlet fragmentation function, 
the $Q$ and $\bar Q$ are projected onto the color-singlet state $Q\bar Q_1$ 
by contracting their color indices $i$ and $j$ 
with $\delta_{ij}/\sqrt{N_c}$.
To determine the color-octet fragmentation function,
the $Q$ and $\bar Q$ are projected onto the color-octet state $Q\bar Q_8$ with color index $a$
by contracting their color indices $i$ and $j$ with $\sqrt{2}\, T^a_{ij}$.
Given the normalizations of the NRQCD operators defined in Ref.~\cite{Bodwin:1994jh}, 
the perturbative approximations to the NRQCD matrix elements are
\begin{subequations}
\begin{eqnarray}
\label{eq:<O1>QQbar}
\langle{\cal O}_1(^1S_0)\rangle^{Q \bar Q_1} &=& 2 N_c,
\\
\langle{\cal O}_8(^1S_0)\rangle^{Q \bar Q_8} &=& N_c^2-1,
\label{eq:<O8>QQbar}
\end{eqnarray}
\label{eq:<O>QQbar}%
\end{subequations}
where $N_c=3$.
On the left side of Eq.~\eqref{eq:<O8>QQbar}, 
there is an implied sum over the $N_c^2-1$ colors of $Q \bar Q_8$.
If dimensional regularization is used to regularize ultraviolet and infrared 
divergences, these matrix elements have no NLO corrections.
By dividing the perturbatively calculated fragmentation function 
$D_{g \to Q \bar Q_8}(z)$ by the perturbative matrix element in Eq.~\eqref{eq:<O8>QQbar}, 
we obtain the fragmentation function multiplying
$\langle{\cal O}_8(^1S_0)\rangle^{\eta_Q}$ in Eq.~\eqref{eq:DNLOeta}.

\subsection{Born fragmentation function}

The fragmentation function for $g \to Q \bar Q$
can be calculated perturbatively from the cut diagrams 
in which the cut lines include $Q$ and $\bar Q$.
At leading order in $\alpha_s$, the cut diagrams are the diagram in Figure~\ref{fig:cutdiagram}
and three other diagrams  obtained 
by interchanging the two gluon vertices on the left side of the cut
and interchanging the two gluon vertices on the right side of the cut.
The final-state $Q$ and $\bar Q$ are on-shell with equal momenta $p$
and total longitudinal momentum fraction $z$.
The final-state gluon is on-shell with a momentum $q$
whose phase space must be integrated over.
The cut through the eikonal line gives a factor of $2 \pi \delta(K.n-(2p+q).n)$.

The amplitude corresponding to the sum of the two diagrams
on the left side of the cut in Figure~\ref{fig:cutdiagram} 
is given in Eq.~(2.10) of Ref.~\cite{Artoisenet:2014lpa}.
The $Q \bar Q$ pair is projected onto the color-singlet state $Q \bar Q_1$ 
in Eq.~(2.11) of Ref.~\cite{Artoisenet:2014lpa}.
If the amplitude is instead projected onto the color-octet state $Q\bar Q_8$ with color index $b$, 
the net effect is the replacement
\begin{equation}
\frac{1}{\sqrt{N_c}} \delta^{ac} \longrightarrow
 \sqrt{2} \, {\rm Tr}\big(\{T^a,T^c\} T^b\big) = \frac{1}{\sqrt{2}} d^{acb},
\label{eq:colorsub-amp}
\end{equation}
where $a$ is the color index of the gluon-field-strength operator 
and $c$ is the color index of the final-state gluon.
After squaring the amplitudes, 
contracting the color indices for the operators,
and summing over the colors of the gluon and the $Q\bar Q$ pair,
the net effect is the substitution
\begin{equation}
\frac{N_c^2-1}{N_c} \longrightarrow
 \frac{(N_c^2-1)(N_c^2-4)}{2N_c}.
\label{eq:colorsub}
\end{equation}
This procedure projects the $Q \bar Q$ pair onto the color-octet states $Q \bar Q_8$.
Thus the LO fragmentation function $D_{g \to Q \bar Q_8}(z)$
differs from the LO fragmentation function $D_{g \to Q \bar Q_1}(z)$
just by the multiplicative color factor $(N_c^2-4)/2$.
The LO fragmentation function $D_{\rm LO}^{(8)}(z)$
in the second term of the factorization formula in Eq.~\eqref{eq:DNLOeta}
differs from the LO fragmentation function $D_{\rm LO}^{(1)}(z)$
in the first term by the product of that color factor 
and the ratio $2N_c/(N_c^2-1)$ of the perturbative NRQCD matrix elements in
Eqs.~\eqref{eq:<O1>QQbar} and \eqref{eq:<O8>QQbar}.
Thus the LO color-octet $^1S_0$ fragmentation function is
\begin{equation}
\label{eq:dLO8-z}
D_{\rm LO}^{(8)}(z) = \frac{ N_c^2-4}{4 N_c(N_c^2-1)m^3} 
\left[2(1-z) \log(1-z) +3z-2z^2 \right].
\end{equation}

In the calculation of the NLO fragmentation function for $g \to Q \bar Q_8$,
it is useful to have the LO  fragmentation function for $g \to Q \bar Q_8$ expressed as an integral
over the gluon phase space in $D=4-2 \epsilon$ dimensions:
\begin{equation}
\label{eq:fragBorn}
D_1(z) = N_{\rm CS}    \int d\phi_{\rm Born} (p,q)
 \mathcal{A}_{\rm Born}(p,q),
\end{equation}
where $N_{\rm CS}$ is the Collins-Soper prefactor in Eq.~\eqref{eq:overalfac}.
The  {\it Born phase-space measure} $d\phi_{\rm Born}$ is the product
of the differential phase space for the final-state gluon
of momentum $q$ and a factor $2\pi \delta(K.n-(2p+q).n)$ from the cut through the eikonal line.
It can be reduced to a single differential in the invariant mass $s$ of the $Q \bar Q g$ system:
\begin{equation}
 d\phi_{\rm Born}(p,q) = 
 \frac{z^{-1+\epsilon} (1-z)^{-\epsilon}}{2 (4\pi)^{1-\epsilon} \Gamma(1-\epsilon) (2p+q).n}
 \left( s-\frac{4m^2}{z} \right)^{-\epsilon} ds,
 \label{eq:BornPSpq}
\end{equation}
where $s$ and $z$ expressed as functions of $p$ and $q$ are
\begin{subequations}
\begin{eqnarray}
s &=& (2p+q)^2,
\label{eq:s-2p+q}
\\
z &=& \frac{(2p).n}{(2p+q).n}.
\label{eq:z-pq}
\end{eqnarray}
\label{eq:z,s-pq}%
\end{subequations}
There is an implied Heavyside theta function that imposes the constraint $s>4m^2/z$.
The {\it Born squared amplitude} $\mathcal{A}_{\rm Born}$ is 
obtained by multiplying the right side of Eq.~(2.15) in Ref.~\cite{Artoisenet:2014lpa} 
by the color factor $(N_c^2-4)/2$:
\begin{eqnarray}
\mathcal{A}_{\rm Born}(p,q) &=&  
\frac{2 (1-2 \epsilon) (N_c^2-1)(N_c^2-4) g_s^4 [(2p+q).n]^2}{N_c m s^2 (s-4m^2)^2}
\nonumber\\ 
&& \times
\left[ (1-2z+2z^2 -\epsilon)s^2   - 8 (z - \epsilon) m^2 s + 16(1-\epsilon) m^4 \right],
\label{eq:ABorn-pq}
\end{eqnarray}
where $z$ and $s$ are expressed as functions of $p$ and $q$ in 
Eqs.~\eqref{eq:z,s-pq}.
The factors of $K.n = (2p+q).n$ cancel between $N_{\rm CS}$, $d\phi_{\rm Born}$,
and $\mathcal{A}_{\rm Born}$ 
in Eqs.~\eqref{eq:overalfac}, \eqref{eq:BornPSpq}, and \eqref{eq:ABorn-pq}.

The LO fragmentation function for $g \to Q \bar Q_8$ in $D$ dimensions is
obtained by inserting the three factors 
in Eqs.~\eqref{eq:overalfac}, \eqref{eq:BornPSpq}, and \eqref{eq:ABorn-pq} into Eq.~\eqref{eq:fragBorn}.
Setting $\epsilon =0$ and integrating over $s$, we obtain the final result for the LO fragmentation function 
for $g \to Q \bar Q_8$ in 4 dimensions:
\begin{equation}
\label{eq:dLO}
D^{\rm (LO)}_{g \rightarrow Q\bar Q_8} (z) = 
\frac{ (N_c^2-4) \alpha_s^2}{4 N_c m^3}
\left[2(1-z) \log(1-z) +3z-2z^2 \right].
\end{equation}
Dividing by the perturbative NRQCD matrix element in Eq.~\eqref{eq:<O8>QQbar},
we obtain the LO fragmentation function $D_{\rm LO}^{(8)}(z)$ in Eq.~\eqref{eq:dLO8-z}
multiplied by $\alpha_s^2$.

\subsection{Born tensors}
\label{sec:Borntensors}

To facilitate the calculation of the real NLO corrections
to the fragmentation function, it is convenient to
generalize the  integration measure $d\phi_{\rm Born}(p,q)$ for the LO
fragmentation function in Eq.~\eqref{eq:BornPSpq}
by allowing $q$ to be a more general light-like 4-vector.
It could be the momentum $q_1$ or $q_2$ of a massless final-state parton,
or it could be another  light-like 4-vector constructed from $q_1$ and $q_2$.
The variables $s=(2p+q)^2$ and $z=(2p.n)/(2p+q).n$ defined in Eqs.~\eqref{eq:z,s-pq}
can be regarded as functions of this more general light-like 4-vector $q$.
The longitudinal momentum of the fragmenting gluon is $(2p+q_1+q_2).n$.
The Collins-Soper prefactor in Eq.~\eqref{eq:overalfac} 
can be generalized to a function of $p$ and $q$:
\begin{equation}
\label{eq:NBorn}
N_{\rm Born}(p,q) = 
\frac{1}{(N_c^2-1)(2-2\epsilon)}  \frac{1}{2\pi (2p+q_1+q_2).n}   
\left( \frac{2p.n}{(2p+q).n} \right)^{1-2\epsilon}.
\end{equation}
The Born phase-space measure in Eq.~\eqref{eq:BornPSpq},
with the factor of $1/(2p+q).n$ replaced by $1/(2p+q_1+q_2).n$
and with its coefficient expressed as a function of $s$ and $z$,
will be denoted by $d\phi_{\rm Born}(s,z)$:
\begin{equation}
 d\phi_{\rm Born}(s,z) = 
 \frac{z^{-1+\epsilon} (1-z)^{-\epsilon}}{2 (4\pi)^{1-\epsilon} \Gamma(1-\epsilon) (2p+q_1+q_2).n}
 \left( s-\frac{4m^2}{z} \right)^{-\epsilon} ds.
 \label{eq:BornPSsz}
\end{equation}
Similarly, the Born squared amplitude $\mathcal{A}_{\rm Born}(p,q)$
in Eq.~\eqref{eq:ABorn-pq}, with the factor of $[(2p+q).n]^2$ replaced by $[(2p+q_1+q_2).n]^2$
and with its coefficient expressed as a function of $s$ and $z$,
will be denoted by $\mathcal{A}_{\rm Born}(s,z)$: 
\begin{eqnarray}
\mathcal{A}_{\rm Born}(s,z) &=&  
\frac{2 (1-2 \epsilon) (N_c^2-1)(N_c^2-4) g_s^4 [(2p+q_1+q_2).n]^2}{N_c m s^2 (s-4m^2)^2}
\nonumber\\ 
&& \times
\left[ (1-2z+2z^2 -\epsilon)s^2   - 8 (z - \epsilon) m^2 s + 16(1-\epsilon) m^4 \right].
\label{eq:ABorn-sz}
\end{eqnarray}
The product of $N_{\rm Born}$, $d\phi_{\rm Born}$, and $\mathcal{A}_{\rm Born}$ 
in Eqs.~\eqref{eq:NBorn},  
\eqref{eq:BornPSsz}, and \eqref{eq:ABorn-sz}
depends only on $s$ and $z$. 
We introduce a more concise notation for this product:
\begin{eqnarray}
\label{eq:NphiABorn}
N d\phi  \mathcal{A}_{\rm Born}(s,z)
&=& \frac{(1-2 \epsilon) (N_c^2-4) (4 \pi)^\epsilon\alpha_s^2}{\Gamma(2-\epsilon) N_c m}
[z (1-z)]^{-\epsilon}  \frac{(s-4m^2/z)^{-\epsilon}}{s^2}
\nonumber \\ 
&& \times 
\left[ 1 - \epsilon - 2 z(1-z) \frac{s (s-4m^2/z)} {(s-4m^2)^2} \right] 
\theta(s-4m^2/z) ds.
\end{eqnarray}
If this measure is multiplied by a function of $s$ 
and integrated over $s$ from $4m^2/z$ to $\infty$, it defines a function of $z$.
If the weight function is simply 1, the integral is
the LO fragmentation function for $g \to Q \bar Q_8$ in $D$ dimensions defined in Eq.~\eqref{eq:fragBorn}:
\begin{eqnarray}
\label{eq:D1-z}
D_1(z)
&=& \frac{(1-2 \epsilon)(N_c^2-4)(4\pi)^\epsilon \alpha_s^2}{\Gamma(2-\epsilon) N_c m}
[z (1-z)]^{-\epsilon} 
\nonumber \\ 
&& \times 
\int_{4m^2/z}^\infty \!\!\!\!\!ds
\frac{(s-4m^2/z)^{-\epsilon}}{s^2}
\left[ 1 - \epsilon - 2 z(1-z) \frac{s (s-4m^2/z)} {(s-4m^2)^2} \right].
\end{eqnarray}

In the calculation of the real NLO corrections to the fragmentation function,
it is convenient to have expressions for the Born squared amplitude
with a pair of uncontracted Lorentz indices.
They will be used to construct subtraction terms 
that cancel the ultraviolet and infrared divergences in the real NLO corrections
point-by-point in the phase space. 
There are two useful choices 
for the uncontracted indices $\mu$ and $\nu$.  One choice is the  Lorentz indices
associated with the ends of the eikonal line.  The other choice is the 
Lorentz indices associated with the polarization vectors of 
the cut gluon line.  We will refer to those expressions as the {\it Born tensors}.

The Born tensor with Lorentz indices associated with the eikonal line is
obtained by multiplying the right side of Eq.~(2.24) in Ref.~\cite{Artoisenet:2014lpa} 
by the color factor $(N_c^2-4)/2$:
\begin{equation}
\mathcal{A}_{\rm eikonal}^{\mu \nu}(p,q) = 
\frac{(1-2\epsilon)(N_c^2-1)(N_c^2-4) g_s^4  [(2p+q).n]^2}{4N_c m [(2p+q)^2]^2 (p.q)^2}
\left[ (2p.q)^2  T^{\mu \nu} - (2p+q)^2 l^\mu l^\nu \right] , 
\label{Borntensor-eikonal}
\end{equation}
where $l^\mu$ and $T^{\mu \nu}$ are
\begin{subequations}
\begin{eqnarray}
l^\mu &= & 2p^\mu - \frac{2p.n}{(2p+q).n} (2p+q)^\mu ,
\\
T^{\mu \nu} & = & - g^{\mu \nu} + \frac{n^\mu (2p+q)^\nu + (2p+q)^\mu n^\nu}{(2p+q).n}.
\end{eqnarray}
\end{subequations}
The tensor $\mathcal{A}_{\rm eikonal}^{\mu \nu}$ is orthogonal to $n_\mu$ and $n_\nu$.
Its contraction with $-g_{\mu \nu}$ is
\begin{equation}
\mathcal{A}_{\rm eikonal}^{\mu \nu}(p,q)\; \left( -g_{\mu \nu} \right) = 
\left(\frac{(2p+q).n}{(2p+q_1+q_2).n} \right)^2 \mathcal{A}_{\rm Born}(s,z) , 
\label{Borntensor-eikonal-gmunu}
\end{equation}
where $\mathcal{A}_{\rm Born}$ is given in Eq.~\eqref{eq:ABorn-sz}.

The Born tensor with  Lorentz indices associated with the final-state gluon is
obtained by multiplying the right side of Eq.~(2.27) in Ref.~\cite{Artoisenet:2014lpa} 
by the color factor $(N_c^2-4)/2$:
\begin{equation}
\mathcal{A}_{\rm gluon}^{\mu \nu}(p,q) =  
\frac{ (N_c^2-1)(N_c^2-4) g_s^4 [(2p+q).n]^2}{N_c m [(2p+q)^2]^2 (p.q)^2}
\sum_{i=1}^{4} C_i(z,p.q) T_{i}^{\mu \nu}(p,q),
\label{Borntensor-gluon}
\end{equation}
where the tensors are
\begin{subequations}
\begin{eqnarray}
T_{1}^{\mu \nu} (p,q)& = & -g^{\mu \nu} +\frac{q^\mu n^\nu +  n^\mu q^\nu}{q.n}, \\
T_{2}^{\mu \nu} (p,q)& = & -g^{\mu \nu} +\frac{q^\mu p^\nu + p^\mu q^\nu }{p.q}, \\
T_{3}^{\mu \nu} (p,q)& = &   \left(p^\mu - \frac{p.q}{q.n} n^\mu \right) \left(p^\nu 
- \frac{p.q}{q.n} n^\nu \right), \\ 
T_{4}^{\mu \nu} (p,q)& = & q^\mu q^\nu.
\end{eqnarray}
\label{eq:Timunu}
\end{subequations}
Their coefficients are
\begin{subequations}
\begin{eqnarray}
C_1(z,p.q) & = & -2(1-z)(m^2 + p.q)\left[z p.q -2  (1-z)m^2 \right], \\
C_2(z,p.q) & = & \left[1-2\epsilon -2z(1-z) \right](p.q)^2-2 z (1-z)m^2p.q,  \\
C_3(z,p.q) & = &  4(1-z)^2(m^2 + p.q),\\
C_4(z,p.q) & = & z^2 p.q + (-1+2\epsilon +z^2)m^2 ,
\end{eqnarray}
\end{subequations}
where $z$ is the momentum fraction in Eq.~\eqref{eq:z-pq}.
The tensor $\mathcal{A}_{\rm gluon}^{\mu \nu}$ is orthogonal to $q_\mu$ and $q_\nu$.
Its contraction with $-g_{\mu \nu}$ is
\begin{equation}
\mathcal{A}_{\rm gluon}^{\mu \nu}(p,q)\; \left( -g_{\mu \nu} \right) = 
\left(\frac{(2p+q).n}{(2p+q_1+q_2).n} \right)^2 \mathcal{A}_{\rm Born}(s,z) , 
\label{Borntensor-gluon-gmunu}
\end{equation}
where $\mathcal{A}_{\rm Born}$ is given in Eq.~\eqref{eq:ABorn-sz}.


\section{Real NLO corrections}
\label{sec:NLOreal}

The real NLO corrections to the perturbative fragmentation function for 
$g \to Q \bar Q$, with the $Q \bar Q$ pair in a color-octet $^1S_0$ state,
come from cut diagrams 
with two real partons in the final state.  The two partons can be 
two gluons ($gg$) or a light quark-antiquark pair ($q \bar q$).
Cut diagrams with two real gluons can be obtained 
from the four LO cut diagrams with a single real gluon,
such as the diagram in Figure~\ref{fig:cutdiagram},
by adding a gluon line that crosses the cut
and runs from any of the 6 colored lines on the left side of the cut 
to any of the 6 colored lines on the right side of the cut.
The additional gluon line can also be attached to the operator vertex, 
in which case the fragmenting gluon is attached to the eikonal line.
The cut diagrams with a light $q \bar q$ pair can be obtained
from the four LO cut diagrams by replacing the real gluon line
that crosses the cut
by a virtual gluon that produces a $q \bar q$ pair that crosses the cut.

Each of the cut diagrams involves an integral over the phase space 
of the two real partons in the final state.  
We denote the equal momenta of the $Q$ and $\bar Q$ by  $p$ 
and the momenta of the final-state partons 
(which can be two gluons or a light quark and antiquark) by $q_1$ and $q_2$.
The real NLO contribution to the fragmentation  function can be expressed as
\begin{equation}
D_{g \to Q \bar Q_8}^{\rm (real)}(z) = 
N_{\rm CS} \int d\phi_{\rm real}(p,q_1,q_2) 
\left( \tfrac{1}{2}\mathcal{A}_{\rm real}^{(gg)}\left(p,q_1,q_2\right)
+ \mathcal{A}_{\rm real}^{(q \bar q)}\left(p,q_1,q_2\right)
\right) ,
\label{eq:real_emission}
\end{equation} 
where $N_{\rm CS}$ is the Collins-Soper prefactor in Eq.~\eqref{eq:overalfac}
and $d\phi_{\rm real}$ is the product of the differential phase space 
for final-state partons with momenta $q_1$ and $q_2$
and the factor $2 \pi \delta(K.n - (2p+q_1+q_2).n)$ from the cut through the eikonal line.
The longitudinal momentum fraction of the $Q \bar Q$ pair is
\begin{equation}
z = \frac{(2p).n}{(2p+q_1+q_2).n}.
\label{eq:defz12}
\end{equation}
In the integrand of Eq.~\eqref{eq:real_emission}, 
$\mathcal{A}_{\rm real}^{(gg)}$ is the squared amplitude
from cut diagrams with two gluons crossing the cut, 
and $\mathcal{A}_{\rm real}^{(q \bar q)}$ is the squared amplitude
from cut diagrams with a light quark and antiquark crossing the cut. The factor 
$\tfrac12$ multiplying $\mathcal{A}_{\rm real}^{(gg)}$ in Eq.~(\ref{eq:real_emission})
 compensates for overcounting the states
by integrating over the entire phase space of the two identical gluons.

\subsection{Anatomy of the poles}

\label{sec:singularlimits}

The phase-space integrals in Eq.~(\ref{eq:real_emission}) 
diverge in several regions, yielding poles in $\epsilon = (4-D)/2$ of both infrared (IR) and ultraviolet (UV) nature. 
The nature of the IR poles can be soft or collinear.

\subsubsection{Soft infrared  limits}

Soft singularities only show up in the squared amplitude 
$\mathcal{A}_{\rm real}^{(gg)}$ from final-state gluons
since a soft (anti)quark does not yield any pole. 
First let us consider the limit $q_2 \rightarrow 0$ (soft limit for the gluon 2). 
The tree-level amplitude is proportional to the color factor $d^{acb}$ in Eq.~\eqref{eq:colorsub-amp},
where $a$, $b$, and $c$ are the color indices of the eikonal line, the $Q \bar Q$ pair, 
and the final-state gluon, respectively.
In the  eikonal approximation associated with the soft limit $q_2 \rightarrow 0$, 
the amplitude for the
emission of the soft-gluon is obtained from the tree amplitude by replacing $d^{acb}$ with
\begin{equation}
i g_s \mu^\epsilon \left( 
\frac{p^\mu }{p.q_2} d_{ace}f_{ebd} 
+ \frac{q_1^\mu }{q_1.q_2} d_{aeb} f_{ecd}
+ \frac{n^\mu }{n.q_2} d_{ecb} f_{ead}
\right) \varepsilon^d_\mu(q_2),
\end{equation}
where $\mu$ and $d$ are the Lorentz index and color index of the gluon with soft momentum $q_2$.
Squaring the amplitude and summing over color and Lorentz indices, one readily
obtains the following approximation for the squared amplitude in the soft limit $q_2 \rightarrow 0$ :
\begin{equation} 
\mathcal{A}_{\rm real}^{(gg)}   
  \longrightarrow   4\pi \alpha_s \mu^{2\epsilon}  N_c \left(   
\frac{p.q_1}{p.q_2 \, q_2.q_1}
+\frac{p.n}{p.q_2 \, q_2.n}
+\frac{q_1.n}{q_1.q_2 \, q_2.n}
- \frac{m^2}{(q_2.p)^2} 
\right)\mathcal{A}_{\rm{Born}} (p,q_1),
\label{limitq2soft}
\end{equation}
where $\mathcal{A}_{\rm Born}$ is the Born squared amplitude defined in Eq.~\eqref{eq:ABorn-pq}.
The analogous result in the soft limit $q_1 \rightarrow 0$ reads
\begin{equation}
\mathcal{A}_{\rm real}^{(gg)}  \longrightarrow  4\pi \alpha_s \mu^{2\epsilon}   N_c \left(   
\frac{p.q_2}{p.q_1 \, q_1.q_2}
+\frac{p.n}{p.q_1 \, q_1.n}
+\frac{q_2.n}{q_2.q_1 \,q_1.n}
- \frac{m^2}{(q_1.p)^2} 
\right)  \mathcal{A}_{\rm{Born}} (p,q_2).
\end{equation}

\subsubsection{Collinear infrared  limits}

Collinear singularities show up in both the squared 
amplitudes $\mathcal{A}_{\rm real}^{(gg)}$ and 
$\mathcal{A}_{\rm real}^{(q \bar q)}$ from the kinematic region 
 in which the light partons 
crossing the cut are collinear.
In order to describe the collinear limit,
it is convenient to define the light-like four-vector
\begin{equation}
\tilde{q}^\mu = q_1^\mu+q_2^\mu - \frac{q_1.q_2}{(q_1+q_2).n} n^\mu  \, ,
\label{eq:qtilde}
\end{equation}
which satisfies $\tilde{q}^2=0$ and $\tilde{q}.n = (q_1+q_2).n$.
In the collinear limit, the four-vector $\tilde{q}$
coincides with $q_1+q_2$, and  the expression for the squared 
amplitude factorizes over the Born amplitude  up to spin correlations. 
In order to account for these spin correlation effects in collinear splittings,
it is convenient to define a 4-vector  ${\breve q}$ by
\begin{equation}
\label{eq:breveq-u}
{\breve q}^\mu =  \frac{q_1.n}{(q_1+q_2).n}{q_1}^\mu - \frac{q_2.n}{(q_1+q_2).n} {q_2}^\mu.
\end{equation}
This 4-vector is orthogonal to $\tilde q$: ${\breve q}.\tilde{q}=0$.
The 4-momenta of the two partons can be expressed as
\begin{subequations}
\begin{eqnarray}
q_1^\mu &=& \frac{q_1.n}{(q_1+q_2).n} \tilde q^\mu + {\breve q}^\mu 
+ \frac{q_1.n\, q_1.q_2}{[(q_1+q_2).n]^2} n^\mu,
\\
q_2^\mu &=& \frac{q_2.n}{(q_1+q_2).n} \tilde q^\mu - {\breve q}^\mu 
+ \frac{q_2.n\, q_1.q_2}{[(q_1+q_2).n]^2} n^\mu.
\end{eqnarray}
\label{eq:q1q2}
\end{subequations}

In the case of $\mathcal{A}_{\rm real}^{(gg)}$,  the factorisation formula when 
the momenta of the two gluons crossing the cut are nearly collinear reads 
\begin{equation} 
\mathcal{A}_{\rm real}^{(gg)}
\longrightarrow   
\frac{4\pi \alpha_s \mu^{2\epsilon}}{q_1.q_2} P^{(gg)}_{\mu \nu}(p,q_1,q_2) 
\mathcal{A}_{\rm gluon}^{\mu \nu} (p, \tilde{q}) ,
\label{limitq1q2coll}
\end{equation}
where the  Born tensor $\mathcal{A}_{\rm gluon}^{\mu \nu}$ is defined in Eq.~(\ref{Borntensor-gluon}).
The tensor $P^{(gg)}_{\mu \nu}$  is 
\begin{equation}
P^{(gg)}_{\mu \nu} (p,q_1,q_2)=N_c \Bigg[
\left(\frac{q_1.p}{q_2.p}+ \frac{q_2.p}{q_1.p} + \frac{q_1.n}{q_2.n}+ \frac{q_2.n}{q_1.n} \right) \left( -g_{\mu \nu} \right)
+2(1-\epsilon) \frac{{\breve q}_\mu {\breve q}_\nu}{q_1.q_2} \Bigg].
\label{eq:P(gg)}
\end{equation}

In the case of $\mathcal{A}_{\rm real}^{(q \bar q)}$, 
the factorisation formula when 
the momenta of the light quark and antiquark crossing the cut are nearly collinear
reads 
\begin{equation} 
\mathcal{A}_{\rm real}^{(q \bar q)}
\longrightarrow   
\frac{4\pi \alpha_s \mu^{2\epsilon}}{q_1.q_2} P^{(q \bar q)}_{\mu \nu}(q_1,q_2) 
\mathcal{A}_{\rm gluon}^{\mu \nu} (p, \tilde{q}) .
\label{limitq1q2collqq}
\end{equation}
The tensor $P^{(q \bar q)}_{\mu \nu}$  is
\begin{equation}
P^{(q \bar q)}_{\mu \nu} (q_1,q_2)  
=  n_fT_F 
\left[  (- g_{\mu \nu})  
- 2  \frac{{\breve q}_\mu {\breve q}_\nu}{q_1.q_2}  \right] ,
\label{eq:P(qq)}
\end{equation}
where $T_F= \frac{1}{2}$ is the trace of the square of a generator
for the fundamental representation and $n_f$ is the number of light flavours.

In the squared amplitude  $\mathcal{A}_{\rm real}^{(gg)}$, 
additional collinear singularities arise from
kinematic regions  in which
one gluon crossing the cut is collinear to the eikonal line.
When the gluon momentum $q_i$ is collinear to the 
four-vector $n$, one has
\begin{equation} 
\mathcal{A}_{\rm real }^{(gg)}
\longrightarrow  
4\pi \alpha_s \mu^{2\epsilon} 4N_c
\frac{K.n}{q_i.n} \frac{1}{Q^2-(Q-q_i)^2} 
\mathcal{A}_{\rm{Born}} (p,\tilde{q}),
\label{limitqjncoll}
\end{equation}
where $Q=2p+q_1+q_2$ is the sum of the momenta of the particles 
crossing the cut.

\subsubsection{Ultraviolet limits}

Ultraviolet singularities in the squared amplitude $\mathcal{A}_{\rm real}^{(gg)}$
arise from the kinemetic region in which the invariant mass $s=(2p+q_1+q_2)^2$ 
of all the final-state particles goes to $\infty$.
The factorisation formula in each of the two  UV limits
$s \gg (2p+q_j)^2$ for  $j=1,2$ also holds up to spin correlations.
In order to account for these spin correlation effects,
it is convenient to introduce the  4-vectors $l_j$  defined by
\begin{subequations}
\begin{eqnarray}
l_1^\mu &=& q_2^\mu - \frac{q_2.n}{(2p+q_1).n}  (2p+q_1)^\mu,
\\
l_2^\mu &=& q_1^\mu -  \frac{q_1.n}{(2p+q_2).n} (2p+q_2)^\mu.
\end{eqnarray}
\label{eq:l1l2}%
\end{subequations}
These 4-vectors are orthogonal to $n$: $l_j.n=0$.
The 4-vector $l_1^\mu$ is the component of $q_2^\mu$ orthogonal to $n$.
It is also convenient to define the longitudinal momentum fraction $y_j$ of
the system consisting of the $Q \bar Q$ pair and the parton of momentum $q_j$:
\begin{equation}
y_j = \frac{(2p+ q_j).n}{(2p+q_1+q_2).n}.
\label{eq:defyi}
\end{equation}

The factorisation formula 
includes a factor of $\mathcal{A}_{\rm eikonal}^{\mu \nu}(p,q_j)$,
where $\mathcal{A}_{\rm eikonal}^{\mu \nu}$ is the Born tensor
defined in Eq.~(\ref{Borntensor-eikonal})
whose Lorentz indices $\mu$ and $\nu$ are associated with the eikonal line.
The factor $\mathcal{A}_{\rm eikonal}^{\mu \nu}$ can be interpreted
as arising from the fragmentation of a gluon with longitudinal momentum
$y_jK.n$ into a $Q \bar Q$ pair with longitudinal momentum $z K.n$
via the radiation of a gluon of momentum $q_j$.
In the limit  $s \gg (2p+q_j)^2$, one has
\begin{equation}
\mathcal{A}_{\rm real }^{(gg)} \longrightarrow
 \frac{8 \pi \alpha_s \mu^{2 \epsilon}}{(2p+q_1+q_2)^2} P^{\rm eik}_{\mu \nu}(y_j,l_j) 
  \frac{1}{y_j^2} \mathcal{A}_{\rm eikonal}^{\mu \nu}(p,q_j ) .
\label{eq:sub1and2}
\end{equation}
The tensor $P^{\rm eik}_{\mu \nu}$ is defined by
\begin{equation}
P^{\rm eik}_{\mu \nu}(y,l) = 2N_c \Bigg[ \left(  \frac{y}{1-y} + 
y \left(1 - y \right) \right) (-g_{\mu \nu})
- 2 (1- \epsilon) 
 \frac{1-y}{y}
 \frac{l_\mu l_\nu }{l^2} \Bigg].
\end{equation}

\begin{figure}
\center
\includegraphics[scale=0.7]{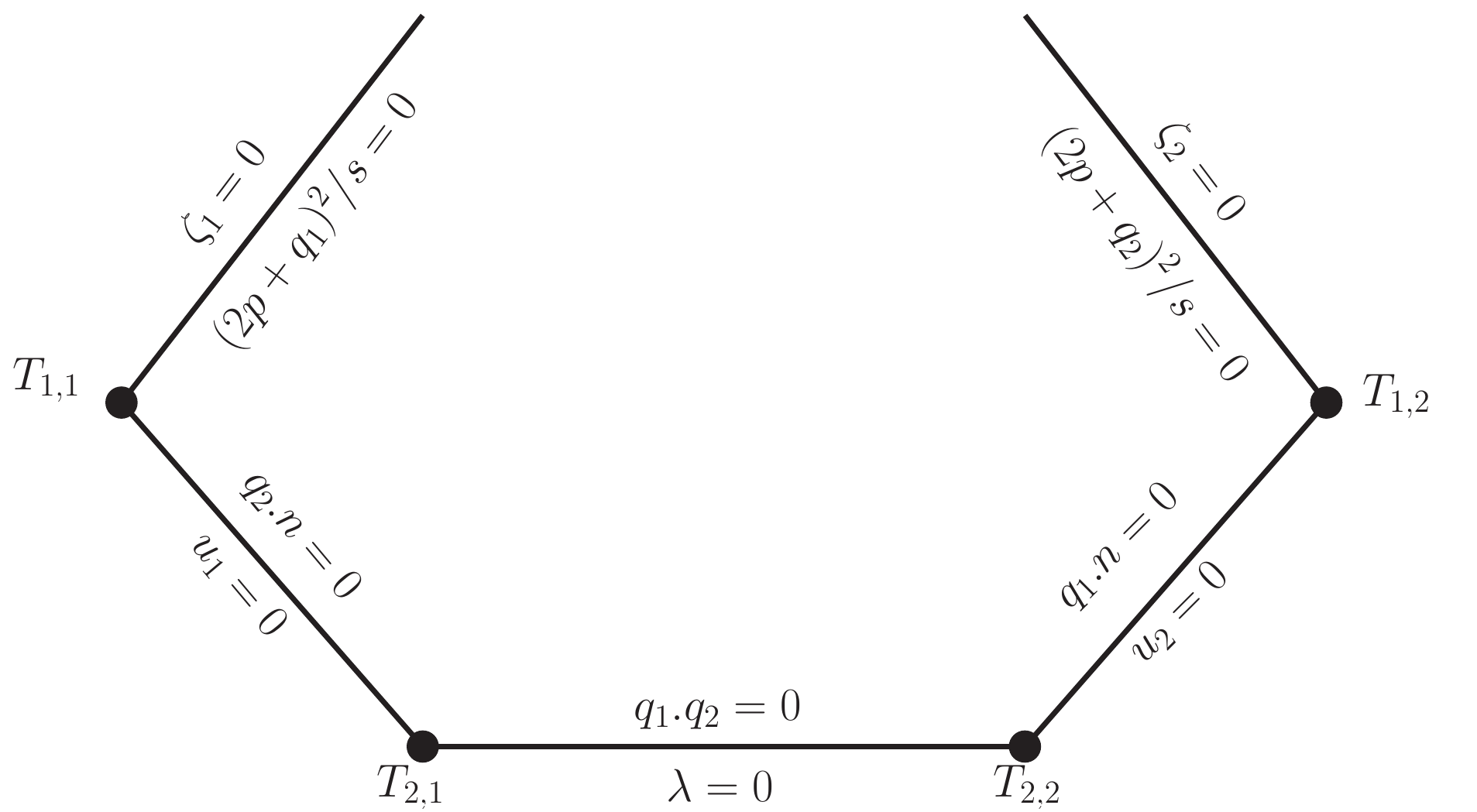}
\caption{Representation of the singular  regions for the integration 
of the real emission amplitude. 
Each line represents a specific limit, which is specified 
in terms of Lorentz invariants in the inner part of the figure.
A line represents a phase-space boundary that gives a single pole in $\epsilon$.
A dot connecting two lines represents a phase-space region 
leading to a double pole in $\epsilon$.    For each dot, there is 
a specific subtraction term $T_{i,j}^{(gg)}$ appearing in Eq.~\eqref{eq:real_emission_subtracted}
whose integral includes the double pole.
The subtraction of  the single pole associated with a line connecting two dots 
is shared among the two subtraction terms associated with these dots.
}
\label{fig:poles}
\end{figure}

\subsection{Subtraction scheme and kinematics}

In order to extract the poles resulting from phase-space integrals 
in the singular regions outlined in the previous section, we adopt in this 
work a subtraction scheme that is inspired by the formalism 
introduced by  Frixione, Kunszt, and Signer (FKS)~\cite{Frixione:1995ms}. 
In the FKS formalism for parton cross sections \cite{Frixione:1995ms},
the phase space is partitioned into kinematic regions that in any singular limit 
give either (a) a finite contribution or (b) a soft IR singularity 
or (c) a collinear IR singularity or (d) the product of a soft IR singularity and a collinear IR singularity.
In the extension of the FKS formalism to fragmentation functions,
a kinematic region in a singular limit may also give (e) a UV singularity 
or (f) the product of a UV singularity and a collinear IR singularity.
For the $(gg)$ term  in Eq.~(\ref{eq:real_emission}), 
the partition of the phase space can be implemented 
by multiplying $\tfrac12 \mathcal{A}_{\rm real}^{(gg)}$ by a partition of unity with four terms:
\begin{equation}
S_{1,1} + S_{1,2}  + S_{2,1}  + S_{2,2} = 1,
\label{eq:partition}
\end{equation}
where the weight functions $S_{i,j}$ are scalar functions of $p$, $q_1$, $q_2$, and $n$.
Each $S_{i,j}$ acts as a damping factor for some singular phase-space regions. 
The second index $j$ is associated with the parton of momentum $q_j$.
For the $(q \bar q)$ term  in Eq.~(\ref{eq:real_emission}), 
 no partition of unity 
is required since there is only one kinematic region yielding a pole, namely 
the singular limit of a collinear $q \bar q$ pair. 
Our strategy to extract the poles in the expression in Eq.~(\ref{eq:real_emission}) 
is to design a subtraction term $T_{i,j}^{(gg)}$ for each of the four terms  
$\tfrac12 \mathcal{A}_{\rm real}^{(gg)}S_{i,j}$ in the integrand
and a subtraction term  $T^{(q \bar q)}$ for the term  $\mathcal{A}_{\rm real}^{(q \bar q)}$.
Hence the fragmentation  function in Eq.~\eqref{eq:real_emission} 
can be decomposed  as
\begin{eqnarray}
D_{g \to Q \bar Q}^{\rm (real)}(z) &=& 
\sum_{i,j} N_{\rm CS} \int d\phi_{\rm real}
\left[ \tfrac{1}{2} \mathcal{A}_{\rm real}^{(gg)}   S_{i,j}  - T_{i,j}^{(gg)} \right]  
+ N_{\rm CS} \int d\phi_{\rm real} 
\left[ \mathcal{A}_{\rm real}^{(q \bar q)}   - T^{(q \bar q)} \right]  
\nonumber \\
&&+ \sum_{i,j} N_{\rm CS} \int d\phi_{\rm real}   T_{i,j}^{(gg)}
+ N_{\rm CS} \int d\phi_{\rm real}   T^{(q \bar q)} ,
\label{eq:real_emission_subtracted}
\end{eqnarray}
where the dependence on the momenta $p$, $q_1$, $q_2$ is implicit in each function in the integrands.
The subtraction terms $T_{i,j}^{(gg)}$ and $T^{(q \bar q)}$ are designed so that
the first and second integrals  on the right side of Eq.~\eqref{eq:real_emission_subtracted}
are finite and can be evaluated in $D=4$ dimensions. 
The third and fourth integrals  are evaluated in $D=4-2\epsilon$ dimensions,
so the UV and IR divergences appear as poles in $\epsilon$.  

Unlike the construction in the FKS formalism, Lorentz invariance is explicitly manifest 
in our construction of the subtraction terms, i.e.\ we do not specify any specific frame. 
Singular regions are characterised by Lorentz-invariant quantities.
We define a dimensionless variable $\lambda$ that vanishes at the boundary 
that gives IR singularities associated with two collinear partons
and/or one soft parton:  
\begin{equation}
 \lambda=\frac{(q_1+q_2)^2}{4m^2}.
\label{eq:lambda}
\end{equation}
[eb,pa]
We define a dimensionless variable $u_1$ ($u_2$) that vanishes  at the boundary
that gives IR singularities when gluon 2 (gluon 1)  is soft and/or collinear to the eikonal line:
\begin{equation}
u_1=\frac{q_2.n}{(q_1+q_2).n}, \quad u_2=\frac{q_1.n}{(q_1+q_2).n}.
\label{eq:uj}
\end{equation}
They satisfy $u_1 + u_2 = 1$.
(The mismatch between the indices of the variable $u_j$ that approaches 0
and the momentum  $q_{3-j}$ that becomes soft or collinear
simplifies the expressions for subtraction terms.) 
The total invariant mass $s$ of the final-state particles
and the invariant mass $s_j$ of the heavy-quark pair and the gluon $j$ are
\begin{subequations}
\begin{eqnarray}
s &=& (2p+q_1+q_2)^2,
\\
s_j &=& (2p+q_j)^2.
\label{eq:si}
\end{eqnarray}
\end{subequations}
We define dimensionless variables $\zeta_j$ that vanish  at the boundaries 
that give UV singularities when $s$ is much larger than $s_j$:
\begin{equation}
 \zeta_1 = \frac{s_1/y_1}{s}, 
 \qquad
  \zeta_2 = \frac{s_2/y_2}{s},
\end{equation} 
where the momentum fraction $y_j$ is defined by Eq.~(\ref{eq:defyi}).
The boundaries of the singular regions are represented in Figure~\ref{fig:poles}.

The subtraction terms $T_{i,j}^{(gg)}$ and $T^{(q \bar q)}$  
introduced in Eq.~\eqref{eq:real_emission_subtracted}
will be defined using cutoff variables, so that the subtraction is applied only
in the vicinity of the singular phase-space boundaries. The variables on which 
these cutoffs apply are selected in such a way to make the integration 
of the poles in $\epsilon$ in the subtraction integrals tractable.

\subsection{Subtraction terms}

\subsubsection{Partition of unity}

To define the weight functions $S_{i,j}$ in the partition of unity in Eq.~\eqref{eq:partition}, 
we first introduce functions $w_{i,j}$:
\begin{subequations}
\begin{eqnarray} 
w_{1,j} &=& \frac{1}{\zeta_j u_j}, 
\\
 w_{2,j} &=&  \frac{1-u_j}{\lambda u_j[u_j^2+(1-u_j)^2]} .
\end{eqnarray} 
\end{subequations}
The sum of these four functions is
\begin{equation}
\sum_{l,k} w_{l,k} = \frac{1}{\zeta_1 u_1} + \frac{1}{\zeta_2 u_2} +\frac{1}{\lambda u_1 u_2}.
\end{equation}
The weight functions in the partition of unity are defined by 
\begin{equation}
S_{i,j} = \frac{w_{i,j}}{\sum_{l,k} w_{l,k}}.
\end{equation}

At the singular  boundaries of phase space,
one or two of the variables $\lambda$, $u_1$, $u_2$, $\zeta_1$, and $\zeta_2$ vanishes.
In the singular limits, some of the weight functions $S_{i,j}$ vanish and the others simplify.
In the collinear IR limit $\lambda \rightarrow 0$, 
 only $S_{2,1}$ and $S_{2,2}$ are nonzero (and they add up to 1):
\begin{eqnarray}
 \lim_{\lambda \rightarrow 0} S_{2,1} = \frac{u_2^2}{u_1^2+u_2^2}\, , 
\qquad
\lim_{\lambda \rightarrow 0} S_{2,2} = \frac{u_1^2}{u_1^2+u_2^2}\, . 
\label{eq:Sij,lambda->0}
\end{eqnarray} 
In the soft IR limit $u_j \rightarrow 0$, 
only $S_{1,j}$ and $S_{2,j}$ are nonzero (and they add up to 1):
\begin{subequations}
\begin{eqnarray}
\lim_{u_j\rightarrow 0} S_{1,j} &=& \frac{\lambda}{\lambda + \zeta_j} \, , 
\label{eq:S1j,u->0}
\\
\lim_{u_j \rightarrow 0} S_{2,j} &=& \frac{\zeta_j }{\lambda  + \zeta_j }\, .
\label{eq:S2j,u->0}
\end{eqnarray} 
\label{eq:Sij,u->0}
\end{subequations}
In the UV  limit $\zeta_j \rightarrow 0$, only $S_{1,j}$ is nonzero:
\begin{equation}
\lim_{\zeta_j\rightarrow 0} S_{1,j} = 1\, . 
\end{equation} 

In the construction of the subtraction terms $T_{i,j}^{(gg)}$ in Eq.~\eqref{eq:real_emission_subtracted},
we use Heavyside theta functions in the variables $\lambda$, $u_j$, and $\zeta_j$
associated with the singular  boundaries:
\begin{equation}
 \theta^{(\lambda)} = \theta(\lambda^{\rm cut} - \lambda), 
 \quad
\theta^{(u_j)} = \theta(u^{\rm cut} - u_j), 
\quad
 \theta^{(\zeta_j)} = \theta(\zeta^{\rm cut} - \zeta_j).
\end{equation}
In the construction of the subtraction terms $T_{2,j}^{(gg)}$ in Eq.~\eqref{eq:real_emission_subtracted},
we also use Heavyside theta functions in variables $\delta$ and $\delta_j$
defined by
\begin{subequations}
\begin{eqnarray}
  \delta &=& 1-\frac{p.\tilde{q}}{p.(q_1+q_2)}, 
    \label{eq:delta}
  \\
  \delta_1&=& 1-\frac{p.q_1+m^2q_2.n/(2p.n)}{p.(q_1+q_2)}, 
    \label{eq:delta1}
  \\
  \delta_2 &=& 1-\frac{p.q_2+m^2q_1.n/(2p.n)}{p.(q_1+q_2)} . 
    \label{eq:delta2}
\end{eqnarray}
\end{subequations}
The theta functions are
\begin{equation}
 \theta^{(\delta)} = \theta(\delta^{\rm cut} -\delta), \quad
 \theta^{( \delta_j)} =  \theta(\delta^{\rm cut} -\delta_j) .
\end{equation} 
In the construction of the subtraction terms $T^{(q \bar q)}$ in Eq.~\eqref{eq:real_emission_subtracted},
we use the Heavyside theta function $ \theta^{(\lambda)}$ in the variable $\lambda$.

\subsubsection{Subtraction  terms  for $T_{1,j}^{(gg)}$ }

The phase-space boundaries that give singularities in the integral of
$\frac{1}{2} \mathcal{A}_{\rm real}^{(gg)} S_{1,j} $ are $u_j=0$ and $\zeta_j=0$. 
The singular piece can be expressed as
\begin{equation}
T_{1,j}^{(gg)} =  
S_{1,j}^{(u_j)} \, D_1^{(u_j)} \, \theta^{(u_j)}   
 \, + \,  D^{(\zeta_j)} \, \theta^{(\zeta_j)}
 \, -\,  D^{(\zeta_j,u_j)} \, \theta^{(\zeta_j)} \, \theta^{(u_j)} .
\label{T1jsingular}
\end{equation} 
The remainder obtained by subtracting $T_{1,j}^{(gg)} $ 
from $\frac{1}{2}\mathcal{A}_{\rm real}^{(gg)} S_{1,j} $ can be integrated numerically in $D=4$ dimensions, 
because the $\theta^{(u_j)}$ term subtracts the singularity when $u_j \rightarrow 0$,
the $\theta^{(\zeta_j)}$ term subtracts the singularity when $\zeta_j \rightarrow 0$,
and the $\theta^{(\zeta_j)} \theta^{(u_j)}$ term  adds back the double singularity 
that is over-subtracted by the other two terms.

The weight function $S_{1,j}^{(u_j)}$ in the first subtraction term in Eq.~\eqref{T1jsingular} is
\begin{equation}
S_{1,j}^{(u_j)} = \frac{\lambda_j}{\lambda_j + (s_j/y_j)/s} \, ,
\label{eq:S1j^uj}
\end{equation}
where $\lambda_j$ is defined by
\begin{equation}
\lambda_j =  \frac{(1-z)(s-s_j/y_j)}{4 m^2}.
\label{eq:lambdaj}
\end{equation}
In the soft limit $q_{3-j} \rightarrow  0$,
this weight function approaches $S_{1,j}$, whose limiting behavior is given in Eq.~\eqref{eq:S1j,u->0}.  
The subtraction term $D_1^{(u_j)}$ is constructed using the 
Born squared amplitude $\mathcal{A}_{\rm Born}$ in Eq.~\eqref{eq:ABorn-pq}
with $s$ replaced by $s_j/y_j$:
\begin{equation}
D_1^{(u_j)}(p,q_1,q_2) = 4\pi \alpha_s \mu^{2\epsilon} \frac{N_c}{2 m^2 \lambda_j u_j} 
\mathcal{A}_{\rm Born} (s_j/y_j,z).
\label{eq:D1uj}
\end{equation} 
The integral of the sum of $S_{1,j}^{(u_j)} \, D_1^{(u_j)}$
and the subtraction term $S_{2,j}^{(u_j)} \, D_2^{(u_j)}$ for  $T_{2,j}^{(gg)}$ 
is evaluated analytically in Sec.~\ref{app:uj} of Appendix~\ref{sec:integralpoles}.
The sum over $j$ of these subtraction integrals is given in Eqs.~\eqref{eq:subint-uj} and \eqref{kernelVu}.

The subtraction term $D^{(\zeta_j)}$  in Eq.~\eqref{T1jsingular}
is constructed using the Born tensor $\mathcal{A}_{\rm eikonal}^{\mu \nu}$ 
in Eq.~\eqref{Borntensor-eikonal} with Lorentz indices associated with the eikonal line:
\begin{equation}
\label{eq:Dzeta}
D^{(\zeta_j)}(p,q_1,q_2) = 4\pi \alpha_s \mu^{2\epsilon} 
V^{(\zeta_j)}_{\mu \nu} (p,q_1,q_2) \mathcal{A}_{\rm eikonal}^{\mu \nu} (p,q_j).
\end{equation}
The tensor $V^{(\zeta_j)}_{\mu \nu}$ is
\begin{equation}
\label{eq:Vzeta}
V^{(\zeta_j)}_{\mu \nu} (p,q_1,q_2) = 
\frac{2N_c}{s} \Bigg[ \left(  \frac{y_j}{1-y_j} + 
y_j \left(1 - y_j \right) \right) (-g_{\mu \nu})
- 2 (1- \epsilon) 
 \frac{1-y_j}{y_j}
 \frac{l_{j,\mu} l_{j,\nu} }{l_j^2} \Bigg] ,
\end{equation}
where the 4-vector $l_j$ is defined in Eqs.~(\ref{eq:l1l2}).
The poles in the integral of $D^{(\zeta_j)}$
are evaluated analytically in Sec.~\ref{app:zetaj} of Appendix~\ref{sec:integralpoles}.
The subtraction integral summed over $j$ is given in Eqs.~\eqref{eq:subint-zetaj} and \eqref{eq:Izeta210}.

The over-subtraction term $D^{(\zeta_j, u_j)}$  in Eq.~\eqref{T1jsingular} is
constructed using the Born squared amplitude $\mathcal{A}_{\rm Born}(s_j/y_j,z)$:
\begin{equation}
D^{(\zeta_j, u_j)}(p,q_1,q_2) = 4\pi \alpha_s \mu^{2\epsilon} 
 \frac{2N_c}{(1-z)u_j s} 
\mathcal{A}_{\rm Born} (s_j/y_j,z).
\label{eq:Dujzetaj}
\end{equation} 
The poles in the integral of $D^{(\zeta_j)}$
are evaluated analytically in Sec.~\ref{app:zetajuj} of Appendix~\ref{sec:integralpoles}.
The over-subtraction integral summed over $j$ is given in Eqs.~\eqref{eq:subint-zetajuj} and \eqref{eq:Izetau210}.

\subsubsection{Subtraction terms for $T_{2,j}^{(gg)}$ }
 
The phase-space boundaries that give singularities in the integral of
$\frac{1}{2}\mathcal{A}_{\rm real}^{(gg)} S_{2,j} $ are $u_j=0$ and $\lambda=0$. 
The singular piece can be expressed as
\begin{eqnarray}
\label{T2jsingular}
T_{2,j}^{(gg)} =  
S_{2,j}^{(u_j)} \, D_2^{(u_j)} \, \theta^{(u_j)}
 \,+\,  S_{2,j}^{(\lambda)} \, D^{(\lambda)} \, \theta^{(\lambda)} 
 \,+ \,  D^{(u_j,\delta_j)} \, \theta^{(u_j)} \,  \theta^{(\delta_j)}
\nonumber
 \\
\, - \,  D^{(u_j,\delta)} \, \theta^{(u_j)} \, \theta^{(\delta)}
 \, -\, D^{(\lambda,u_j)} \, \theta^{(\lambda)} \theta^{( u_j)}.
\end{eqnarray}
The remainder obtained by subtracting $T_{2,j}^{(gg)} $ 
from $\frac{1}{2}\mathcal{A}_{\rm real}^{(gg)} S_{2,j} $ can be integrated numerically in $D=4$ dimensions, 
because the $\theta^{(u_j)}$ term subtracts the singularity when $u_j \rightarrow 0$,
the $\theta^{(\lambda)}$ term subtracts the singularity when $\lambda \rightarrow 0$,
the $\theta^{(u_j)} \,  \theta^{(\delta_j)}$ term subtracts additional singularities 
when both $u_j \rightarrow 0$ and $\delta_j \rightarrow 0$, and 
the last two terms add back singularities that are over-subtracted by the other terms.

The weight function $S_{2,j}^{(u_j)}$  in the first subtraction term  in Eq.~\eqref{T2jsingular} is
\begin{equation}
S_{2,j}^{(u_j)} = \frac{\tilde{s}/s}{\lambda+\tilde{s}/s},
\label{eq:S2j^uj}
\end{equation}
where $\tilde{s} = (2p+\tilde{q})^2$.
In the soft  limit $q_{3-j} \rightarrow  0$,
this weight function approaches $S_{2,j}$,
whose limiting behavior is given in Eq.~\eqref{eq:S2j,u->0}.  
The subtraction term $D_2^{(u_j)}$ is
constructed using the Born squared amplitude $\mathcal{A}_{\rm Born}$ 
in Eq.~\eqref{eq:ABorn-pq}
with $s$ replaced by $\tilde{s} = (2p+\tilde q)^2$: 
\begin{equation}
D_2^{(u_j)}(p,q_1,q_2) = 4\pi \alpha_s \mu^{2\epsilon} 
\frac{N_c}{2 m^2 \lambda u_j}  \mathcal{A}_{\rm Born}(\tilde{s},z).
\label{eq:D2uj}
\end{equation}
The integral of the sum of $S_{2,j}^{(u_j)} \, D_2^{(u_j)}$ 
and the subtraction term $S_{1,j}^{(u_j)} \, D_1^{(u_j)}$ for  $T_{1,j}^{(gg)}$ 
is evaluated analytically in Sec.~\ref{app:uj} of Appendix~\ref{sec:integralpoles}.
The sum over $j$ of the subtraction integrals is given in Eqs.~\eqref{eq:subint-uj} and \eqref{kernelVu}.

The weight functions $S_{2,j}^{(\lambda)}$ in the second subtraction term in Eq.~\eqref{T2jsingular} are
\begin{equation}
S_{2,j}^{(\lambda)} = \frac{(1-u_j)^2}{u_j^2+(1-u_j)^2}.
\label{eq:S2j^lambda}
\end{equation}
Since $u_1+u_2 = 1$, they satisfy
\begin{equation}
S_{2,1}^{(\lambda)} + S_{2,2}^{(\lambda)} = 1.
\label{eq:S21+S22}
\end{equation}
In the collinear IR limit $\lambda \rightarrow  0$, this weight function approaches $S_{2,j}$,
whose limiting behavior is given in Eq.~\eqref{eq:Sij,lambda->0}.
The subtraction term $D^{(\lambda)}$ is constructed from the Born tensor 
$\mathcal{A}_{\rm gluon}^{\mu \nu}$ in Eq.~\eqref{Borntensor-gluon} with Lorentz indices 
associated with the final-state gluon:
\begin{equation}
D^{(\lambda)} (p,q_1,q_2)= 4\pi \alpha_s \mu^{2\epsilon} \frac{1}{4 m^2 \lambda} P^{(gg)}_{\mu \nu}(p,q_1,q_2) 
\mathcal{A}_{\rm gluon}^{\mu \nu} (p, \tilde{q}),
\label{eq:Dlambda}
\end{equation}
where the tensor $P^{(gg)}_{\mu \nu}$  is given in Eq.~\eqref{eq:P(gg)}.
The poles in the integral of $D^{(\lambda)}$
are evaluated analytically in Sec.~\ref{app:lambda} of Appendix~\ref{sec:integralpoles}.
The subtraction integral summed over $j$ is given in Eqs.~\eqref{eq:subint-lambda},
\eqref{eq:V1lambda}, and \eqref{eq:Ilambda210}.

The subtraction term $D^{(u_j,\delta_j)}$ is constructed using the Born squared amplitude 
$\mathcal{A}_{\rm Born}(\breve s_j,z)$,
where the variable $\breve s_j $ is defined by
\begin{equation}
\breve s_j = s_j + \frac{1-z}{z}u_j( 4 m^2) \, ,
\label{eq:brevesj}
\end{equation}
with $s_j = (2p+q_j)^2$ and $u_j$ defined in Eq.~\eqref{eq:uj}.
The subtraction term is
\begin{equation}
D^{(u_j,\delta_j)}(p,q_1,q_2) = 4\pi \alpha_s \mu^{2\epsilon} 
\frac{N_c}{2m^2} \left[ \frac{z}{2(1-z)u_j w_j} -\frac{1}{w_j^2} \right] \mathcal{A}_{\rm Born}(\breve s_j,z),
\label{eq:Duj,deltaj}
\end{equation}
where $w_j$ is defined by
\begin{equation}
w_1 = \frac{p.q_2}{m^2}, \quad w_2 = \frac{p.q_1}{m^2} .
   \label{eq:wj}
\end{equation}
The poles in the integral of $D^{(u_j,\delta_j)}$
are evaluated analytically in Sec.~\ref{app:uj,deltaj} of Appendix~\ref{sec:integralpoles}.
The subtraction integral summed over $j$ is given in Eqs.~\eqref{eq:subint-uj,deltaj} and \eqref{eq:Iudelta'210}.

The over-subtraction terms $D^{(u_j,\delta)}$ and $D^{(\lambda,u_j)}$ 
in Eq.~\eqref{T2jsingular} are the same.
They are constructed using the Born squared amplitude $\mathcal{A}_{\rm Born}(\tilde{s} ,z)$: 
\begin{subequations}
\begin{eqnarray}
D^{(u_j,\delta)}(p,q_1,q_2) &=& 
4\pi \alpha_s \mu^{2\epsilon}\frac{N_c}{4m^2 \lambda  u_j} \mathcal{A}_{\rm Born}(\tilde{s},z),
\label{eq:Dujdelta}
\\
D^{(\lambda,u_j)}(p,q_1,q_2) &=& 
4\pi \alpha_s \mu^{2\epsilon}  \frac{N_c}{4m^2 \lambda u_j}  \mathcal{A}_{\rm Born}(\tilde{s},z).
\label{eq:Dlambdauj}
\end{eqnarray}
\end{subequations}
The poles in the integral of  $D^{(u_j,\delta)}$
are evaluated analytically in Sec.~\ref{app:ujdelta} of Appendix~\ref{sec:integralpoles}.
The over-subtraction integral summed over $j$ is given in Eqs.~\eqref{eq:subint-ujdelta} and \eqref{eq:Iudelta210}.
The over-subtraction integral for $D^{(\lambda,u_j)}$
is evaluated analytically in Sec.~\ref{app:lambdauj} of Appendix~\ref{sec:integralpoles}.
The over-subtraction integral summed over $j$ is given in Eqs.~\eqref{eq:subint-lambdauj} and \eqref{kernelVlambdau}.

\subsubsection{Subtraction term for  $T^{(q \bar q)}$ }

The only phase-space boundary that gives singularities in the integral of
$\mathcal{A}_{\rm real}^{(q\bar q)}$  is $\lambda=0$. 
The singular piece can be expressed as
\begin{equation}
T^{(q \bar q)} =  D^{(q \bar q)} \, \theta^{(\lambda)}.
\label{Tqq}
\end{equation}
The remainder obtained by subtracting $T^{(q \bar q)}$ 
from $\mathcal{A}_{\rm real}^{(q\bar q)}$ can be integrated numerically in $D=4$ dimensions, 
because the the $\theta^{(\lambda)}$ term subtracts the singularity when $\lambda \rightarrow 0$.

The subtraction term $D^{(q \bar q)}$ is constructed from the Born tensor 
$\mathcal{A}_{\rm gluon}^{\mu \nu}$ in Eq.~\eqref{Borntensor-gluon} with Lorentz indices 
associated with the final-state gluon:
\begin{equation}
D^{(q \bar q)} (p,q_1,q_2)= 4\pi \alpha_s \mu^{2\epsilon} \frac{1}{q_1.q_2} P^{(q \bar q)}_{ \mu \nu}(q_1,q_2) 
\mathcal{A}_{\rm gluon}^{\mu \nu} (p, \tilde{q}).
\label{Dqq}
\end{equation}
The tensor $P^{(q \bar q)}_{\mu \nu}$  is given in Eq.~\eqref{eq:P(qq)}.
The subtraction integral for $D^{(q\bar q)}$
is evaluated analytically in Sec.~\ref{app:qqbar} of Appendix~\ref{sec:integralpoles}.
It is given in Eqs.~\eqref{Dqq_integrated} and \eqref{kernelVqq}.

\subsection{Insensitivity to cut parameters}

\label{sec:valudation}

The finite   parts of the real NLO corrections to the fragmentation function for $g \rightarrow Q \bar Q_8$
can be obtained by adding  two contributions:
\begin{itemize}
\item
the subtracted real NLO corrections,
which are given by the first two  integrals on the right side of Eq.~\eqref{eq:real_emission_subtracted}.
The integrals over the phase space
of the two final-state partons in 4 dimensions are evaluated numerically.
\item
the finite parts of the subtractions for the real NLO corrections,
which are given by the  last two  integrals on the right side of Eq.~\eqref{eq:real_emission_subtracted}.
They are the difference between the sum of the finite parts of the subtraction integrals 
in Eqs.~\eqref{eq:subint-zetaj}, \eqref{eq:subint-uj,deltaj},
\eqref{eq:subint-lambda}, and \eqref{Dqq_integrated}
and the sum of the finite parts of the over-subtraction integrals 
in Eqs.~\eqref{eq:subint-zetajuj}, \eqref{eq:subint-ujdelta}, \eqref{eq:subint-uj}, and \eqref{eq:subint-lambdauj}.
Most of these finite parts include one-dimensional or two-dimensional integrals that are evaluated numerically.
\end{itemize}
 Both contributions depend
on the cut parameters $u^{\rm cut}$, $\lambda^{\rm cut}$, $\delta^{\rm cut}$, and $\zeta^{\rm cut}$ that
define the phase-space regions where the subtractions are applied.
The dependence  on the cut parameters must cancel 
 between the two contributions.
The fact that the overall sum of finite pieces 
should be independent of the cut parameters can be used as a sanity check 
for the subtraction procedure. 

\begin{figure}
\center
\includegraphics[scale=0.64]{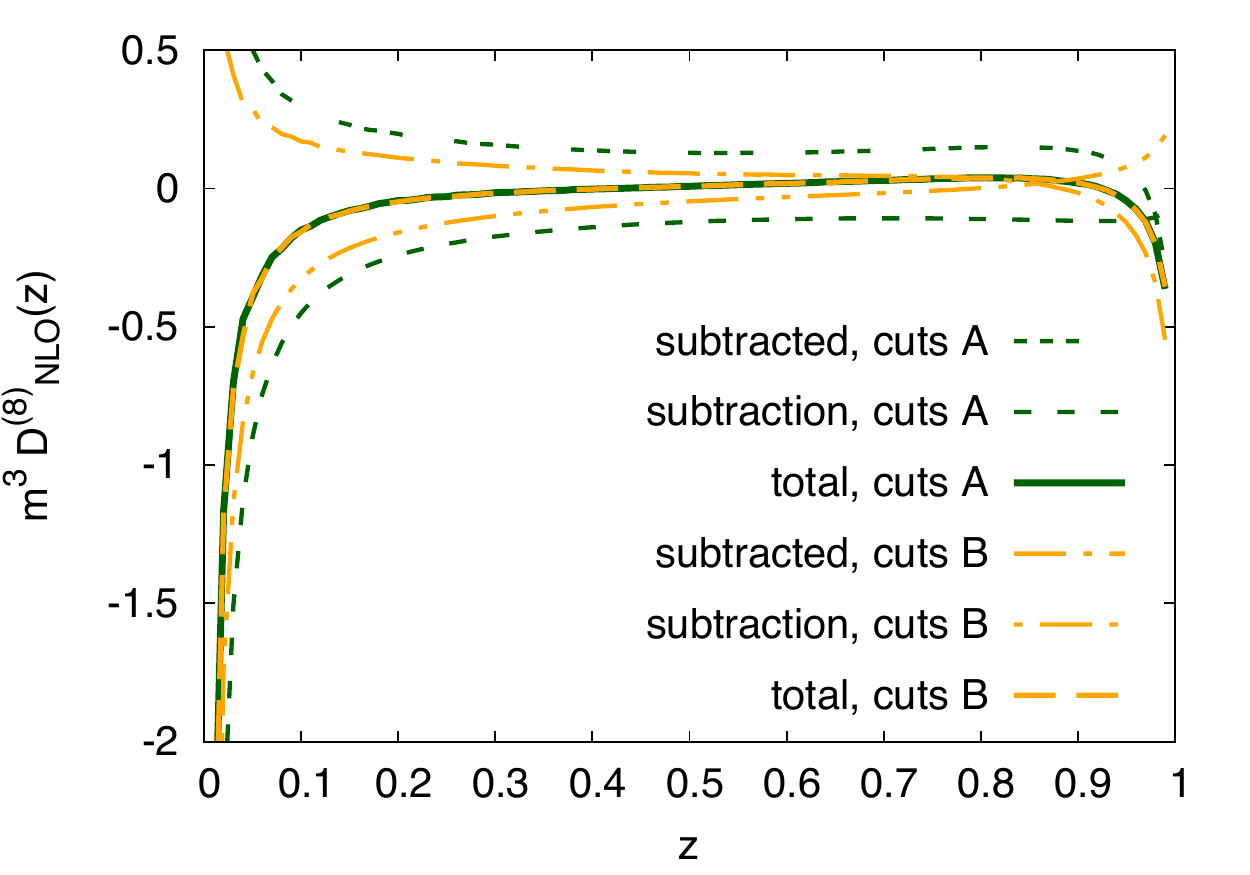}
\includegraphics[scale=0.64]{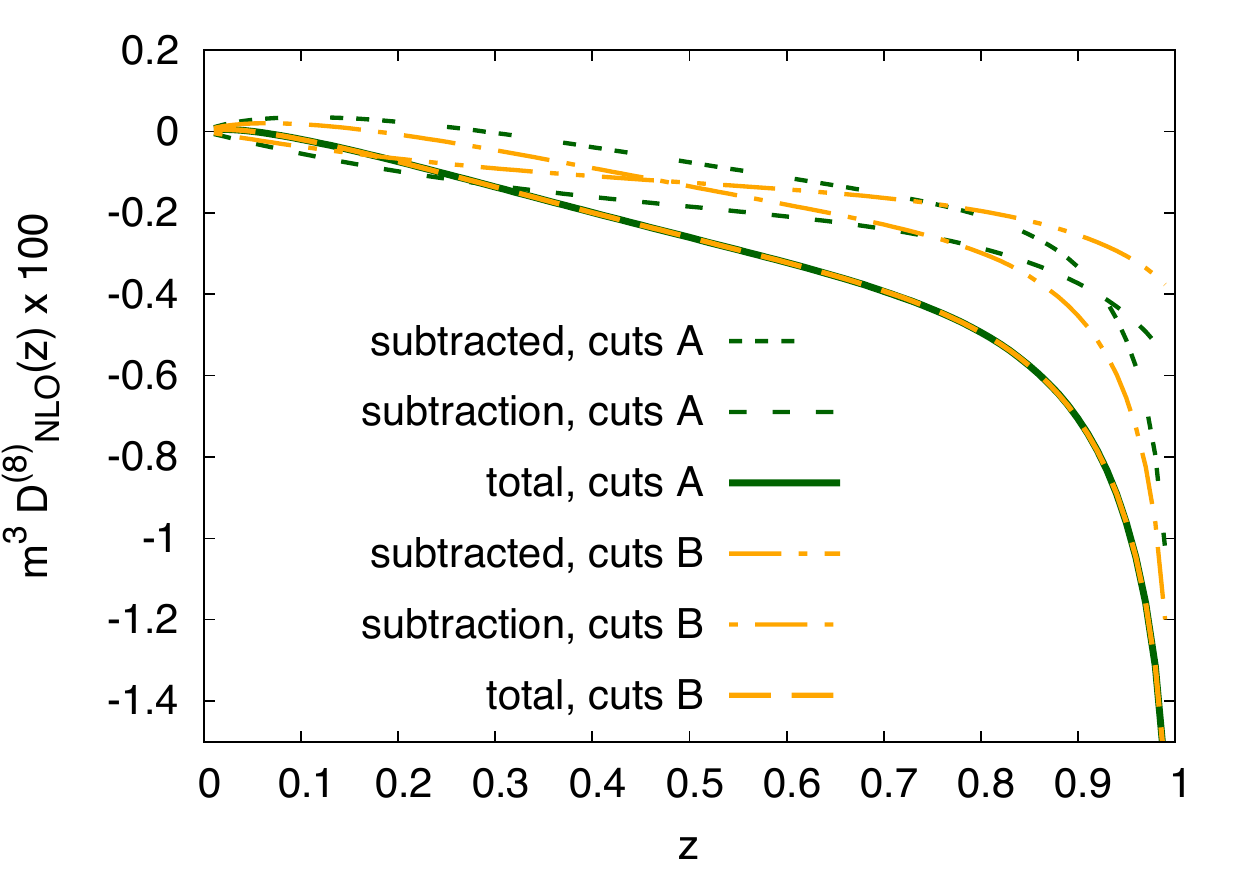}
\caption{
The  finite parts of the real contributions to the NLO fragmentation function $D^{(8)}_{\rm NLO}(z)$.
The fragmentation function with $n_f=0$ and the coefficient of $n_f$ in the fragmentation function
 are shown in the left panel and in the right panel, respectively.
The \textit{subtracted} and \textit{subtraction} contributions 
are associated with the decomposition in Eq.~(\ref{eq:real_emission_subtracted}).
Cuts A and B refer to the two sets of cut parameters specified in Eqs.~(\ref{eq:cutsA})
and (\ref{eq:cutsB}). The ``total'' curves for cuts A and cuts B are almost indistinguishable. }
\label{fig:NumericalValidation}
\end{figure}

The numerical  integrations 
are performed with the use of the
 adaptive Monte Carlo integrator Vegas~\cite{Lepage:1980dq}.
To validate the subtraction procedure, we use two sets of cut parameters:
\begin{subequations}
\begin{eqnarray}
\textrm{cuts A: } && u^{\rm cut}=0.1, \quad \lambda^{\rm cut}=0.6, \quad
\delta^{\rm cut} = 0.25, \quad \zeta^{\rm cut} = 0.25,
\label{eq:cutsA}
\\
\textrm{cuts B: } && u^{\rm cut}=0.2, \quad \lambda^{\rm cut}=1.2, \quad
\delta^{\rm cut} = 0.50, \quad \zeta^{\rm cut} = 0.50.
\label{eq:cutsB}
\end{eqnarray}
\end{subequations}
The fragmentation function with $n_f=0$ comes
from cut diagrams with two gluons crossing the cut.
The coefficient of $n_f$ in the fragmentation function comes
from cut diagrams with a  light quark  and antiquark crossing the cut.
These two contributions are shown in Figure~\ref{fig:NumericalValidation}.
For both contributions, 
the sum of the subtracted terms and  the finite  parts 
of the subtraction terms is seen to be independent of the choice of values 
for the cut parameters.


\section{Virtual NLO corrections}
\label{sec:NLOvirtual}

The virtual NLO corrections to the perturbative fragmentation function for 
  $g \rightarrow Q \bar Q_8$, where
the $Q \bar Q$ pair is in a  color-octet $^1S_0$ state, 
come from cut diagrams with one loop on either 
the right side or the left side of the cut. 
Loop diagrams on one side of the cut can be obtained 
from the LO diagrams, like the one in Figure.~\ref{fig:cutdiagram},
by adding a gluon line connecting any pair 
of the 6 colored lines, by adding a loop correction to the propagator 
of the fragmenting gluon, or by adding a loop correction to the propagator 
of the virtual heavy quark.
There are additional loop diagrams in which the heavy-quark line is attached 
to the eikonal line by both the fragmenting gluon that attaches to the 
 operator vertex and by a second gluon line, 
with the gluon that crosses the cut attached to either the fragmenting gluon 
or the eikonal line.

As in the LO cut diagrams,
we denote the equal momenta of the $Q$ and $\bar Q$ by $p$
and the momentum of the final-state gluon by $q$.
The sum of the virtual one-loop cut diagrams at order $\alpha_s^3$ defines a 
scalar function $\mathcal{A}_{\rm virtual}(p,q,l,n)$
that is integrated over  the loop momentum $l$.
The virtual NLO contribution to the fragmentation function can be expressed as
\begin{equation}
D_{g \to Q \bar Q_8}^{\rm (virtual)}(z) = 
N_{\rm CS} \int d\phi_{\rm Born}(p,q)
 \int \frac{d^Dl}{(2\pi)^D} \mathcal{A}_{\rm virtual}(p,q,l,n) \, ,
\label{eq:virtual_emission}
\end{equation}  
where $N_{\rm CS}$ is the Collins-Soper prefactor in Eq.~(\ref{eq:overalfac}) 
and $d\phi_{\rm Born}$
is the phase-space measure in Eq.~\eqref{eq:BornPSpq}. 
The integral of $\mathcal{A}_{\rm virtual}$ over the loop momentum $l$ 
is a homogeneous function of $n$ of degree 2.  It can be expressed as 
the product of $[(2p+q).n]^2/s^2$, where $s =(2p+q)^2$, 
and a dimensionless function of $s/m^2$ and the momentum fraction 
$z$ defined in Eq.~\eqref{eq:z-pq}.

By means of standard tensor reduction techniques,
the integral of $\mathcal{A}_{\rm virtual}$ over $l$ 
can be reduced to $[(2p+q).n]^2/s^2$ multiplied by a linear combination of one-loop scalar master integrals
whose numerators are simply 1 and whose coefficients are functions of $s$, $z$, 
the number of dimensions $D = 4 - 2 \epsilon$, and the number of colors $N_c$.
The denominators of the master integrals come from at most three Feynman propagators 
with mass $m$ or 0 and at most one eikonal propagator  
of the form $1/[(l+P).n + i \epsilon]$, where $P$ is a linear combination of $p$ and $q$.
The set of master integrals is the same as in the NLO calculation of the fragmentation function for 
$g \rightarrow Q \bar Q_1$  in Ref.~\cite{Artoisenet:2014lpa}.
Each master integral can be
expanded as a Laurent expansion in $\epsilon$ to order $\epsilon^0$.
The coefficients of the poles in $\epsilon$ can be evaluated analytically.
Analytic expressions for the poles in $\epsilon$ for the master integrals with an eikonal propagator 
are given in Appendix B of Ref.~\cite{Artoisenet:2014lpa}.
 The poles and the finite parts of  the virtual NLO correction can be readily obtained 
from the  calculation  of the fragmentation function for 
$g \rightarrow Q \bar Q_1$ in Ref.~\cite{Artoisenet:2014lpa}
by changing the coefficients of the master integrals.

In all the poles in $\epsilon$ from the loop integral in 
Eq.~(\ref{eq:virtual_emission}), the Born squared amplitude 
$\mathcal{A}_{\textrm{Born}}(p,q)$ in Eq.~\eqref{eq:ABorn-pq}
appears as a multiplicative factor.  
The virtual NLO corrections to the fragmentation function
can therefore be expressed as
\begin{eqnarray}
\label{virtual_correction0}
 D_{g \to Q \bar Q_8}^{\rm (virtual)}(z)
 = \frac{\alpha_s}{2 \pi}  
 \Gamma(1+\epsilon) \left(\frac{\pi \mu^2}{m^2}\right)^\epsilon 
  N_\textrm{CS} \int d\phi_{\rm Born}(p,q)   
  \big[f_\textrm{pole}(p,q) \mathcal{A}_{\textrm{Born}}(p,q) 
\nonumber \\
+ \mathcal{A}_{\textrm{finite}}(p,q) \big],
\end{eqnarray} 
where $f_\textrm{pole}(p,q)$ has only poles in $\epsilon$
and  $\mathcal{A}_{\textrm{finite}}(p,q)$ is a finite function of 
$s =  (2p+q)^2$ and $z$.
The terms in $f_\textrm{pole}(p,q)$ can be organized
to make their cancellation against the poles from other contributions 
of the NLO correction more transparent:  
\begin{equation}
\label{one-loop-div}
f_\textrm{pole}(p,q)  = 
\mathcal{U}^g + \mathcal{U}^Q(s) + 2\mathcal{U}^{gQ\bar Q}+ \mathcal{U}^{\rm eikonal}
+ \mathcal{M}(s) + \mathcal{S}_1 + \mathcal{S}_2(s,z) . 
\end{equation}
There are four terms in Eq.~(\ref{one-loop-div}) with only UV poles:
\begin{subequations}
\label{eq:virtual-div}
\begin{eqnarray}
\mathcal{U}^g & = & \left( \frac{5}{3} N_c
- \frac{4}{3} n_f T_F \right)\frac{1}{\epsilon_\textrm{UV}} ,
\\
\mathcal{U}^Q(s) & = &  C_F\frac{1}{\epsilon_\textrm{UV} }
 \left(  \frac{12m^2}{s-4m^2}  -1 \right),
 \\
\mathcal{U}^{gQ \bar Q} & = &  \left( N_c + C_F  \right)\frac{1}{\epsilon_\textrm{UV}} , 
\\
 \mathcal{U}^{\rm eikonal} & = &   
\label{eq:virtualeikonal-div}
N_c \frac{1}{\epsilon_\textrm{UV} } ,
\end{eqnarray}
\end{subequations}
where $C_F=(N_c^2-1)/(2N_c)$ is the Casimir for the fundamental representation.
In Feynman gauge, the terms  $\mathcal{U}^g$, $\mathcal{U}^Q(s)$,  
$\mathcal{U}^{gQ\bar Q}$, and $\mathcal{U}^{\rm eikonal}$ arise from 
virtual-gluon propagator corrections, virtual-quark propagator corrections, 
quark-gluon vertex corrections, and eikonal line corrections, 
respectively. 
There is one term in Eq.~(\ref{one-loop-div}) with mixed UV and IR poles:
\begin{equation}
\label{eq:calMmixed}
\mathcal{M}(s) = 2N_c \frac{1}{\epsilon_\textrm{UV}\epsilon_\textrm{IR}} 
\left[ 1-\epsilon \log(s/4m^2) \right] .
\end{equation}
In Feynman gauge, this term comes from loop correction to the operator vertex.
There are two terms in Eq.~(\ref{one-loop-div}) with only IR poles:
\begin{subequations}
\label{eq:S12IR}
\begin{eqnarray}
\label{eq:S1IR}
\mathcal{S}_1 &=&  \left( 2C_F - N_c \right)  \frac{1}{\epsilon_\textrm{IR}} ,
\\
\label{eq:S2IR}
\mathcal{S}_2(s,z) &=& N_c
\left[ -\frac{2}{\epsilon_\textrm{IR}^2}
+ \frac{1}{\epsilon_\textrm{IR}}
 \left( 2\log \frac{s}{4m^2} + \log \frac{s-4m^2}{4m^2}  + \log\left(z(1-z)\right) - 1 \right) \right].
\end{eqnarray}
\end{subequations}
The infrared poles originate from loop-momentum configurations 
with partons that can be soft and/or collinear. 
The term $\mathcal{S}_1$ is a soft pole that in Feynman gauge
comes from one-loop diagrams obtained from the four LO cut diagrams 
by exchanging a gluon between the on-shell heavy quarks.
All other infrared poles are included in the term $\mathcal{S}_2(s,z)$.

The virtual NLO corrections in Eq.~\eqref{virtual_correction0}
can be expressed as
\begin{eqnarray}
\label{virtual_correction}
 D_{g \to Q \bar Q_8}^{\rm (virtual)}(z)
 &=& \frac{\alpha_s}{2 \pi}  
 \Gamma(1+\epsilon) \left(\frac{\pi \mu^2}{m^2}\right)^\epsilon 
 \bigg[
\Big( \mathcal{U}^g + 2\mathcal{U}^{gQ\bar Q}
+ \mathcal{U}^{\rm eikonal} + \mathcal{S}_1 \Big)
 D_1(z)
\nonumber \\
&& \hspace{4cm}
+ \int Nd\phi \mathcal{A}_{\textrm{Born}}(s,z)
\Big( \mathcal{U}^Q(s)+ \mathcal{M}(s) + \mathcal{S}_2(s,z) \Big)
\nonumber  \\
&&\hspace{4cm}
+  N_\textrm{CS} \int d\phi_{\rm Born}(p,q)   
   \mathcal{A}_{\textrm{finite}}(p,q) \bigg],
\end{eqnarray}
where  $D_1(z)$ is the LO fragmentation function 
in $4-2 \epsilon$ dimensions in Eq.~\eqref{eq:D1-z} and $N d\phi \mathcal{A}_{\rm Born}$ is the 
LO differential fragmentation function in Eq.~\eqref{eq:NphiABorn}.

Many of the IR poles cancel against terms in the real NLO corrections,
which are given by the integrals of the subtraction terms in Eq.~\eqref{eq:real_emission_subtracted}.
The poles in the real NLO corrections are contained in the
difference between the sum of the subtraction integrals in 
Eqs.~\eqref{eq:subint-zetaj},  \eqref{eq:subint-uj,deltaj},
 \eqref{eq:subint-lambda}, and \eqref{Dqq_integrated}
and the sum of the over-subtraction integrals in 
Eqs.~\eqref{eq:subint-zetajuj}, \eqref{eq:subint-ujdelta}, \eqref{eq:subint-uj}, and \eqref{eq:subint-lambdauj}.
The mixed double pole in $\mathcal{M}(s)$ cancels 
against the $1/(\epsilon_\textrm{UV} \epsilon_\textrm{IR})$ terms in Eqs.~\eqref{eq:Idouble},
 \eqref{eq:Izetau2},  and \eqref{kernelVu}.
The double IR pole in $\mathcal{S}_2(s,z)$
cancels against the $1/\epsilon_\textrm{IR}^2$ terms
in Eqs.~\eqref{eq:Iudelta2}, \eqref{eq:Iudelta'2}, 
\eqref{eq:V1lambda}, \eqref{eq:Ilambda2}, \eqref{kernelVu}, and \eqref{kernelVlambdau}.
The  single pole proportional to $\log(z(1-z))$ in $\mathcal{S}_2(s,z)$
cancels against logarithmic terms in Eqs.~\eqref{eq:Iudelta1}  and  \eqref{eq:Ilambda2}. 
The  single poles proportional to $\log(s/4m^2)$   in $\mathcal{M}(s)$ and $\mathcal{S}_2(s,z)$ 
cancel against terms in Eqs.~\eqref{eq:Idouble} and \eqref{eq:Izetau2},
leaving a single  pole proportional to $\log u^{\rm cut}$.
The  single pole proportional to $\log((s-4m^2)/4m^2)$  in $\mathcal{S}_2(s,z)$  
cancels against terms in Eqs~\eqref{eq:Iudelta2},  \eqref{eq:Iudelta'2}, and \eqref{eq:subint-lambda},
leaving single  poles proportional to $\log u^{\rm cut}$ and $\log \lambda^{\rm cut}$.
The single poles proportional to $\log u^{\rm cut}$ and $\log \lambda^{\rm cut}$
that are left over from the cancellations of the logarithmic functions of $s$
cancel against additional single poles
from Eqs.~\eqref{eq:V1lambda}, \eqref{eq:Ilambda1}, \eqref{kernelVu}, and \eqref{kernelVlambdau}.
After these cancellations between the real NLO corrections 
and the virtual NLO corrections, the only poles that remain are single IR poles proportional to 
 $D_1(z)$  and single UV poles.


\section{Renormalization}
\label{sec:NLOrenorm}

The calculation of the fragmentation function is performed in terms of 
the renormalized fields $\Psi_r$ and $A_r$, 
the renormalized coupling constant $g$, 
and the physical mass $m$ of the heavy quark.
Their relations with the corresponding bare quantities 
involve renormalization constants $\delta_2$, 
$\delta_3$, $\delta_g$, and $\delta_m$:
\begin{equation}
\Psi =  (1+\delta_2)^{1/2} \Psi_r, \quad 
A^\mu = (1+\delta_3)^{1/2} A_r^\mu, \quad
g_0 = \mu^\epsilon (1+\delta_g) g, \quad
m_0 = m(1+\delta_m). 
\end{equation}
The renormalization of the  coupling constant is performed in
the $\overline{\textrm{MS}}$ scheme, whereas the renormalization
of the heavy-quark mass is performed in the on-shell mass scheme.
In the resulting expressions for the renormalization constants 
$\delta_2$, $\delta_3$, $\delta_g$, and $\delta_m$, it is convenient to 
pull out a common factor:
\begin{equation}
\delta_i = \frac{\alpha_s}{2 \pi} \Gamma(1+\epsilon) \left(\frac{\pi \mu^2}{m^2}  \right)^\epsilon \tilde{\delta}_i .
\end{equation} 
The rescaled renormalization constants $\tilde{\delta}_i$
in the schemes specified above  
read 
\begin{subequations}
\begin{eqnarray}
\tilde{\delta}_2 & = & - \frac{C_F}{2}  \left[\frac{1}{\epsilon_\textrm{UV}}
+\frac{2}{\epsilon_{\textrm{IR}}} + 4 + 6 \log 2  \right] ,
\\  
\tilde{\delta}_3 & = &  
\left(\frac56 N_c - \frac23  n_f T_F \right)
 \left[ \frac{1}{\epsilon_\textrm{UV}}-\frac{1}{\epsilon_\textrm{IR}}  \right] ,
\\ 
\label{eq:deltag}
\tilde{\delta}_g & = & -  \frac{b_0}{2} 
\left[ \frac{1}{\epsilon_\textrm{UV}} + \log \frac{4m^2}{\mu_R^2}  \right] ,
\\
\tilde{\delta}_m &=& - \frac{3 C_F}{2} 
\left[ \frac{1}{\epsilon_\textrm{UV}} + \frac{4}{3}+ 2 \log 2  \right] ,
\end{eqnarray}
\end{subequations}
where  $b_0=(11N_c-4 n_fT_F)/6$ 
is the coefficient of $-\alpha_s^2/\pi$ in the beta function  $(\mu d/d\mu)\alpha_s(\mu)$.
In the counterterm for $g$ in Eq.~\eqref{eq:deltag},
we have allowed for  the renormalization scale $\mu_R$ of $\alpha_s$
to be different from the scale $\mu$ introduced through dimensional regularization.
 
The contributions to the NLO fragmentation function from inserting propagator counterterms 
and vertex counterterms into the LO cut diagrams are
\begin{eqnarray}
\label{ct-def}
D_{g \to Q \bar Q_8}^{\rm (counter)} (z) &=& 
\frac{\alpha_s}{2 \pi} \Gamma(1+\epsilon) \left(\frac{\pi \mu^2}{m^2}  \right)^\epsilon 
 \int N d\phi \mathcal{A}_{\textrm{Born}}(s,z)   
 \left[  \mathcal{C}^g + \mathcal{C}^Q(s) + 2\mathcal{C}^{gQ\bar Q}+ \mathcal{C}^{\rm{eikonal}}   \right].
 \nonumber \\
\end{eqnarray} 
The terms with the coefficients $\mathcal{C}^g$, $\mathcal{C}^Q(s)$, $\mathcal{C}^{gQ\bar Q}$, 
and $\mathcal{C}^{\rm{eikonal}}$  are associated with the virtual-gluon propagator, the virtual-quark propagator,
the quark-gluon vertices, and the eikonal-gluon vertex
in the LO cut diagrams, respectively. The expressions for these coefficients 
in terms of the rescaled renormalization constants $\tilde{\delta}_i$'s are
\begin{subequations}
\label{eq:ct-div}
\begin{eqnarray}
\mathcal{C}^g &=& -2\tilde{\delta}_3,
\\
\mathcal{C}^Q(s) &=&  \frac{8m^2}{s-4m^2} \tilde{\delta}_m 
-2  \tilde{\delta}_2 \, ,
\\
\mathcal{C}^{gQ \bar Q} &=& 2 \tilde{\delta}_g+2\tilde{\delta}_2 + \tilde{\delta}_3 ,
\\  
\mathcal{C}^{\rm{eikonal}} &=& \tilde{\delta}_3.
\label{eq:cteikonal-div}
\end{eqnarray}
\end{subequations}
The counterterm contributions to the NLO fragmentation function can be reduced to
\begin{eqnarray}
\label{eq:NLOcount}
D_{g \to Q \bar Q_8}^{\rm (counter)} (z) &=& 
\frac{\alpha_s}{2 \pi} \Gamma(1+\epsilon) \left(\frac{\pi \mu^2}{m^2}  \right)^\epsilon 
\bigg[ \left( 4 \tilde{\delta}_g + 2\tilde{\delta}_2 + \tilde{\delta}_3  \right)
D_1(z)
\nonumber \\
&& \hspace{4cm}
+  2\tilde{\delta}_m \int N d\phi \mathcal{A}_{\textrm{Born}}(s,z) \frac{4m^2}{s-4m^2} \bigg].
\end{eqnarray}

The field renormalization constants $\tilde{\delta}_2$ and $\tilde{\delta}_3$
 have single IR poles that must cancel the single IR poles that remain after adding 
the real NLO corrections and the virtual NLO corrections.
The linear combination 
$2 \tilde{\delta}_2 + \tilde{\delta_3}$ in Eq.~\eqref{eq:NLOcount}
has infrared poles proportional to  $C_F$, $n_fT_F$, and $N_c$.
The IR pole proportional to  $C_F$ in $\tilde{\delta}_2$ cancels the $C_F$ term 
in the IR pole in $\mathcal{S}_1$  in Eq.~\eqref{eq:S1IR}.
The IR pole proportional to  $n_fT_F$ in $\tilde{\delta}_3$ cancels the IR pole  in Eq.~\eqref{kernelVqq}. 
The IR pole proportional to  $N_c$ in $\tilde{\delta}_3$ cancels 
single IR poles in $\mathcal{S}_1$ and $ \mathcal{S}_2(s,z)$ and single IR poles  in 
Eqs.~\eqref{eq:Iudelta'1},  \eqref{eq:V1lambda}, and \eqref{eq:Ilambda1}.
This completes the verification of the cancellation of the IR poles.

The renormalization of the operator defining the fragmentation function 
introduces an additional counterterm.
Its expression in the $\overline{\textrm{MS}}$ scheme reads\footnote{
There is a typographical error in the analogous expression in Eq.~(5.8) 
of Ref.~\cite{Artoisenet:2014lpa}:
$D_{g \to Q \bar Q}^{\rm (LO)}(z)$ should be $D_{g \to Q \bar Q}^{\rm (LO)}(z/y)$.}
\begin{equation}
D_{g \to Q \bar Q_8}^{\rm (operator)} (z) =
-\frac{\alpha_s}{2 \pi} \Gamma(1+\epsilon)
\left( \frac{\pi \mu^2}{m^2} \right)^\epsilon 
\left[ \frac{1}{\epsilon_\textrm{UV}} + \log \frac{4 m^2}{\mu_F^2}  \right] 
\int_z^1 \frac{dy}{y} P_{gg} (y) D_1(z/y)  ,
\label{operator_ren1}
\end{equation}
where $P_{gg} (y)$ is the Altarelli-Parisi splitting function:
\begin{equation}
\label{eq:Pgg}
P_{gg}(z)= 2N_c \left[ \frac{z}{(1-z)_+} + \frac{1-z}{z} + z(1-z)  \right]   
+ b_0 \delta(1-z) .
\end{equation}
We have allowed for  the factorization scale $\mu_F$
to be different from the scale $\mu$ introduced through dimensional regularization.

The UV poles in the sum of the real NLO corrections in Eq.~(\ref{eq:real_emission_subtracted})
and the virtual NLO corrections  in Eq.~(\ref{virtual_correction})
must be canceled by the counterterms  in Eqs.~(\ref{ct-def}) and (\ref{operator_ren1}).
The UV poles in the NLO virtual corrections from the  coefficients 
$\mathcal{U}^g$, $\mathcal{U}^Q(s)$, and $\mathcal{U}^{gQ \bar Q}$ 
in Eqs.~\eqref{eq:virtual-div} are canceled by the corresponding counterterms
$\mathcal{C}^g$, $\mathcal{C}^Q(s)$, and  $ \mathcal{C}^{gQ \bar Q}$ in Eqs.~(\ref{eq:ct-div}).
The UV pole from real NLO corrections  in Eq.~(\ref{eq:Isingle}) is a convolution integral over $y$.
It is canceled by the contribution to the operator counterterm in Eq.~(\ref{operator_ren1}) from the region  $y<1$.
The contribution to the virtual NLO corrections from the  
coefficient $\mathcal{U}^{\rm{eikonal}}$ in Eq.~(\ref{eq:virtualeikonal-div})
is canceled by the sum of  the eikonal counterterm $\mathcal{C}^{\rm{eikonal}}$ in Eq.~(\ref{eq:cteikonal-div})
and the contribution to the operator counterterm  in Eq.~(\ref{operator_ren1}) from the endpoint $y=1$.
This completes the verification of the cancellation of the UV poles.

The complete NLO term $D_{g \to Q \bar Q_8}^{\rm (NLO)} (z)$ in the fragmentation function
for $g \to Q \bar Q_8$ is obtained by adding the real NLO corrections in Eq.~(\ref{eq:real_emission_subtracted}),
the virtual NLO corrections  in Eq.~(\ref{virtual_correction}), 
and the counterterms in Eqs.~(\ref{ct-def}) and (\ref{operator_ren1}), and then taking the limit $\epsilon \to 0$.
After dividing by the perturbative NRQCD matrix element in Eq.~\eqref{eq:<O8>QQbar},
we obtain the NLO fragmentation function $D_{\rm NLO}^{(8)}(z)$ in Eq.~\eqref{eq:DNLOeta}
multiplied by $\alpha_s^3$.


\section{Numerical results}
\label{sec:numres}

Beyond leading order in $\alpha_s$,
the fragmentation function $D_{g \to \eta_Q}(z)$ depends on a 
factorization scale $\mu_F$ and on a renormalization scale $\mu_R$.
If we make those scales explicit, the expression for the color-octet $^1S_0$ term in
the NLO fragmentation function for $g \to \eta_Q$ in Eq.~\eqref{eq:DNLOeta} is
\begin{equation}
\label{eq:DNLOeta-scales}
D_{g \to \eta_Q}(z) =
\langle{\cal O}_8(^1S_0)\rangle^{\eta_Q} 
\left[ \alpha_s^2(\mu_R) \, D_{\rm LO}^{(8)}(z) + \alpha_s^3(\mu_R) \, D_{\rm NLO}^{(8)}(z;\mu_R,\mu_F) 
+ \ldots \right].
\end{equation}
The scales $\mu_F$ and $\mu_R$ were introduced through renormalization.
The Altarelli-Parisi evolution equation for the fragmentation function can be used to sum up 
large logarithms of $\mu_F/m$ to all orders in $\alpha_s$.  The solution to the evolution equation 
is sensitive to the behavior of the NLO fragmentation function near the upper endpoint $z \to 1$.
The fragmentation function includes terms with endpoint singularities 
 of the form $\log(1-z)$ and $\log^2(1-z)$.
It is worthwhile to make those singularities explicit.
Our final result for the NLO term in the  fragmentation function has the form
\begin{eqnarray}
\label{eq:NLOFF}
D_{\rm NLO}^{(8)}(z;\mu_R,\mu_F) = 
 \frac{b_0}{ \pi}  \log \frac{\mu_R^2}{4 m^2}  D^{(8)}_{\rm LO}(z)
+ \frac{1}{2 \pi}\log \frac{\mu_F^2}{4 m^2}  \int_z^1 \frac{dy}{y} P_{gg} (y) D^{(8)}_{\rm LO}(z/y)
\nonumber \\
+ D^{(8)}_{\rm sing}(z)+ D^{(8)}_{\rm finite}(z) ,
\end{eqnarray}
where $D^{(8)}_{\rm LO}(z)$ is the LO fragmentation function in Eq.~\eqref{eq:dLO8-z},
$D^{(8)}_{\rm finite}(z)$ is a function with a smooth limit as $z \to 1$,
and $D^{(8)}_{\rm sing}(z)$ is a function with logarithmic singularities in that limit:
\begin{equation}
D^{(8)}_{\rm sing}(z) = \frac{c^{(8)}_2 \log^2(1-z) +  c^{(8)}_1 \log(1-z)}{m^3}.
\label{eq:D8sing}
\end{equation}
The coefficients $c^{(8)}_1 $ and $c^{(8)}_2 $ and the  function 
$D^{(8)}_{\rm finite}(z)$  can be calculated numerically.
 The numerical values  $c^{(8)}_1 $ and $c^{(8)}_2 $
are estimated by applying  linear regression in the region $10^{-5}<1-z <10^{-3}$:
\begin{subequations}
\label{eq:fit}
  \begin{eqnarray}
c^{(8)}_1 &=&  (-0.012 \pm 0.002) + ( 0.00278  \pm 0.00001 )  n_f   \, ,    
\label{eq:fitA}\\
c^{(8)}_2  &=& (-0.0472 \pm 0.001)  \, . 
 \label{eq:fitB}  
\end{eqnarray}
\end{subequations}

\begin{figure}
\center
\includegraphics[scale=0.64]{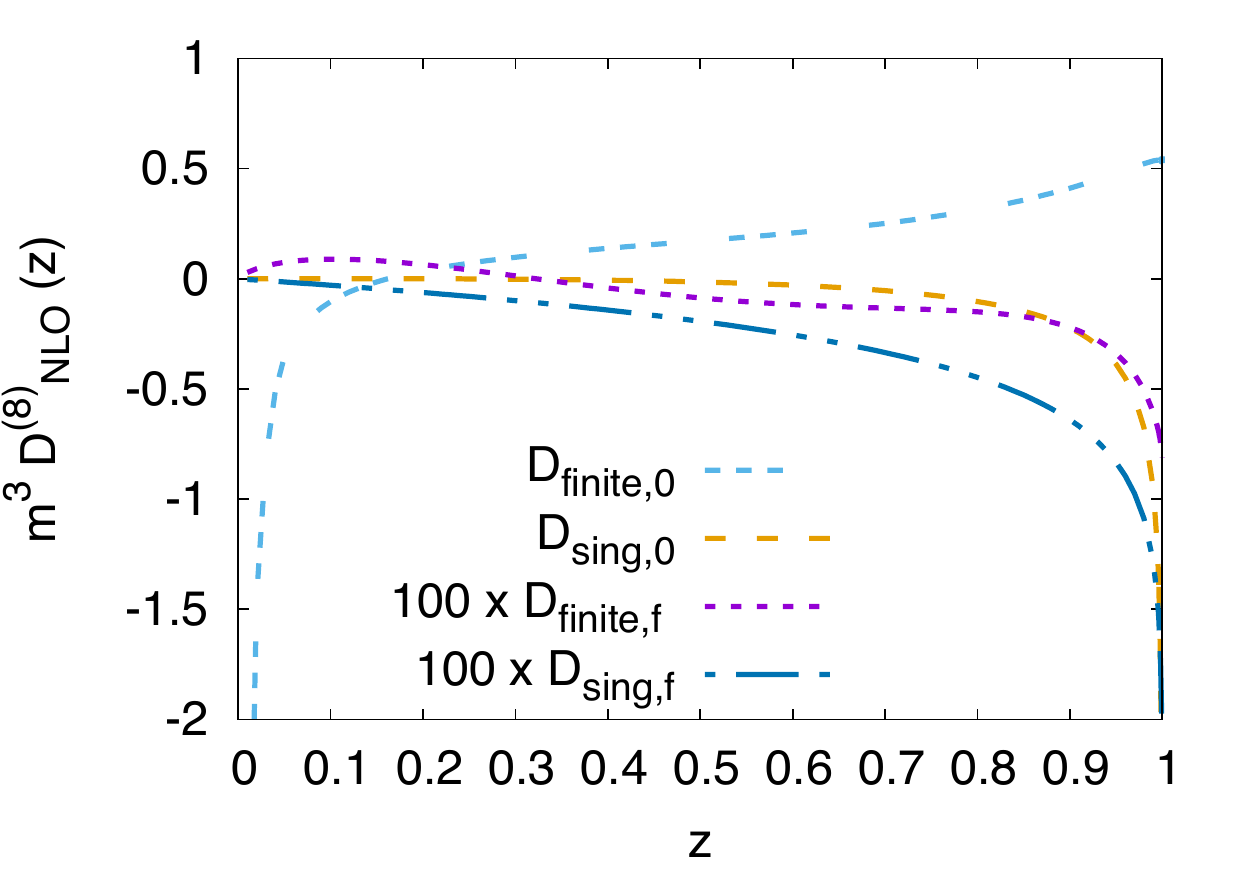}
\includegraphics[scale=0.64]{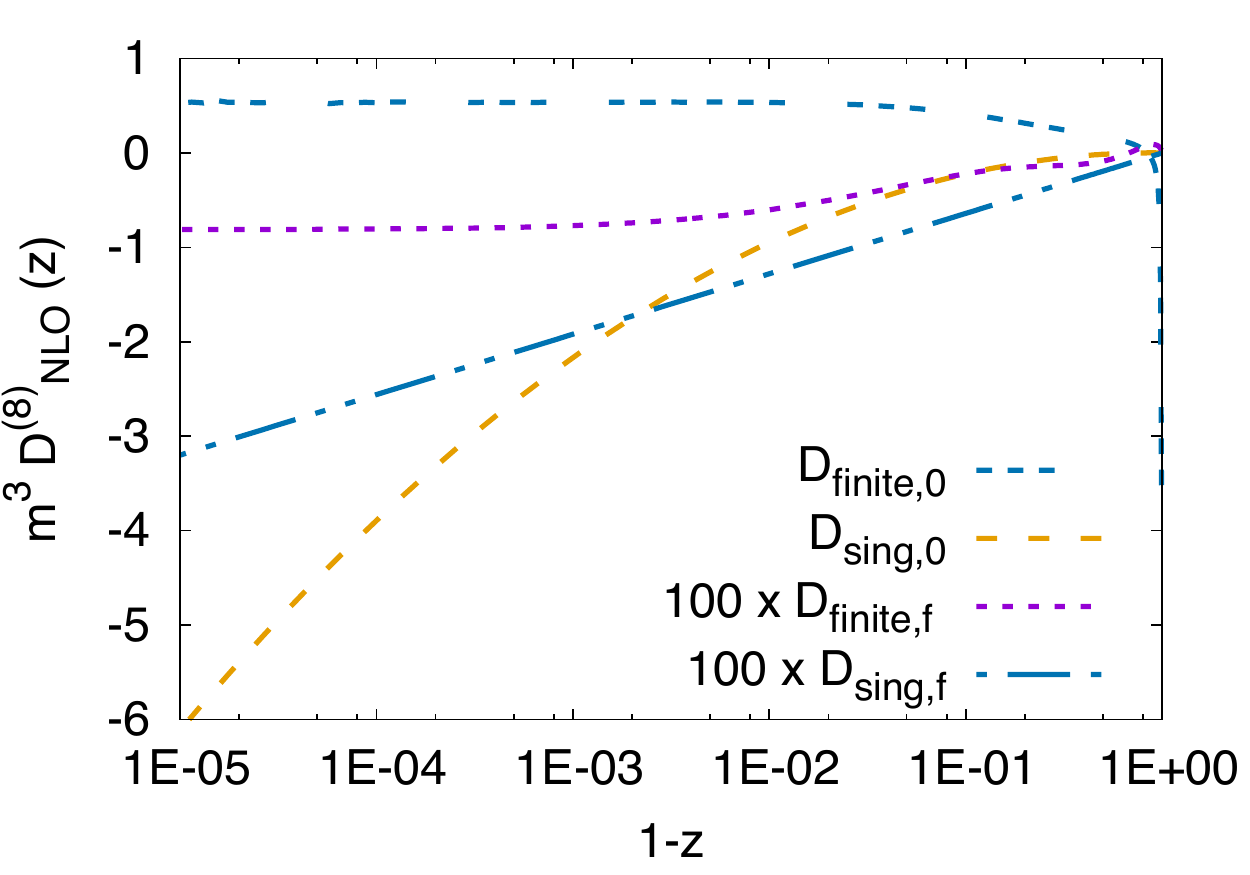}
\caption{The singular contributions $D^{(8)}_{\rm sing}(z)$ and the remaining finite contributions 
$D^{(8)}_{\rm finite}(z)$ to the NLO fragmentation function.
The same curves are shown as a function of $z$ with a linear scale (left panel)
and as a function of $1-z$ with a logarithmic scale (right panel). 
In the labels for the curves, a subscript 0 indicates the $n_f=0$ term
and a subscript $f$ indicates the term from one additional light-quark flavor.
}
\label{fig:numerical}
\end{figure}

The uncertainties are the estimated standard deviations of the parameter estimators, 
assuming that the errors in the regression model are independent and normally distributed.  
We found no numerical evidence for a $\log^2(1-z)$ contribution proportional to $n_f$.
In the numerical analysis presented in this section, we use the  
central values in the estimates of $c^{(8)}_1$ and $c^{(8)}_2$ in Eqs.~\eqref{eq:fit}.
Numerical values for the function $D^{(8)}_{\rm finite}(z)$ are calculated by subtracting
the singular term $D^{(8)}_{\rm sing}(z)$ defined by Eq.~\eqref{eq:D8sing}
from the fragmentation function $D_{\rm NLO}^{(8)}(z)$ with $\mu_R=\mu_F=2m$:
\begin{equation}
D^{(8)}_{\rm finite}(z) = \frac{1}{m^3} \big[   D_0(z)  + n_f D_f(z)  \big] .
\label{eq:DecomposeDfinite}
\end{equation}
The resulting curves are shown in Fig.~\ref{fig:numerical}, both as a function of $z$
with a linear scale  (left panel) and as a function of $1-z$ with a 
logarithmic scale   (right panel). By construction, 
the curves for $D^{(8)}_{\rm finite}(z)$ with $n_f=0$
and for the contribution to $D^{(8)}_{\rm finite}(z)$ from one additional light-quark flavor
approach plateaus when $1-z$ gets very close to 1.

\begin{figure}
\center
\includegraphics[scale=0.64]{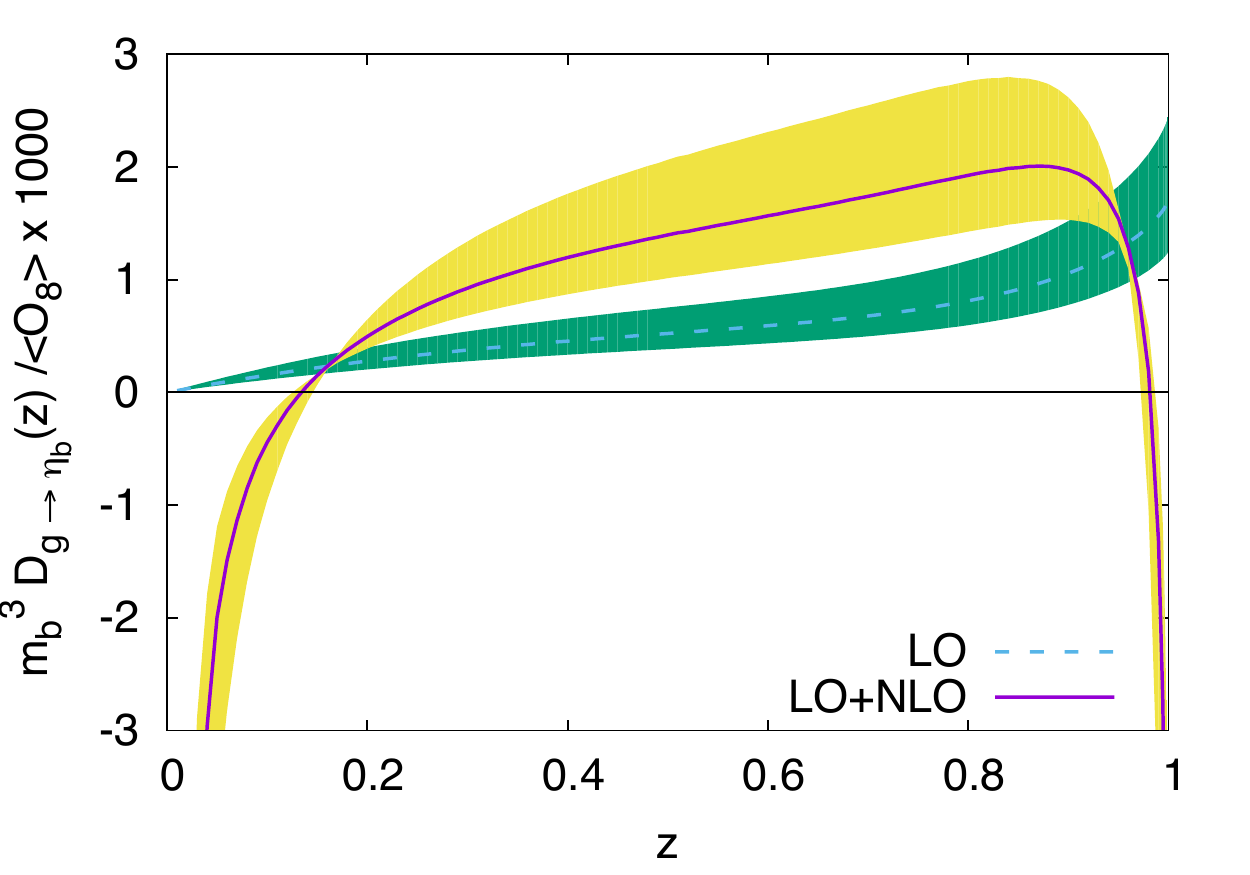}
\includegraphics[scale=0.64]{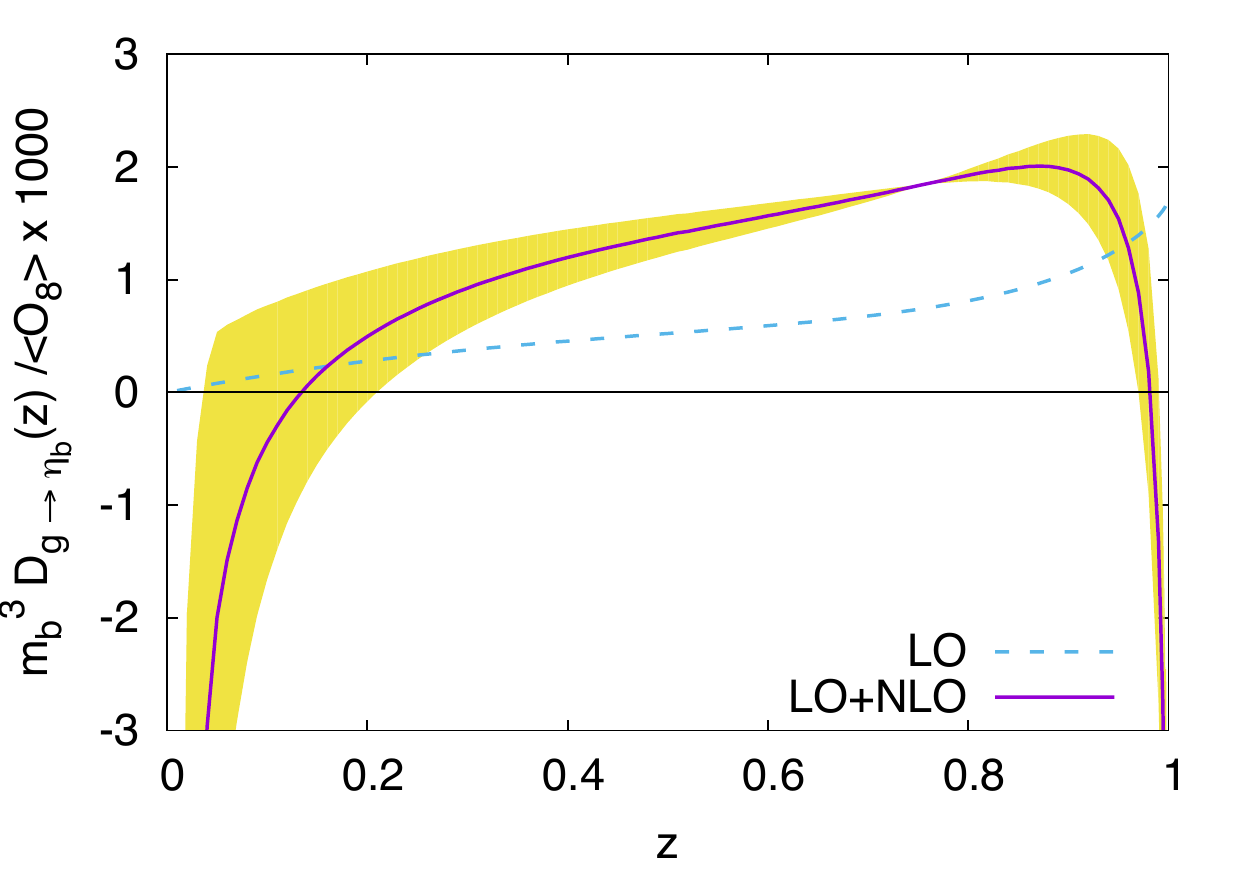}
\caption{ 
The color-octet $^1S_0$ contribution to the fragmentation function $D_{g \to \eta_b}(z)$ at LO and at NLO.
The curves are the LO fragmentation function (dashed line) and 
the sum of the LO and NLO terms in the fragmentation function
(solid line) for the scale choices $\mu_R=\mu_F = 2 m_b$.
The bands are obtained by varying the renormalization scale $\mu_R$ by a factor of 2 
(left panel) and varying the factorization scale $\mu_F$ by a factor of 2 (right panel).
}
\label{fig:NumericalResultsR}
\end{figure}

The NLO fragmentation function is compared with the LO fragmentation function
in Figure~\ref{fig:NumericalResultsR} for the case of bottomonium.
The LO fragmentation function is proportional to $\alpha_s^2(\mu_R) D_\textrm{LO}^{(8)}(z)$,
where $D_\textrm{LO}^{(8)}$ is given in Eq.~\eqref{eq:dLO8-z}.
The NLO fragmentation function is proportional to the sum of $\alpha_s^2(\mu_R) D_\textrm{LO}^{(8)}(z)$ and
$\alpha_s^3(\mu_R)D_\textrm{NLO}^{(8)}(z;\mu_R,\mu_F)$,
where $D_\textrm{NLO}^{(8)}$ is given in Eq.~\eqref{eq:NLOFF}.
We set  $m_b=4.75$ GeV and $n_f=4$, and we use the value 
$\alpha_s(\mu_R=2m_b)=0.181$ for the strong coupling constant. 
For the central values of the renormalization and factorization scales,
we choose twice the mass of the heavy quark:
$\mu_R=\mu_F = 2 m_b$. 
The LO term $\alpha_s^2D_\textrm{LO}^{(8)}(z)$
increases monotonically from 0 to $5\alpha_s^2/(96 m_b^3)$
as $z$ increases from 0 to 1.
The NLO term $\alpha_s^3D_\textrm{NLO}^{(8)}(z)$
increases from $-\infty$ as $z\to 0$
to a  maximum at  $z\approx 0.8$,
and then decreases to $-\infty$ as $z \to 1$.
For $\mu_R=\mu_F = 2 m_b$, its maximum is  $1.1 \times 10^{-3}/m_b^3$ at $z=0.80$.
At $z=0.5$, the NLO fragmentation function is larger than the
LO fragmentation function by a factor of 2.67.
The NLO term is large and negative in the $z \to 1$ region because of the $\log^2(1-z)$ term.
To determine the fragmentation function accurately as a function of $z$ in this region,
it would be necessary to sum the leading logarithms of $1-z$ to all orders.
This is not essential because the logarithmic singularities as $z \to 1$ are integrable. 
The NLO term is also large and negative in the $z \to 0$ region.
This may arise from a singular term of the form $\log z$.
The large NLO corrections in this region may not be a problem,
because kinematics generally excludes contributions from the small-$z$
region of the fragmentation function.

The sensitivity of the LO and NLO fragmentation functions  to the renormalization scale $\mu_R$ 
and to the factorization scale $\mu_F$ is also illustrated in Figure~\ref{fig:NumericalResultsR}.
The bands are obtained by varying  $\mu_R$ or $\mu_F$ up or down
by a factor of 2 around the central value $2 m_b$
(with  the other scale held fixed). 
The left panel of Figure~\ref{fig:NumericalResultsR} shows the NLO band 
from varying $\mu_R$ (with $\mu_F=2m_b$). 
The NLO band is significantly wider than the  LO band in the central region of $z$.
One might have expected the sensitivity to $\mu_R$ 
to be decreased  by adding NLO corrections, but this is not the case
simply because the NLO term in the fragmentation function is
larger than the LO term in the central region of $z$.
The right panel of Figure~\ref{fig:NumericalResultsR}
shows the band from varying  $\mu_F$ (with $\mu_R=2m_b$). 
In the central region of $z$, the width of the band from varying $\mu_F$ is much narrower
than that from varying $\mu_R$, partly because the function multiplying $\log (\mu_F^2/4m^2)$ 
in Eq.~\eqref{eq:NLOFF} vanishes near $z \approx 0.75$.
The width of the band  increases at larger $z$ and near the $z \to 0$ endpoint.


\section{Summary}
\label{sec:summary}

In this paper, we have presented the calculation of the NLO QCD corrections to the fragmentation function 
for a gluon into a color-octet $^1S_0$ $Q \bar Q$ pair at leading order in $v$. 
To calculate the real NLO correction, we developed a generalization of the FKS subtraction scheme
that can be applied to fragmentation functions. 
The UV and IR poles in the real NLO corrections arise from boundaries in the phase space for the final-state partons. 
We constructed subtraction terms for the integrand of the real NLO correction 
that cancel the singular contributions from each of these boundaries
and for which the poles in the subtraction integrals can be calculated analytically.
Our development of an FKS subtraction scheme for fragmentation functions 
paves the way toward the automated calculation of the NLO fragmentation functions in other NRQCD channels.

We found that the NLO QCD corrections have a dramatic effect on the fragmentation function. 
Instead of increasing monotonically with $z$ as at LO, 
the NLO fragmentation function has a maximum in the central region of $z$.
For $\mu_F = \mu_R = 2 m$, the NLO fragmentation function is larger than
the LO fragmentation function  by about a factor of 3. 
As a consequence, the NLO fragmentation function displays strong sensitivity to the renormalization scale. 
These results suggest that QCD corrections to the color-octet $^1S_0$ fragmentation function 
could have a significant impact on the production of the spin-singlet quarkonium states $\eta_c$ and $\eta_b$
 at large transverse momentum.

\begin{acknowledgments}
P.A. would like to thank Fabio Maltoni for enlightening discussions.
We thank Geoffrey Bodwin for useful comments.
P.A.~was funded in part by the F.R.S.-FNRS Fonds de la Recherche Scientifique (Belgium), 
and by the Belgian Federal Science Policy Office through the Interuniversity Attraction Pole P7/37.
E.B.~was supported in part by the Department of Energy under grant DE-SC0011726.
\end{acknowledgments}


\appendix

\section{Two-parton phase-space measures}
\label{sec:PhaseSpace}

The real NLO corrections to the fragmentation function
involve integrals over the phase space for two massless partons
whose longitudinal momenta are constrained to add up to $(K - 2p).n$.  
The dimensionally regularized phase-space measure is
\begin{equation}
\label{eq:dphiq1q2}
d \phi_{\rm real}(p,q_1,q_2) = 
[dq_1]\;  [dq_2] \; 2 \pi \delta\big(K.n - (2p + q_1 + q_2).n\big),
\end{equation}
where the single-particle phase-space measure in $D$ dimensions is 
\begin{equation}
\label{eq:dk}
[d k]
=\frac{d^{D-1} k}{(2 \pi)^{D-1} 2 k_0}.
\end{equation}
Explicit parametrizations of this phase-space measure 
are required in order to extract the poles in $\epsilon$ in the integrals of the
subtraction terms for the real NLO corrections.

In Appendix A of Ref.~\cite{Artoisenet:2014lpa},
two parametrizations of the two-parton phase-space measure were derived.
The parametrization in Eq.~(A.10) of Ref.~\cite{Artoisenet:2014lpa}
is used to extract the poles in the subtraction integral in Sec.~\ref{app:zetaj}.
Simple changes of variables are then used to obtain the parametrizations
used to extract the poles in the over-subtraction integral in Sec.~\ref{app:zetajuj}
and the first subtraction integral in Sec.~\ref{app:uj}.
The parametrization in Eq.~(A.17) of Ref.~\cite{Artoisenet:2014lpa}
is used to extract the poles in the over-subtraction integrals in 
Secs.~\ref{app:ujdelta} and \ref{app:lambdauj}, the subtraction integral in Sec.~\ref{app:qqbar},
the second subtraction integral in Sec.~\ref{app:uj},
and the first subtraction integral in Sec.~\ref{app:lambda}.
To extract the poles in the subtraction integral in Sec.~\ref{app:uj,deltaj}
and the second subtraction integral in Sec.~\ref{app:lambda},
different parametrizations of the two-parton phase-space measure are required.
These parametrizations are derived in this Appendix.

\subsection{Phase-space measure for $D^{(u_j,\delta_j)}$ }
\label{app:dphi2(uj,deltaj)}

The subtraction term  $D_1^{(u_j,\delta_j)}$ in Eq.~\eqref{eq:Duj,deltaj}
is  the product of the Born squared amplitude $\mathcal{A}_{\rm Born}(\breve s_j,z)$,
where $\breve s_j$ is defined in Eq.~\eqref{eq:brevesj},
and a simple function of two dimensionless variables:
\begin{equation}
\label{eq:uj,wj}
w_j = \frac{p.(q_1+q_2)-p.q_j}{m^2}, \qquad u_j = \frac{(q_1+q_2).n-q_j.n}{(q_1+q_2).n} .
\end{equation}
To obtain an expression for the phase-space measure  in Eq.~\eqref{eq:dphiq1q2}
that is differential in $w_j$ and $u_j$, we begin with the expression for   
the phase-space measure in Eq.~(A.6) of Ref.~\cite{Artoisenet:2014lpa}, 
which depends on two light-like 4-vectors $k_1$ and $k_2$ 
that define polar axes for the parton momenta $q_1$ and $q_2$:
\begin{equation}
\label{eq:dphiuk1k2}
d \phi_{\rm real}(p,q_1,q_2) =
\frac{2^{-2\epsilon}[(1-z)K.n]^{1-2\epsilon}}{(4 \pi)^{4-3 \epsilon}\Gamma(1-\epsilon)}\, 
\rho_1^{-\epsilon} d \rho_1\; 
\rho_2^{-\epsilon} d \rho_2\; 
[u_j(1-u_j)]^{-\epsilon} du_j\; d \Omega_{2\perp} ,
\end{equation}
where the scalar variables $\rho_j$ are defined by
\begin{equation}
\label{eq:rhoj-def}
\rho_j = 2k_j.q_j/k_j.n .
\end{equation}
The first light-like 4-vector $k_1$ should be a linear combination of $p^\mu$ and $n^\mu$
with coefficients that are scalar functions of $p$ and $n$.
The second light-like 4-vector $k_2$ should be a linear combination of $p^\mu$, $n^\mu$,  
and the first parton momentum $q_j^\mu$
with coefficients that are scalar functions of $p$, $n$,  and $q_j$.
The transverse angular measure $d \Omega_{2\perp}$ is that for the second parton momentum.

The appropriate choices for the two light-like 4-vectors in this case  are
\begin{equation}
\label{eq:lk1k2}
k_1^\mu = k_2^\mu = 2p^\mu -\frac{m^2}{p.n}n^\mu.
\end{equation}
The  first scalar variable $\rho_j$ can be expressed as
\begin{equation}
\label{eq:rhoj}
\rho_j = \frac{\breve s_j - 4m^2/z}{z K.n} .
\end{equation}
By using the identity $q_{3-j}=(q_1+q_2) - q_j$,
the  second scalar variable $\rho_{3-j}$ can be expressed as
\begin{equation}
\label{eq:rho3-j}
\rho_{3-j} = \frac{[w_j - (1-z)u_j/z]4m^2}{z K.n} .
\end{equation}
After changing variables from  $\rho_j$ and $\rho_{3-j}$ to $\breve s_j$ and $w_j$,
we obtain our final result for the two-parton phase-space measure:
\begin{equation}
\label{eq:dphiq1q2lambda}
d \phi_{\rm real}(p,q_1,q_2) =
d\phi_{\rm Born}(\breve s_j,z) \,d \phi_2^{(u_j,\delta_j)}(p,q_1,q_2),
\end{equation}
where $d\phi_{\rm Born}$ is the Born phase-space measure defined in Eq.~\eqref{eq:BornPSsz}
and $d \phi_2^{(u_j,\delta_j)}$ is the parton-emission measure
\begin{equation}
\label{eq:dphi2(uj,deltaj)}
d \phi^{(u_j,\delta_j)}(p,q_1,q_2) =  
\frac{(2\pi)^{2\epsilon}}{32\pi^3} \left( \frac{1-z}{z} 4m^2\right)^{1-\epsilon} 
\left[ w_j - \frac{1-z}{z} u_j \right]^{-\epsilon} dw_j \;
\big[ u_j (1-u_j)\big]^{-\epsilon} du_j\;  
 d\Omega_\perp .
\end{equation}
The range of $w_j$ is from $(1-z)u_j/z$ to $\infty$.  The range of $u_j$ is from 0 to 1.

\subsection{Phase-space measure for $D^{(\lambda)}$ }
\label{app:dphi2(lambda)}

After averaging over transverse angles,
the subtraction term  $D^{(\lambda)}$ in Eq.~\eqref{eq:Dlambda}
can be expressed as the product of the Born squared amplitude $\mathcal{A}_{\rm Born}(\tilde s,z)$,
where $\tilde s =(2 p + \tilde q)^2$ and $\tilde q$ is the 4-vector defined in Eq.~\eqref{eq:qtilde},
and the sum of two functions of dimensionless variables.
In the second function, the two dimensionless variables are
\begin{equation}
\label{eq:lambda,v}
\lambda = \frac{(q_1+q_2)^2}{4m^2}, \qquad v = \frac{p.q_2}{p.(q_1+q_2)}.
\end{equation}
To obtain an expression for the phase-space measure  in Eq.~\eqref{eq:dphiq1q2}
that is differential in $\lambda$ and $v$, we begin by expressing 
the phase-space measure in an iterated form
involving the product of the phase-space measure for $Q = q_1+q_2$
and the differential phase space for the decay of a particle with 4-momentum $Q$ 
into particles with 4-momenta $q_1$ and $q_2$:
\begin{equation}
\label{eq:dphiq1q2iterate}
d \phi_{\rm real}(p,q_1,q_2) = \frac{d(Q^2)}{2 \pi} [dQ] \,
2 \pi \delta\big(K.n - (2p + Q).n\big) \, d \Phi[Q,q_1,q_2] ,
\end{equation}
where the differential 2-particle phase space is
\begin{equation}
\label{eq:dphiQq1q2}
d \Phi[Q,q_1,q_2] = 
(2 \pi)^4 \delta^4(Q-q_1-q_2)\;  [dq_1]\;  [dq_2].
\end{equation}

The product of the phase-space measure for $Q$ and the delta function in Eq.~\eqref{eq:dphiq1q2iterate}
is most easily simplified in the rest frame of $P$.  The phase-space measure for $Q$ is
\begin{equation}
\label{eq:[dQ]}
 [dQ] =
 \frac{(4 \pi)^{2 \epsilon}}{8 \pi^2 \Gamma(1-\epsilon)} |\bm{Q}|^{1-2\epsilon} dQ_0 
 ( 1 - \cos^2\theta)^{- \epsilon} d \cos \theta,
\end{equation}
where $\theta$ is the angle between $\bm{Q}$ and $\bm{n}$ in that frame.
The energy $Q_0$ and the magnitude $|\bm{Q}|$ of the 3-momentum can be expressed 
in terms of Lorentz invariants:
\begin{equation}
\label{eq:Q0,|Q|}
Q_0 = \frac{s-Q^2 - 4 m^2}{4m} , \qquad
|\bm{Q}| = \frac{\lambda^{1/2}(s,Q^2,4 m^2)}{4m},
\end{equation}
where $\lambda(a,b,c) = a^2 + b^2+c^2-2(ab+bc+ca)$.
After dividing each term by $2p.n$, the constraint provided by 
the delta function in Eq.~\eqref{eq:dphiq1q2iterate} can be expressed as
\begin{equation}
\label{eq:ndot=0}
\frac{1}{z}  -1 -\frac{Q_0 - |\bm{Q}| \cos \theta}{2m} = 0.
\end{equation}
The delta function can be used to integrate over $\cos \theta$.
It is convenient to change variables  from $s=(2p+Q)^2$ to $\tilde s = (2p+\tilde q)^2$:
\begin{equation}
\label{eq:stilde-s}
\tilde s = s - Q^2/(1-z). 
\end{equation}
The resulting expression for the product of the phase-space measure and the delta function is
\begin{equation}
\label{eq:[Q]delta}
[dQ] \, 2 \pi \delta\big(K.n - (2p + Q).n\big)  =
\frac{(4 \pi)^\epsilon z^{-1+ \epsilon} (1-z)^{- \epsilon}}{8 \pi \Gamma(1-\epsilon) K.n} 
(\tilde s - 4m^2/z)^{-\epsilon} d \tilde s.
\end{equation} 

The differential 2-body phase-space measure $d \Phi(Q,q_1,q_2)$ is most easily simplified in the rest frame of $Q$.
It can be reduced to 
\begin{equation}
\label{eq:dphi2}
d \Phi[Q,q_1,q_2] =
\frac{(4\pi)^{2 \epsilon} }{32 \pi^2} (Q^2)^{- \epsilon} 
( 1 -  \cos^2 \theta)^{- \epsilon} d \cos \theta\, d\Omega_\perp,
\end{equation}
where $\theta$ is the angle between $\bm{P}$ and $\bm{q}_2$ in that frame
and $d\Omega_\perp$ is the transverse angular measure.
The dimensionless variable $v$ defined in Eq.~\eqref{eq:lambda,v}
can be expressed in terms of  $\cos \theta$ and Lorentz invariants:
\begin{equation}
\label{eq:vcostheta}
v = \frac12 \left(1 - \frac{\lambda^{1/2}(s,Q^2,4m^2)}{s-Q^2-4m^2} \cos \theta \right).
\end{equation}
The differential phase space then reduces to
\begin{equation}
\label{eq:dphi2simp}
d \Phi[Q,q_1,q_2] =
\frac{(2\pi)^{2 \epsilon}}{16 \pi^2} (Q^2)^{- \epsilon} 
\left(\frac{\lambda^{1/2}(s,Q^2,4m^2)}{s-Q^2-4m^2} \right)^{-1+2\epsilon}
\big[(v_+ - v)(v - v_-) \big]^{- \epsilon} dv\, d\Omega_\perp.
\end{equation}

Inserting Eqs.~\eqref{eq:[Q]delta} and \eqref{eq:dphi2simp} into Eq.~\eqref{eq:dphiq1q2iterate},
we obtain our final result for the two-parton phase-space measure:
\begin{equation}
\label{eq:dphiq1q2(lambda)}
d \phi_{\rm real}(p,q_1,q_2) =
d\phi_{\rm Born}(\tilde s,z) \,d \phi_2^{(\lambda)}(p,q_1,q_2),
\end{equation}
where $d\phi_{\rm Born}$ is the Born phase-space measure defined in Eq.~\eqref{eq:BornPSsz}
and $d \phi_2$ is the parton-emission measure
\begin{equation}
\label{eq:dphi2(lambda)}
d \phi_2^{(\lambda)}(p,q_1,q_2) =  
\frac{(2\pi)^{2\epsilon}}{32\pi^3} 
(4m^2)^{1-\epsilon} \lambda^{-\epsilon} d\lambda \;   
\frac{\big[(v_+-v) (v-v_-)\big]^{-\epsilon}}{(v_+ - v_-)^{1-2\epsilon}} dv \;
d\Omega_\perp .
\end{equation}
The variables $v_+$ and $v_-$ depend on $\lambda$ and $\tilde s$: 
\begin{equation}
v_\pm = \frac{1}{2} \left[1 \pm 
\frac{\lambda^{1/2}\big( 1+ \tilde r+ \lambda/(1-z), \lambda,1 \big)}
       {\tilde r +  z \lambda/(1-z)} \right],
\end{equation}
where $\tilde r = (\tilde s - 4 m^2)/4m^2$.
The range of $\lambda$ is from 0 to $\infty$.
The range of $v$ is from $v_-$ to $v_+$.


\section{Subtraction Integrals}
\label{sec:integralpoles}

Explicit expressions for the poles $\epsilon$ in the integrals of the subtraction terms 
can be obtained by carrying out the integration over a
$(3-2\epsilon)$-dimensional slice of phase-space.
The  choice of this slice depends on the subtraction term under consideration, 
which is labelled by a superscript $(A)$ that indicates the variables that vanish in the singular region.
The  general  decomposition of the phase-space measure reads 
\begin{equation}
\label{eq:Ndphi_general}
N_{\rm CS} d\phi_{\rm real}(p,q_1,q_2) = 
N_{\rm CS} \, d\phi_{\rm Born}(s^{(A)}, z^{(A)}) \, d \phi^{(A)} (p,q_1,q_2).
\end{equation}
The Collins-Soper prefactor $N_{\rm CS}$ is
defined in Eq.~(\ref{eq:overalfac}).
The  Born phase-space measure $d\phi_{\rm Born}$,
which is differential in  $s^{(A)}$, is defined in Eq.~\eqref{eq:BornPSsz}.
The {\it parton-emission measure} $d \phi^{(A)}$ can be reduced to a differential 
in two variables multiplied by a transverse angular measure $d\Omega_\perp$,
whose integral is 
\begin{equation}
\int d\Omega_\perp =2 \pi^{1-\epsilon}/\Gamma(1-\epsilon).
\end{equation}
For each subtraction term $(A)$, the choices of  the  arguments $s^{(A)}$ and $z^{(A)}$ of $d\phi_{\rm Born}$
and the choice of the measure $d \phi^{(A)}$ are designed to ease the extraction of the poles in $\epsilon$.

\subsection{Subtraction term proportional to $\bm{\theta^{(\zeta_j)}}$ }
\label{app:zetaj}

In the subtraction term in $T_{1,j}^{(gg)}$ that is proportional to $\theta^{(\zeta_j)} $,
the function $D^{(\zeta_j)}$ is given in Eqs.~\eqref{eq:Dzeta}, 
where the tensor $V^{(\zeta_j)}_{\mu\nu}$ is given in Eq.~\eqref{eq:Vzeta}.
To extract the poles in the integral of the subtraction term,
we use the decomposition of the phase-space measure in
Eqs.~(3.7) and (3.8) of Ref.~\cite{Artoisenet:2014lpa}.
The factor $N(p,q_j)$ is $N_{\rm CS} y_j^{-2+2\epsilon}$, where
$y_j$ is the momentum fraction defined in Eq.~(\ref{eq:defyi}).
The  factor $d\phi_{\rm Born}(p,q_j)$ 
can be expressed in terms of the measure defined in Eq.~\eqref{eq:BornPSsz}
as  $(1/y_j)d\phi_{\rm Born}(s_j,z/y_j)$, where $s_j=(2p+q_j)^2$.
In the decomposition of the phase-space measure in Eq.~(\ref{eq:Ndphi_general}), 
the Born phase-space measure is $d\phi_{\rm Born}(s_j,z/y_j)$ and
the  parton-emission measure is
\begin{equation}
d \phi^{(\zeta_j)}(p,q_1,q_2) =  
\frac{1}{4(2\pi)^{3-2\epsilon}} (s - s_j/y_j)^{-\epsilon}  ds \; 
y_j^{-2+\epsilon} (1-y_j)^{-\epsilon} dy_j\;  d\Omega_\perp .
\label{eq:def_dphi1_UV}
\end{equation}

The extraction of the poles follows the same approach 
as in Section~3.3 of Ref.~\cite{Artoisenet:2014lpa}. 
The only difference is  that the phase-space boundary $s> s_j/y_j$
is replaced by the cut  $s> (s_j/y_j)/\zeta^{\rm cut}$.
The tensor $V^{(\zeta_j)}_{\mu\nu}$ can be averaged over transverse angles using results 
in Section~3.3 of Ref.~\cite{Artoisenet:2014lpa}.  The angular average of $D^{(\zeta_j)}$ is
\begin{equation}
\Big\langle D^{(\zeta_j)}(p,q_1,q_2) \Big \rangle_{\Omega_\perp} = 
4 \pi \alpha_s \mu^{2 \epsilon} \frac{2N_c}{s} 
\left[ \frac{y_j}{1-y_j} + \frac{1-y_j}{y_j} + y_j(1-y_j) \right] 
y_j^2 {\cal A}_{\rm Born}(s_j, z/y_j).
\end{equation}
The integral over $s$  gives a UV pole.  Up to terms of relative order $\epsilon^3$, 
it can be expressed as
\begin{equation}
\int ds \frac{(s-s_j/y_j)^{-\epsilon}}{s} \theta\big(s-(s_j/y_j)/\zeta^{\rm cut} \big) = 
\Gamma(\epsilon)\Gamma(1- \epsilon)
\left( \frac{s_j/y_j}{\zeta^{\rm cut}} \right)^{-\epsilon} 
\left[ 1 - \left( \li(\zeta^{\rm cut}) - \frac{\pi^2}{6} \right) \epsilon^2 \right].
\end{equation}
The integral over $y_j$ gives an IR pole.
After renaming the differential variable in $d\phi_{\rm Born}$ as 
 $s_j \to s$, 
the subtraction integral is the same for $j=1$ and $j=2$.  
Their sum is
\begin{equation}
\sum_{j=1}^{2} N_{\rm CS} \int d\phi_{\rm real} D^{(\zeta_j)} \theta^{(\zeta_j)} =
\frac{\alpha_s}{2\pi} \Gamma(1+\epsilon) \left(\frac{\pi \mu^2}{m^2} \right)^\epsilon
\left[ I_2^{(\zeta)}(z)+ I_1^{(\zeta)}(z) + I_0^{(\zeta)}(z) \right].
\label{eq:subint-zetaj}
\end{equation}
The functions $I_n^{(\zeta)}$ are 
\begin{subequations}
\begin{eqnarray}
 I_2^{(\zeta)}(z) &=&
- \frac{2N_c}{\epsilon_{\textrm{UV}}\epsilon_{\textrm{IR}}} 
\left[ D_1(z) - \epsilon D_{\log}(z, 1/\zeta^{\rm cut}) + \frac12 \epsilon^2 D_{\log^2}(z,1/\zeta^{\rm cut}) \right]  ,
 \label{eq:Idouble} 
 \\
 I_1^{(\zeta)}(z)  &=&  \frac{1}{\epsilon_{\textrm{UV}}}   \int_z^1 \frac{dy}{y}   
P_{gg}^{\rm (real)}(y) \left[D_1(z/y)  - \epsilon D_{\log}(z/y, 1/\zeta^{\rm cut}) \right], 
\label{eq:Isingle}  
\\
\label{eq:remainder}
I_0^{(\zeta)}(z) & =&    2N_c \bigg(
- \int_z^1 \frac{dy}{y}    
\bigg[  \left( \frac{\log(1-y)}{1-y} \right)_+
+ \left(\frac{1}{y}+y(1-y)-2 \right)\log(1-y) \bigg] D_1(z/y) 
\nonumber
\\ & &   \hspace{1cm}
- \left[ \li(\zeta^{\rm cut}) - \frac{\pi^2}{6} \right]   D_1(z) \bigg) ,
\end{eqnarray}
 \label{eq:Izeta210}%
\end{subequations}
 where  $P_{gg}^{\rm (real)}(y)$  is the real-gluon
contribution to the Altarelli-Parisi splitting function for $g \rightarrow g$:
\begin{equation}
 P_{gg}^{\rm (real)}(y)= 2N_c
 \left[ \frac{y}{(1-y)_+} + \frac{1-y}{y}  + y(1-y)  \right].
\end{equation}
The function $D_1$ is defined in Eq.~\eqref{eq:D1-z}.
The functions $D_{\log}$ and $D_{\log^2}$  are defined by
\begin{subequations}
\label{eq:Dlog12}
\begin{eqnarray}
D_{\log}(z,A^{\rm cut}) &=& \int N d\phi  \mathcal{A}_{\rm Born}( s, z)  \,
\log \left(A^{\rm cut} \frac{s}{4m^2}\right) ,
\\
D_{\log^2}(z, A^{\rm cut}) &=& \int N d\phi  \mathcal{A}_{\rm Born}( s, z)  \,
\log^2 \left(A^{\rm cut} \frac{s}{4m^2}\right) ,
\end{eqnarray}
\end{subequations}
where the measure $N d\phi  \mathcal{A}_{\rm Born}( s, z)$
is given in Eq.~\eqref{eq:NphiABorn}.
Note that $D_{\log}(z,A^{\rm cut})$ depends on $A^{\rm cut}$
through the term $\log(A^{\rm cut}) D_1(z)$.

\subsection{Over-subtraction term proportional to $\bm{\theta^{(\zeta_j)} \theta^{(u_j)}}$}
\label{app:zetajuj}

In the over-subtraction term proportional to $\theta^{(\zeta_j)} \theta^{(u_j)}$ in $T_{1,j}^{(gg)}$,
the function $D^{(\zeta_j,u_j)}$ is given in Eq.~\eqref{eq:Dujzetaj}.
To extract the poles in the integral of the over-subtraction term,
we use the decomposition of the phase-space measure in Eq.~(\ref{eq:Ndphi_general}) with the
Born phase-space measure $d\phi_{\rm Born}(s_j/y_j,z)$,
where $s_j = (2p+q_j)^2$ and $y_j$ is defined in Eq.~(\ref{eq:defyi}),
and with the parton-emission measure
\begin{equation}
d \phi^{(\zeta_j,u_j)}(p,q_1,q_2) =  
\frac{(1-z)^{1-\epsilon}}{4(2\pi)^{3-2\epsilon}} 
(s-s_j/y_j)^{-\epsilon} ds \;
 u_j^{-\epsilon} (1-u_j)^{-\epsilon} du_j\;  
 d\Omega_\perp ,
\label{eq:def_dphi2_UVIR}
\end{equation}
where $u_j$ is defined in Eq.~\eqref{eq:uj}.
This parton-emission measure can be derived from $d \phi^{(\zeta_j)}$ in Eq.~\eqref{eq:def_dphi1_UV} 
by changing variables from $y_j$ to $u_j=(1-y_j)/(1-z)$.

The integral over $s$ gives a UV pole.  The integral over $u_j$ gives an IR pole.
After renaming the differential variable in $d\phi_{\rm Born}$ as $s_j/y_j \to \bar s$, 
the over-subtraction integral is the same for $j=1$ and $j=2$.  
Their sum is
\begin{equation}
\sum_{j=1}^{2}
N_{\rm CS} \int d\phi_{\rm real} D^{(\zeta_j, u_j)} \theta^{(\zeta_j)} \theta^{ (u_j)}
 = \frac{\alpha_s}{2\pi} \Gamma(1+\epsilon) \left(\frac{\pi \mu^2}{m^2} \right)^{\epsilon} 
 \left[ I^{(\zeta, u)}_2(z) + I^{(\zeta, u)}_1(z) + I^{(\zeta, u)}_0(z) \right] .
 \label{eq:subint-zetajuj}
\end{equation}
The functions $I^{(\zeta, u)}_n$ are 
\begin{subequations}
\begin{eqnarray}
I^{(\zeta, u)}_2(z) &=&
- \frac{2N_c}{\epsilon_{\textrm{UV}}\epsilon_{\textrm{IR}}} 
\left[ D_1(z) - \epsilon D_{\log}(z, u^{\rm cut}/\zeta^{\rm cut}) 
+ \frac12 \epsilon^2 D_{\log^2}(z,u^{\rm cut}/\zeta^{\rm cut}) \right]  ,
 \label{eq:Izetau2} 
 \\
I^{(\zeta, u)}_1(z)  &=&  \frac{2N_c}{\epsilon}  \log(1-z)
\left[ D_1(z) - \epsilon D_{\log}(z, u^{\rm cut}/\zeta^{\rm cut})  \right], 
\label{eq:Izetau1}  
\\
\label{eq:Izetau0}  
I^{(\zeta, u)}_0(z) & =&    2N_c 
\left[  -\frac{1}{2}\log^2 (1-z)
 +\li(u^{\rm cut})-\li (\zeta^{\rm cut}) +\frac{\pi^2}{6} \right] D_1(z). 
\end{eqnarray}
\label{eq:Izetau210}%
\end{subequations}
The function $D_1$ is defined in Eq.~\eqref{eq:D1-z}, and 
the functions $D_{\log}$ and $D_{\log^2}$  are defined in Eqs.~\eqref{eq:Dlog12}.

\subsection{Over-subtraction term proportional to $\bm{\theta^{(u_j)} \theta^{(\delta)}}$ }
\label{app:ujdelta}

In the over-subtraction term proportional to $\theta^{(u_j)} \theta^{(\delta)}$ in $T_{2,j}^{(gg)}$,
the function $D^{(u_j,\delta)}$ is given in Eq.~\eqref{eq:Dujdelta}.
To extract the poles in the integral of the subtraction term,
we use the decomposition of the phase-space measure in Eqs.~(3.26) and (3.27) of Ref.~\cite{Artoisenet:2014lpa}.
The factor $N_{\rm Born}(p,\tilde q)$ coincides with $N_{\rm CS}$.
The  factor $d\phi_{\rm Born}(p,\tilde q)$ can be expressed in terms of the measure 
defined in Eq.~\eqref{eq:BornPSsz} as $d\phi_{\rm Born}(\tilde s,z)$, where $\tilde s=(2p+\tilde q)^2$.
In the decomposition of the phase-space measure
in Eq.~(\ref{eq:Ndphi_general}),
the Born phase-space measure is $d\phi_{\rm Born}(\tilde{s},z)$
and the parton-emission measure is 
\begin{equation}
d \phi^{(u_j,\delta)}(p,q_1,q_2) =  
\frac{(4m^2)^{1-\epsilon}}{4(2\pi)^{3-2\epsilon}} 
\lambda^{-\epsilon} d\lambda \;
 u_j^{-\epsilon} (1-u_j)^{-\epsilon} du_j\;  
 d\Omega_\perp ,
\label{eq:def_dphi1_uj_B}
\end{equation}
where $\lambda$ is defined in Eq.~(\ref{eq:lambda}) and 
 $u_j$ is defined in Eq.~\eqref{eq:uj}.
The cut variable $\delta$ defined in Eq.~\eqref{eq:delta} can be expressed as
\begin{equation}
\delta =  \frac{z \lambda/(1-z)}{z \lambda/(1-z) + (\tilde s - 4 m^2)/4m^2}.
\end{equation}
The cut $\delta<\delta^{\rm cut}$ can therefore be expressed as a cut on $\lambda$
that depends on $\tilde s$ and is proportional to $\delta^{\rm cut}/(1-\delta^{\rm cut})$.

The integrals over $\lambda$ and  over $u_j$ give IR poles.
The over-subtraction integrals are the same for $j=1$ and $j=2$.  
Their sum  is
\begin{equation}
\sum_{j=1}^{2} N_{\rm CS} \int d\phi_{\rm real} D^{(u_j,\delta)} \, \theta^{(u_j)} \, \theta^{(\delta)}
=  \frac{\alpha_s}{2\pi} \Gamma(1+\epsilon) \left(\frac{\pi \mu^2}{m^2} \right)^{\epsilon}  
\left[ I^{(u,\delta)}_2(z) + I^{(u,\delta)}_1(z) + I^{(u,\delta)}_0(z) \right] .
\label{eq:subint-ujdelta}
\end{equation}
The functions $I^{(u,\delta)}_n$ are 
\begin{subequations}
\begin{eqnarray}
I^{(u,\delta)}_2(z) &=&
\frac{N_c}{\epsilon_{\textrm{IR}}^2} 
\left[ D_1(z) - \epsilon D'_{\log}\big(z, \delta^{\rm cut}u^{\rm cut}/(1-\delta^{\rm cut})\big) 
+ \frac12 \epsilon^2 D'_{\log^2}\big(z,\delta^{\rm cut}u^{\rm cut}/(1-\delta^{\rm cut})\big) \right]  ,
\nonumber\\
 \label{eq:Iudelta2} 
 \\
I^{(u,\delta)}_1(z)  &=& -  \frac{N_c}{\epsilon_{\textrm{IR}}}  \log\frac{1-z}{z}
\Big[ D_1(z) - \epsilon D'_{\log}\big(z,\delta^{\rm cut}u^{\rm cut}/(1-\delta^{\rm cut})\big)  \Big], 
\label{eq:Iudelta1}  
\\
\label{eq:Iudelta0}  
I^{(u,\delta)}_0(z) & =&    N_c 
\left[  \frac{1}{2}\log^2\frac{1-z}{z}
 - \li (\zeta^{\rm cut}) - \frac{\pi^2}{6} \right] D_1(z). 
\end{eqnarray}
 \label{eq:Iudelta210}%
\end{subequations}
The function $D_1$ is defined in Eq.~\eqref{eq:D1-z}.
The functions $D'_{\log}$ and $D'_{\log^2}$  are defined by
\begin{subequations}
\label{eq:Dlog12'}
\begin{eqnarray}
D'_{\log}(z,A^{\rm cut}) &=& \int N d\phi  \mathcal{A}_{\rm Born}( s, z)  \,
\log \left(A^{\rm cut}  \frac{s-4m^2}{4m^2} \right),
\\
D'_{\log^2}(z,A^{\rm cut}) &=& \int N d\phi  \mathcal{A}_{\rm Born}( s, z)  \,
\log^2 \left(A^{\rm cut}  \frac{s-4m^2}{4m^2} \right),
\end{eqnarray}
\end{subequations}
where the measure $N d\phi  \mathcal{A}_{\rm Born}( s, z)$,
which is differential in $s$, is given in Eq.~\eqref{eq:NphiABorn}.
Note that $D'_{\log}(z,A^{\rm cut})$ depends on $A^{\rm cut}$
through the term $\log(A^{\rm cut}) D_1(z)$.

\subsection{Subtraction term proportional to $\bm{\theta^{(u_j)} \theta^{(\delta_j)}}$ }
\label{app:uj,deltaj}

In the subtraction term proportional to $\theta^{(u_j)} \theta^{(\delta_j)}$ for $T_{2,j}^{(gg)}$,
the function $D^{(u_j,\delta_j)}$ is given in Eq.~\eqref{eq:Duj,deltaj}.
To extract the poles in the integral of the subtraction term,
we use the decomposition of the phase-space measure in Eq.~(\ref{eq:Ndphi_general}), with
Born phase-space measure $d\phi_{\rm Born}( \breve s_j,z)$,
where $\breve s_j $ is defined in Eq.~(\ref{eq:brevesj}),
and with the parton-emission measure
$d \phi^{(u_j,\delta_j)}$ derived in Section~\ref{app:dphi2(uj,deltaj)} and given in Eq.~\eqref{eq:dphi2(uj,deltaj)}.
This parton emission measure is differential in the variables $w_j$ and $u_j$ given in Eqs.~\eqref{eq:uj,wj} 
as well as in  the transverse angles.
The cut variable $\delta_j$ defined in Eqs.~(\ref{eq:delta1}) and Eq.~(\ref{eq:delta2}) 
can be expressed as
\begin{equation}
\delta_j =  \frac{w_j - (1-z)u_j/z}{(w_j - (1-z)u_j/z) + (\breve s_j - 4 m^2)/4m^2}.
\end{equation}
The cut $\delta_j<\delta^{\rm cut}$ can therefore be expressed as a cut on $w_j - (1-z)u_j/z$
that depends on $\breve s_j$ and is proportional to $\delta^{\rm cut}/(1-\delta^{\rm cut})$.

The integrals over $w_j$ and $u_j$ give a double infrared pole.
After relabelling the differential variable in $d\phi_{\rm Born}$ by $\breve s_j \to s$,
we obtain the same subtraction integral for $j=1$ and $j=2$. Their sum is
\begin{equation}
\sum_{j=1}^{2} N_{\rm CS} \int d\phi_{\rm real} D^{(u_j,\delta_j)} \theta^{(u_j)} \theta^{(\delta_j)}=  \frac{\alpha_s}{2\pi} \Gamma(1+\epsilon) \left(\frac{\pi \mu^2}{m^2} \right)^{\epsilon}  
\left[ I^{(u,\delta)'}_2(z) + I^{(u,\delta)'}_1(z) + I^{(u,\delta)'}_0(z) \right] .
\label{eq:subint-uj,deltaj}
\end{equation}
The functions $I^{(u,\delta)'}_n$ are 
\begin{subequations}
\begin{eqnarray}
I^{(u,\delta)'}_2(z) &=&
\frac{N_c}{\epsilon_{\textrm{IR}}^2} 
\left[ \frac12 D_1(z) - \epsilon D'_{\log}\big(z, \delta^{\rm cut}/(1-\delta^{\rm cut})\big) 
+ \frac12 \epsilon^2 D'_{\log^2}\big(z,\delta^{\rm cut}/(1-\delta^{\rm cut})\big) \right]  ,
\nonumber\\
 \label{eq:Iudelta'2} 
 \\
I^{(u,\delta)'}_1(z)  &=&  \frac{N_c}{\epsilon_{\textrm{IR}}}  
\Big[ D_1(z) + \epsilon \log\frac{(1-z)u^{\rm cut}}{z} D'_{\log}\big(z,\delta^{\rm cut}/(1-\delta^{\rm cut})\big)  \Big], 
\label{eq:Iudelta'1}  
\\
\label{eq:Iudelta'0}  
I^{(u,\delta)'}_0(z) & =&    N_c \bigg[ 
\left( - \frac12 \log^2 \frac{(1-z)u^{\rm cut}}{z} -2  \log \frac{(1-z)u^{\rm cut}}{z} - \frac{\pi^2}{6} \right)D_1(z)
+ F^{(u,\delta)'}(z) \bigg].
\nonumber\\
\end{eqnarray}
 \label{eq:Iudelta'210}%
\end{subequations}
The function $D_1$ is defined in Eq.~\eqref{eq:D1-z}, 
the functions $D'_{\log}$ and $D'_{\log^2}$  are defined in Eqs.~\eqref{eq:Dlog12'},
and the function $F^{(u,\delta)'}$ is given by
\begin{eqnarray}
F^{(u,\delta)'}(z) &=& \int N d\phi \mathcal{A}_{\rm Born}(s,z) 
\bigg[ -\li \!\left(- \frac{(1-z)u^{\rm cut} (1-\delta^{\rm cut})}{z\delta^{\rm cut} r} \right) 
\nonumber\\
&& \hspace{4cm} 
+ 2 \log \!\left( 1 + \frac{(1-z)u^{\rm cut} (1-\delta^{\rm cut})}{z\delta^{\rm cut} r} \right)
\bigg] ,
\end{eqnarray}
where $r= (s - 4m^2)/4m^2$.

\subsection{Subtraction term proportional to $\bm{\theta^{(\lambda)}}$ }

\label{app:lambda}

In the subtraction term proportional to $\theta^{(\lambda)}$ in $T_{2,j}^{(gg)}$,
the function $D^{(\lambda)}$ is given in Eq.~\eqref{eq:Dlambda},
and it is multiplied by the weight function $S^{(\lambda)}_{2,j}$ in Eq.~\eqref{eq:S2j^lambda}.
The function $D^{(\lambda)}$ is proportional to the contraction of the Born tensor $\mathcal{A}_{\rm gluon}^{\mu \nu}$
 in Eq.~(\ref{Borntensor-gluon}) and the tensor $P^{(gg)}_{\mu \nu}$ in Eq.~\eqref{eq:P(gg)}.
The sum over $j$ of the subtraction terms can be simplified by using the identity in Eq.~\eqref{eq:S21+S22}:
\begin{equation}
\left( S_{2,1}^{(\lambda)}  + S_{2,2}^{(\lambda)}\right) D^{(\lambda)} \theta^{(\lambda)} = 
D^{(\lambda)} \theta^{(\lambda)} .
\end{equation}
The tensor $P^{(gg)}_{\mu \nu}$  in Eq.~\eqref{eq:P(gg)}
can be averaged over the transverse angles using results in Section~3.5 of Ref.~\cite{Artoisenet:2014lpa}.
The angular average of $ D^{(\lambda)}$ is
\begin{equation}
\Big\langle D^{(\lambda)}(p,q_1,q_2) \Big \rangle_{\Omega_\perp} = 
4 \pi \alpha_s \mu^{2 \epsilon} \Big( V_1^{(\lambda)} + V_2^{(\lambda)} \Big)   
{\cal A}_{\rm Born}(\tilde s, z).
\end{equation}
The functions $V_1^{(\lambda)} $ and $V_2^{(\lambda)}$ are
\begin{subequations}
\begin{eqnarray}
 V_1^{(\lambda)}  & = & \frac{N_c}{4m^2 \lambda} \left(  \frac{u_1}{u_2} + \frac{u_2}{u_1} +2u_1u_2 \right) ,
 \\
 V_2^{(\lambda)}  & = &  \frac{N_c}{4m^2 \lambda} \left(  \frac{1-v}{v} + \frac{v}{1-v} \right),
\end{eqnarray}
\end{subequations}
where $v$ is defined in Eq.~\eqref{eq:lambda,v}.
To extract the poles in the integrals of the subtraction terms,
the $ V_1^{(\lambda)}$ and $ V_2^{(\lambda)}$ terms are integrated using 
the same Born phase-space measure $d\phi_{\rm Born}(\tilde{s},z)$ but different parton-emission measures.
The $V_1^{(\lambda)}$ term is integrated using  
the parton-emission measure   $d \phi_1^{(\lambda)} =d \phi^{(u_1,\delta)}$ defined in Eq.~(\ref{eq:def_dphi1_uj_B}).
The $V_2^{(\lambda)}$ term is integrated using 
 the parton-emission measure $d \phi_2^{(\lambda)}$ 
derived in Section~\ref{app:dphi2(uj,deltaj)} and given in Eq.~\eqref{eq:dphi2(lambda)}.

For the $V_1^{(\lambda)}$ term, the integrals over $\lambda$ and $u$ both give  IR poles. 
For  the $V_2^{(\lambda)}$ term, the integrals over $\lambda$ and $v$ give a double IR pole.
The complete subtraction integral is
\begin{eqnarray}
&&N_{\rm CS} \int d\phi_{\rm real} 
\left( S_{2,1}^{(\lambda)}  + S_{2,2}^{(\lambda)}\right) D^{(\lambda)} \theta^{(\lambda)}  
\nonumber\\
&& \hspace{1cm}
= \frac{\alpha_s}{2\pi} \Gamma(1+\epsilon) \left(\frac{\pi \mu^2}{m^2} \right)^{\epsilon}  
\left( \mathcal{V}_1^{(\lambda)} D_1(z) 
+\left[ I_2^{(\lambda)}(z) + I^{(\lambda)}_1(z) + I^{(\lambda)}_0(z) \right]  \right).
\label{eq:subint-lambda}
\end{eqnarray}
The constant $ \mathcal{V}_1^{(\lambda)}$ is
\begin{equation}
\mathcal{V}_1^{(\lambda)}  =  
\frac{N_c}{\epsilon_{\textrm{IR}}^2} (\lambda^{\rm cut})^{-\epsilon_{\textrm{IR}}}
\Bigg[  1 +\frac{5}{6} \epsilon_{\textrm{IR}}
+ \left( - \frac{\pi^2}{3} +\frac{31}{18} \right) \epsilon_{\textrm{IR}}^2 \Bigg] .
\label{eq:V1lambda}
\end{equation}
The functions $I^{(\lambda)}_n$ are 
\begin{subequations}
\begin{eqnarray}
I^{(\lambda)}_2(z) &=&
\frac{N_c}{\epsilon_{\textrm{IR}}^2} 
\left[ \frac12 D_1(z) - \epsilon D'_{\log}\big(z, 1/\lambda^{\rm cut}\big) 
- \epsilon^2 D'_{\log^2}\big(z,1/\lambda^{\rm cut}\big) \right]  ,
 \label{eq:Ilambda2} 
 \\
I^{(\lambda)}_1(z)  &=&  \frac{N_c}{\epsilon_{\textrm{IR}}} 
(1- \log \lambda^{\rm cut}) \Big[ D_1(z) - \epsilon  D'_{\log}\big(z,1/\lambda^{\rm cut}\big)  \Big], 
\label{eq:Ilambda1}  
\\
\label{eq:Ilambda0}  
I^{(\lambda)}_0(z) & =&    N_c \Bigg[ 
\left(  \frac{1}{2} \log^2\lambda^{\rm cut} -\log \lambda^{\rm cut}  +2 - \frac{\pi^2}{3} \right) D_1(z)
+ F^{(\lambda)}(z) \Bigg].
\end{eqnarray}
\label{eq:Ilambda210}%
\end{subequations}
The function $D_1$ is defined in Eq.~\eqref{eq:D1-z}, 
the functions $D'_{\log}$ and $D'_{\log^2}$  are defined in Eqs.~\eqref{eq:Dlog12},
and the function $F^{(\lambda)}$ is 
\begin{equation}
F^{(\lambda)}(z) = \int N d\phi \mathcal{A}_{\rm Born}(s,z)  
\int_0^{\lambda^{\rm cut}} \frac{d\lambda}{\lambda}
\left[ \frac{1}{1 - 2v_-} \log \frac{1-v_-}{v_-}  + \log \frac{\lambda}{r^2}  \right]  ,
\end{equation}
where  $r= (s - 4m^2)/4m^2$ and $v_-$ is
\begin{equation}
v_- = \frac12 \left( 1 
- \frac{\lambda^{1/2}\big( 1 + r + \lambda/(1-z),\lambda,1\big)}{r+z\lambda/(1-z)} \right).
\end{equation}
Its limiting behavior as $\lambda \to 0$ is $v_- \to \lambda/r^2$.

\subsection{Subtraction terms proportional to $\bm{\theta^{( u_j)}}$}
\label{app:uj}

In the subtraction term proportional to $\theta^{( u_j)}$ in $T_{1,j}^{(gg)}$,
the function $D_1^{(u_j)}$ is given in Eq.~\eqref{eq:D1uj}
and it is multiplied by the weight function $S_{1,j}^{( u_j)}$ in Eq.~\eqref{eq:S1j^uj}.
In the subtraction term proportional to $\theta^{( u_j)}$ in $T_{2,j}^{(gg)}$,
the function $D_1^{(u_j)}$ is given in Eq.~\eqref{eq:D2uj}
and it is multiplied by the weight function $S_{2,j}^{( u_j)}$ in Eq.~\eqref{eq:S2j^uj}.
If the weight functions $S_{1,j}^{( u_j)}$ and $S_{2,j}^{( u_j)}$ were set to 1,
the integrals of the $D_1^{(u_j)}$ and $D_2^{(u_j)}$ subtraction terms would be  equal. 
This  will be shown below by considering appropriate 
choices for $s^{(A)}$ and $d \phi^{( A)}$ for each subtraction term.
It is therefore convenient to consider these two subtraction terms together.

To extract the poles in the integral of the $D_1^{(u_j)}$ subtraction term  
 given in Eq.~\eqref{eq:D1uj},
we use the decomposition of the phase-space measure in 
 Eq.~(\ref{eq:Ndphi_general}) with the 
Born phase-space measure $d\phi_{\rm Born}(s_j/y_j,z)$,
where $s_j = (2p+q_j)^2$ and $y_j$ is defined in Eq.~(\ref{eq:defyi}), and
with the  parton-emission measure 
\begin{equation}
d \phi_1^{( u_j)}(p,q_1,q_2) =  
\frac{(4m^2)^{1-\epsilon}}{4(2\pi)^{3-2\epsilon}} 
\lambda_j^{-\epsilon} d\lambda_j \;
 u_j^{-\epsilon} (1-u_j)^{-\epsilon} du_j\;  
 d\Omega_\perp ,
\label{eq:def_dphi1_uj_A}
\end{equation}
where $u_j$ is defined in Eq.~\eqref{eq:uj} and $\lambda_j$ is defined in Eq.~\eqref{eq:lambdaj}.
The parton-emission measure $d \phi_1^{( u_j)}$ in Eq.~\eqref{eq:def_dphi1_uj_A}
can be derived from $d \phi^{(\zeta_j,u_j)}$ in Eq.~\eqref{eq:def_dphi2_UVIR}
by changing variables from $s$ to $\lambda_j$.
To extract the poles in the integral of the $D_2^{(u_j)}$ subtraction term 
 given in Eq.~\eqref{eq:D2uj},
we use the decomposition of the phase-space measure 
 in Section~\eqref{app:ujdelta}
with the Born phase-space measure $d\phi_{\rm Born}(\tilde{s},z)$ and  with the parton-emission measure 
$d \phi_2^{(u_j)} = d \phi^{(u_j,\delta)}$ in Eq.~\eqref{eq:def_dphi1_uj_B}.

The integral of the subtraction term $S^{(u_j)}_{2,j} D_2^{(u_j)}$ over the Born phase space 
and over the variable $\lambda$ is
\begin{eqnarray}
\int \!\! d\phi_{\rm Born}(\tilde s,z)\,  \lambda^{-\epsilon} d\lambda \, S^{(u_j)}_{2,j} D_2^{(u_j)} 
=  4\pi \alpha_s \mu^{2\epsilon} \int \!\! d\phi_{\rm Born}(\tilde s,z)\, 
\frac{N_c  \lambda^{-\epsilon} d\lambda}{2 m^2 u_j \lambda} \mathcal{A}_{\rm Born} (\tilde s,z) S^{(u_j)}_{2,j}  .
\label{eq:intS2D2}
\end{eqnarray}
The integral of the subtraction term $S^{(u_j)}_{1,j} D_1^{(u_j)}$ over the Born phase space 
and over the variable $\lambda_j$  can be expressed in a similar form
by renaming integration variables $s_j/y_j \to \tilde s$ and  $\lambda_j \to \lambda$:
\begin{eqnarray}
\int \!\!  d\phi_{\rm Born}(s_j/y_j,z)\,  \lambda_j^{-\epsilon} d\lambda_j\, S^{(u_j)}_{1,j} D_1^{(u_j)}
=  4\pi \alpha_s \mu^{2\epsilon} \int \!\! d\phi_{\rm Born}(\tilde s,z)\, 
\frac{N_c  \lambda^{-\epsilon} d\lambda}{2 m^2 u_j \lambda}\mathcal{A}_{\rm Born} (\tilde s,z) S^{(u_j)}_{1,j}  .
\nonumber \\
\label{eq:intS1D1}
\end{eqnarray}
The integrands in Eqs.~\eqref{eq:intS2D2} and \eqref{eq:intS1D1} differ only in the factors
 $S^{(u_j)}_{2,j}$ and $S^{(u_j)}_{1,j}$.
But $S^{(u_j)}_{1,j}$ in Eq.~(\ref{eq:S1j^uj}), with the substitutions $s_j/y_j \to \tilde s$ and  $\lambda_j \to \lambda$,
and $S^{(u_j)}_{2,j}$ in Eq.~(\ref{eq:S2j^uj})  are weight functions  that add up to 1.
The sum of the integrals in Eqs.~\eqref{eq:intS2D2} and \eqref{eq:intS1D1} 
therefore factors into a Born phase-space integral and an integral over $\lambda$.
With dimensional regularization, the integral over $\lambda$ is 0, but it can be expressed as the difference
between a UV pole and an IR pole.
The  integral over $u_j$ gives an IR pole.
The sum of the two subtraction integrals is the same for $j=1$ and $j=2$.
The sum over $j$ of the  subtraction  integrals is
\begin{equation}
\sum_{j=1}^{2} N_{\rm CS} \int d\phi_{\rm real}(p,q_1,q_2)  
\left( S^{(u_j)}_{1,j} D_1^{(u_j)} + S^{(u_j)}_{2,j} D_2^{(u_j)} \right)  \theta^{(u_j)}
= \frac{\alpha_s}{2\pi}\Gamma(1+\epsilon) \left(\frac{\pi \mu^2}{m^2} \right)^{\epsilon}   \mathcal{V}^{(u)} 
D_1(z),
\label{eq:subint-uj}
\end{equation}
where the constant $\mathcal{V}^{(u)}$ is
\begin{equation}
\mathcal{V}^{(u)} = 2N_c \left( u^{\rm cut}\right)^{-\epsilon}
\left( \frac{1}{\epsilon_{\mathrm{IR}}} -\frac{1}{\epsilon_{\mathrm{UV}}} \right)  
\left[ \frac{1}{\epsilon_{\mathrm{IR}}} - \left(\li( u^{\rm cut}) + \frac{\pi^2}{6} \right) \epsilon  \right].
\label{kernelVu}
\end{equation}
This constant vanishes if $\epsilon_{\mathrm{UV}}  = \epsilon_{\mathrm{IR}}$.

\subsection{Over-subtraction term proportional to $\bm{\theta^{(\lambda)} \theta^{(u_j)}}$ }
\label{app:lambdauj}

In the over-subtraction term proportional to $\theta^{(\lambda)} \theta^{(u_j)}$ in $T_{2,j}^{(gg)}$,
the function $D^{(\lambda,u_j)}$ is given in Eq.~\eqref{eq:Dlambdauj}.
To extract the poles in the integral of the over-subtraction term,
we use the same decomposition of the phase-space measure 
as in Section~\eqref{app:ujdelta},
with the Born phase-space measure $d\phi_{\rm Born}(\tilde{s},z)$  
and the parton-emission measure   
$d \phi^{(\lambda, u_j)} = d \phi^{(u_j,\delta)}$ in Eq.~(\ref{eq:def_dphi1_uj_B}).

The integrals over $\lambda$  and over $u_j$ give IR poles.
The over-subtraction integral is the same  for $j=1$ and $j=2$.  Their sum is
\begin{equation}
\sum_j N_{\rm CS} \int d\phi_{\rm real} D^{(\lambda, u_j)} \theta^{(\lambda)} \theta^{(u_j)} =  
\frac{\alpha_s}{2\pi} \Gamma(1+\epsilon) \left(\frac{\pi \mu^2}{m^2} \right)^{\epsilon}   
\mathcal{V}^{(\lambda, u)} D_1(z),
\label{eq:subint-lambdauj}
\end{equation}
where the constant $\mathcal{V}^{(\lambda,u)} $ is
\begin{equation}
\mathcal{V}^{(\lambda,u)}  = 
\frac{N_c}{\epsilon_{\textrm{IR}}^2} (\lambda^{\rm cut}  u^{\rm cut}) ^{-\epsilon_{\textrm{IR}}}
\left[1
+ \left(  -\li(u^{\rm cut}) -\frac{\pi^2}{6 } \right) \epsilon_{\textrm{IR}}^2 \right].
 \label{kernelVlambdau}
\end{equation}

\subsection{Subtraction term involving light quarks }
\label{app:qqbar}

In the subtraction term  $T^{(q \bar q)}$ in Eq.~\eqref{Tqq},
which is proportional to $\theta^{(\lambda)}$,
the function $D^{(q \bar q)}$ is given in Eqs.~(\ref{Dqq}). 
It is proportional to the contraction of the Born tensor $\mathcal{A}_{\rm gluon}^{\mu \nu}$ in Eq.~(\ref{Borntensor-gluon})
and the tensor $P^{(q \bar q)}_{\mu \nu}$ in Eq.~\eqref{eq:P(qq)}.
To extract the poles in the integral of the subtraction term,
we use the same decomposition of the phase-space measure as in Section~\eqref{app:ujdelta},
with the Born phase-space measure $d\phi_{\rm Born}(\tilde{s},z)$ 
and the parton-emission measure 
 $d \phi^{(q \bar q)} = d \phi^{(u_1,\delta)}$ defined  in Eq.~(\ref{eq:def_dphi1_uj_B}).

The tensor $P^{(q \bar q)}_{\mu \nu}$  in Eq.~\eqref{eq:P(qq)}
can be averaged over the transverse angles using results in Section~3.5 of Ref.~\cite{Artoisenet:2014lpa}.
The angular average of $D^{(q \bar q)}$ is
\begin{equation}
\Big\langle D^{(q \bar q)}(p,q_1,q_2) \Big \rangle_{\Omega_\perp} = 
4 \pi \alpha_s \mu^{2 \epsilon} \frac{ n_f T_F}{2 m^2 \lambda} \left[ 1 - 2\frac{u_1(1-u_1)}{1 - \epsilon}  \right]   
{\cal A}_{\rm Born}(\tilde s, z).
\end{equation}
The integral over $\lambda$ gives an IR pole. The final result for the subtraction integral is
\begin{equation}
N_{\rm CS} \int d\phi_{\rm real} D^{(q \bar q)} \theta^{(\lambda)}  =  
\frac{\alpha_s}{2\pi} \Gamma(1+\epsilon) \left(\frac{\pi \mu^2}{m^2} \right)^{\epsilon}   \mathcal{V}^{(q \bar q )} D_1(z),
\label{Dqq_integrated}
\end{equation}
where the constant $\mathcal{V}^{(q \bar q)} $ is
\begin{equation}
\mathcal{V}^{(q \bar q)}  = - \frac{2n_f T_F}{3\epsilon_{\mathrm{IR}}} (\lambda^{\rm cut})^{-\epsilon_{\mathrm{IR}}}
\left[ 1  + \frac{5}{3}\epsilon_{\mathrm{IR}} \right].
\label{kernelVqq}
\end{equation}


\section{Helicity amplitudes with MadGraph}
\label{app:madgraph}

We use \madgraph \cite{Alwall:2011uj} to generate the amplitude associated with the sum of 
cut diagrams with an S-wave color-octet heavy-quark pair and two light partons 
in the final state. The squared amplitudes generated by \madgraph are expressed 
in the helicity basis, and they are summed over the sets of helicities that give non-zero
contributions. Such a computation in terms of helicity amplitudes is 
particularly well suited for numerical purposes.

The \madgraph generator is primarily aimed at generating 
matrix elements and events for scattering or decay processes.
The input is a UFO model~\cite{Degrande:2011ua} 
that encodes all the Feynman rules associated with the process.
The Feynman rules associated with fragmentation processes can  be 
translated into a UFO model, so the \madgraph can also be used to generate
the matrix elements associated with fragmentation cut diagrams. 
A convenient precedure is to append the following additional (fictious)
particles and interactions to the UFO model associated with the Standard Model:
\begin{description}

\item[Particles:] the new \textit{states} have the following properties: 
\begin{itemize}
\item the state \texttt{eik} is a zero-mass, scalar, color-octet  state;
\item the state \texttt{source} is a zero-mass, vector, color-singlet state;
\item the state \texttt{co} is a duplicate of a heavy-quark pair state (either charm or bottom) 
that is projected onto a color-octet state.
\end{itemize}

\item[Interactions:] the new interactions specify the Lorentz structure
of the coupling of the gluon field with the source and/or the eikonal lines. 
Two vertices must be defined:
\begin{itemize}
\item  the \texttt{source-eik-g} vertex,
\item the \texttt{eik-g-eik} vextex.
\end{itemize}

\end{description}
The additional Feynman rules for fragmentation functions include
a delta function for the cut through the eikonal line.  That delta function is included in
the phase-space measure, so it does not appear in the UFO model.  

The relabelling of the heavy-quark+antiquark state $\texttt{co}$ simply acts as a trigger 
to force the heavy quark and antiquark in the final state into a color-octet configuration, 
a projection that is applied by means of an ad-hoc
modification of the color treatment in \madgraph\!\!.  Given that we are only interested 
in a final-state heavy-quark pair projected onto a scalar state in this work, 
we also add a filter 
to remove diagrams where a single gluon is attached to the 
$\texttt{co}$ fermion line.

\begin{figure}
\includegraphics[scale=0.5]{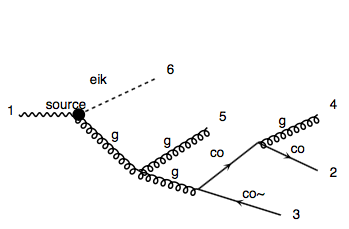}
\hspace{0.5cm}
\includegraphics[scale=0.5]{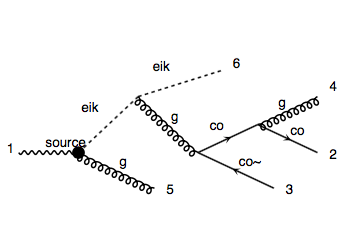}
\includegraphics[scale=0.5]{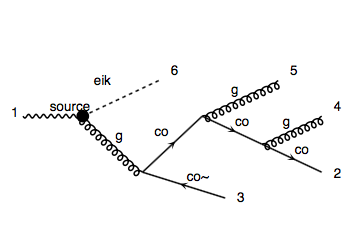}
\hspace{0.5cm}
\includegraphics[scale=0.5]{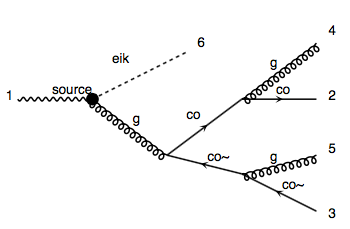}
\includegraphics[scale=0.5]{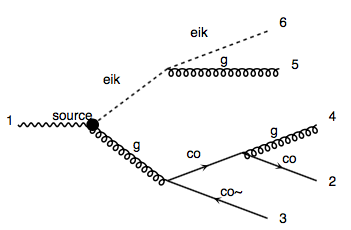}
\hspace{0.5cm}
\includegraphics[scale=0.5]{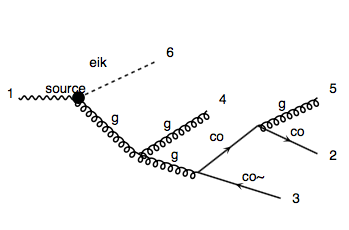}
\includegraphics[scale=0.5]{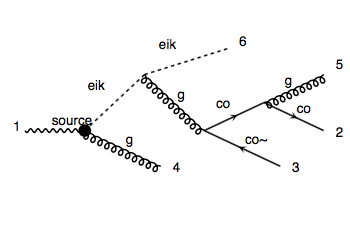}
\hspace{0.5cm}
\includegraphics[scale=0.5]{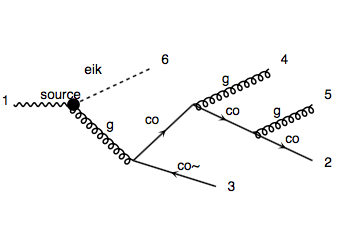}
\includegraphics[scale=0.5]{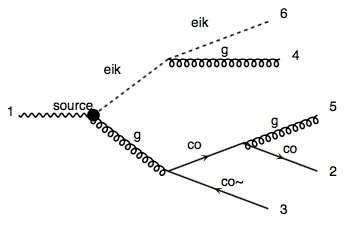}
\hspace{2.1cm}
\includegraphics[scale=0.5]{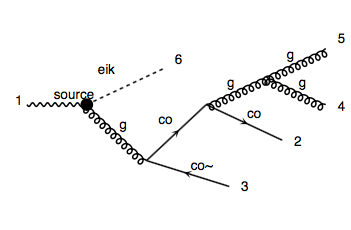}
\caption{Diagrams generated by \madgraph for the process $g \rightarrow 
Q\bar Q_8(^1S_0) \, + \, g \, +\,  g$. Only half the Feynman diagrams are shown.
The other half are obtained by reversing the directions of the arrows on the quark lines. 
There are no diagrams 
with ghosts in the final state since the amplitudes are 
calculated with physical helicity states. 
\label{fig:diagrams}}
\end{figure}

With the above UFO model at hand, \madgraph can be used to generate (i) 
the diagrams  associated with a fragmentation process and (ii) a Fortran code
for the numerical evaluation of the amplitude associated with the cut diagrams.
The left-hand side of the cut of each generated diagram for gluon fragmentation with two gluons crossing the cut  
 in addition to the $^1S_0$ $Q\bar Q_8$ state is displayed in 
figure~\ref{fig:diagrams}. 

The projection of the heavy-quark pair
onto a spin-singlet state has not yet been applied.
Moreover, with the above extension of the UFO model
for SM processes, \madgraph has treated 
the \texttt{eik} and \texttt{source}
states as if they were regular particles. 
This calls for several modifications in 
the default Fortran code generated by \madgraph\!\!:
\begin{itemize}

\item The projection onto a spin-singlet state can be 
easily achieved by combining amplitudes with different
heavy-quark helicities in the appropriate configurations.

\item \madgraph automatically decomposes the calculation of 
helicity amplitudes into generic building blocks, 
and writes the routines associated with these blocks
using a module called \aloha\cite{deAquino:2011ub}. 
In the two routines associated with the two vertices involving 
the eikonal line, \madgraph identifies the four-vector $n^\mu$ 
associated with the eikonal line with the momentum flowing along the 
eikonal line. Thus the two routines must be modified so that  
the four-vector $n^\mu$ is properly set to the (fixed) eikonal four-vector.  

\item A cut diagram with Lorentz index $\mu$ at the operator
on the left side of the cut and 
Lorentz index $\nu$ at the operator on the right side of the cut 
must be contracted with $-g_{\mu \nu}$. 
Instead, \madgraph contracts each index with the polarization 
vector $\epsilon_\mu(\lambda)$ [or $\epsilon^*_\nu(\lambda)$]
built upon the helicity state $\lambda$ of the source. 
Summing over the helicity states of the source 
will not yield the correct result in general, 
unless we define the momentum $K$ associated with the source
with some care. For a fixed value of $K.n$, 
we define light-like four-vectors $K_+$ and $K_-$
whose spacial components are  equal and opposite
and such that $K_+.n = K.n$ and $K_-^\mu$
is proportional to the eikonal four-vector $n^\mu$. 
The tensor that is used by \madgraph to contract indices 
$\mu$ and $\nu$ at the sources on both sides of the cut
can be decomposed as follows: 
\begin{equation}
\sum_{\lambda} \epsilon^{\mu}(\lambda) \epsilon^{*\nu} (\lambda)= 
- g^{\mu \nu} + \frac{K_+^\mu K_-^\nu + K_-^\mu K_+^\nu} {2K_+ . K_-}.
\label{eq:polarization}
\end{equation}
However by gauge invariance,  any 
amplitude associated with the sum of cut diagrams 
with Lorentz index $\mu$ and $\nu$ at the sources 
on both sides of the cut
vanishes if it is contracted with either 
the eikonal four-vector $n_\mu$ or the eikonal four-vector $n_\nu$.
Since $K_-^\nu$ is porportional to $n^\mu$, the second term 
in Eq.~(\ref{eq:polarization}) will not contribute 
after contraction of the Lorentz indices.

\end{itemize}

After implementing the above modifications, the code generated 
by \madgraph can be used to evaluate the matrix element
of processes involving gluon fragmentation into an S-wave 
color-octet heavy-quark pair plus other light partons.  
Gauge invariance and Lorentz invariance have been checked 
numerically. We also checked that the above procedure adapted 
to the case of fragmentation into a color-singlet heavy-quark pair 
reproduces numerically the squared amplitudes calculated analytically 
in Ref.~\cite{Artoisenet:2014lpa}.



%

\end{document}